\documentclass[lettersize,onecolumn]{IEEEtran}

\usepackage[utf8]{inputenc} % Gère l'encodage des caractères
\usepackage[T1]{fontenc}    % Gère l'encodage des polices
\usepackage[english]{babel} % Gère la langue du document (ici, l'anglais)
\usepackage{csquotes}       % Pour les guillemets intelligents, souvent utilisé avec babel

\usepackage{cite}           % Pour la gestion des citations
\usepackage{graphicx}       % Pour l'inclusion d'images
\usepackage{url}            % Pour les URL, utile même si hyperref le redéfinit
\usepackage{textcomp}       % Pour des symboles textuels supplémentaires (déjà présent dans votreliste)
\usepackage{manyfoot}       % Pour les notes de bas de page (déjà présent dans votre liste)
\usepackage{float}          % Fournit l'option de placement 'H' pour les figures/tableaux
\usepackage{stfloats}       % Pour un placement plus fin des flottants sur plusieurs colonnes
\usepackage{array}
\usepackage{float}
\usepackage{placeins}

\usepackage{amsmath,amssymb,amsfonts} % Essentiel pour les mathématiques
\usepackage{bm}             % Pour les symboles gras en mathématiques
\usepackage{braket}         % Pour la notation bra-ket de Dirac
\usepackage{mleftright}     % Pour un espacement correct avec \left et \right
\usepackage{blkarray}       % Pour la composition de matrices/tableaux par blocs

\usepackage{amsthm}         % Pour la définition des environnements de théorèmes
\newtheorem{theorem}{\textbf{Theorem}}[section]
\newtheorem{proposition}[theorem]{\textbf{Proposition}}
\newtheorem{lemma}[theorem]{\textbf{Lemma}}
\newtheorem{definition}[theorem]{\textbf{Definition}}
\newtheorem{remarks}[theorem]{\textbf{Remark}}
\newtheorem{corollary}[theorem]{\textbf{Corollary}}

\usepackage{multirow}       % Pour les cellules multi-lignes dans les tableaux
\usepackage{makecell}       % Pour les retours à la ligne dans les cellules de tableau (y compris les en-têtes)
\usepackage{booktabs}       % Pour des règles de tableau plus esthétiques
\usepackage{tabularx}       % Pour les tableaux avec des largeurs de colonnes ajustables
\usepackage{array}          % Pour des fonctionnalités améliorées dans les environnements tabular
\usepackage{longtable}      % Si la table est trop longue pour une seule page
\usepackage{subcaption}     % Pour les sous-figures et sous-tableaux

\usepackage{tikz}

\usepackage{algorithmic}    % Pour la pseudo-code algorithmique
\usepackage{algorithm}      % Pour l'environnement flottant des algorithmes
\usepackage{listings}       % Pour les listings de code

\usepackage{hyperref}

\def\BibTeX{{\rm B\kern-.05em{\sc i\kern-.025em b}\kern-.08em
    T\kern-.1667em\lower.7ex\hbox{E}\kern-.125emX}}

\newcommand{\C}{\mathbb{C}}
\newcommand{\F}{\mathbb{F}}
\newcommand{\gG}{\mathbb{G}}
\newcommand{\LL}{\mathbb{L}}

\newcommand{\N}{\mathbb{N}}
\newcommand{\R}{\mathbb{R}}
\newcommand{\Z}{\mathbb{Z}}

\newcommand{\bA}{\mathbf{A}}
\newcommand{\bB}{\mathbf{B}}
\newcommand{\bC}{\mathbf{C}}
\newcommand{\bG}{\mathbf{G}}
\newcommand{\bH}{\mathbf{H}}
\newcommand{\bI}{\mathbf{I}}

\newcommand{\bO}{\mathbf{O}}
\newcommand{\bP}{\mathbf{P}}
\newcommand{\bQ}{\mathbf{Q}}
\newcommand{\bR}{\mathbf{R}}
\newcommand{\bS}{\mathbf{S}}
\newcommand{\bU}{\mathbf{U}}
\newcommand{\bX}{\mathbf{X}}
\newcommand{\bY}{\mathbf{Y}}
\newcommand{\bZ}{\mathbf{Z}}

\newcommand{\bb}{\mathbf{b}}
\newcommand{\bc}{\mathbf{c}}

\newcommand{\bT}{\mathbf{T}}
\newcommand{\bu}{\mathbf{u}}
\newcommand{\bv}{\mathbf{v}}
\newcommand{\bw}{\mathbf{w}}

\newcommand{\wt}{\mathsf{wt}}

\newcommand{\oalpha}{\overline{\alpha}}
\newcommand{\oa}{\overline{a}}
\newcommand{\ob}{\overline{b}}
\newcommand{\oun}{\overline{1}}
\newcommand{\ozero}{\overline{0}}

\newcommand{\ok}{\overline{k}}
\newcommand{\op}{\overline{p}}
\newcommand{\oT}{\overline{T}}
\newcommand{\zeroL}{0_{\Z^2/\LL}}

\newcommand{\G}{ (\Z/n\Z, \, \,  \oun , \, \, \oalpha)}

\newcommand{\calC}{\mathcal{C}}
\newcommand{\cQ}{\mathcal{Q}}
\begin{document}

\title{(2,2)-GB Codes: Classification and Comparison with weight-4 Surface Codes}

\author{ \IEEEauthorblockN{François Arnault\IEEEauthorrefmark{1}, Philippe Gaborit\IEEEauthorrefmark{1}, Nicolas Saussay\IEEEauthorrefmark{1}} 

\IEEEauthorblockA{\IEEEauthorrefmark{1} XLIM, UMR 7252, Université de Limoges\\ 123, Avenue Albert Thomas, 87000 Limoges, France\\ \{arnault, philippe.gaborit, nicolas.saussay\}@unilim.fr}}

\maketitle

\begin{abstract}
Generalized Bicycle (GB) codes offer a compelling alternative to surface codes for quantum error correction. This paper focuses on (2,2)-Generalized Bicycle codes, constructed from pairs of binary circulant matrices with two non-zero elements per row. Leveraging a lower bound on their minimum distance, we construct three novel infinite families of optimal (2,2)-GB codes with parameters $ [[ 2n^2, 2, n ]] $, $ [[ 4r^2, 2, 2r ]] $, and $ [[(2t  + 1)^2 + 1, 2, \ 2t + 1 ]] $. These families match the performance of Kitaev’s toric code and the best 2D weight-4 surface codes, reaching known theoretical limits. In particular, the second family breaks a long-held belief by providing optimal even-distance GB codes, previously deemed impossible.

All  are CSS codes derived from Cayley graphs. Recognizing that standard equivalence relations do not preserve their CSS structure, we introduce a CSS-preserving equivalence relation for rigorous comparison of Cayley graph-based CSS codes. Under this framework, the first two families are inequivalent to all previously known optimal weight-4 2D surface codes, while the third family is equivalent to the best-known odd-distance 2D surface code. 

Finally, we classify all extremal, non-equivalent $(2,2)$-GB codes with length below 200 and present a comparison table with existing notable 2D weight-4 surface codes.
\end{abstract}


\section{Introduction}

\subsection{Motivation}
Quantum computers hold the promise of revolutionizing computation by efficiently solving problems that are intractable for standard computers, as demonstrated Shor's algorithm \cite{Shor95}. However, the inherent fragility of quantum information (decoherence) and the presence of environmental noise make quantum computers susceptible to errors during computation. These errors accumulate over time, potentially leading to incorrect syndrome measurements and erroneous results.

To enable long-duration quantum computation, quantum error correction is essential. 
This process involves employing quantum codes, syndrome measurement circuits, and decoding algorithms to detect and correct errors as they occur. As errors can accumulate even during syndrome measurement and decoding, the error correction process must be both fast and reliable. To ensure reliable quantum computation, the physical error rate must be kept below a critical threshold, allowing for effective error suppression \cite{NC2010}.

\subsection{Previous results:}
Kitaev codes, introduced in 2003, were the first quantum low-density parity-check (qLDPC) codes with minimum distance scaling linearly with the square root of the code length \cite{K03}. Since then, various generalizations of Kitaev codes have been developed in an effort to improve their parameters. Notable examples include Kitaev's optimized 45-degree rotated codes \cite{BD07}, which offer enhanced minimum distance, and surface codes, which provide higher rates.
However, neither of these generalizations has managed to surpass the square root scaling of the minimum distance.

Until recently, all known qLDPC codes were limited by a square root scaling of the minimum distance. However, in recent years, significant progress has been made, with the theoretical existence of asymptotically good qLDPC codes, possessing positive rate and linear minimum distance scaling, being established \cite{LZ22, PK22}.
While these codes offer superior theoretical performance, their practical implementation remains challenging compared to surface codes, which exhibit high error thresholds and efficient decoding algorithms.

\subsection{Contributions}

Currently, Generalized Bicycle (GB) codes, denoted as $(a,b)$-GB when constructed from circulant matrix pairs with $a$ and $b$ non-zero elements per row, stand out as one of the most promising alternatives to surface codes for practical implementation. A recent breakthrough by the IBM research team \cite{IBM24} highlighted their potential by introducing a family of (3,3)-GB codes that match the error thresholds of surface codes while significantly reducing encoding overhead. 

Within this framework, $(2,2)$-GB codes form a particularly appealing subclass. As generalizations of Kitaev codes and their optimized variants \cite{KP13}, they combine structural simplicity, planar layouts, and local weight-four stabilizers, features highly attractive for physical implementation. However, despite the identification of optimal odd-distance (2,2)-GB codes by Kovalev and Pryadko \cite{KP13}, the literature has largely lacked (2,2)-GB codes with a minimum distance scaling linearly with the square root of the code length, especially for even distances. \\

Drawing inspiration from Haah, Hastings, and O'Donnell's suggestion \cite{HHO21} that toric code geometry alterations could yield codes with improved minimum distance scaling, this paper introduces a novel approach for constructing high-performing (2,2)-GB codes. 

Our approach blends graph theory and arithmetic, viewing these codes as CSS codes whose stabilizer generator matrices are incidence matrices of undirected Cayley graphs over abelian groups with two generators. In this interpretation, $(2,2)$-GB codes correspond to Cayley graphs of the form $ (\Z/n\Z, \overline{a}, \overline{b}) $.

Within this perspective, we derive a general lower bound on the minimum distance of these codes, based on the shortest Manhattan norm of non-zero vectors in associated $ \Z^2$ lattices. This bound proved to be an indispensable tool, enabling the explicit construction of three infinite families of optimal weight-4 2D surface codes:
\begin{itemize}
    \item GB$(1 + X, 1 + X^{n}, n^2)$ of parameters $[[ 2n^2, 2, n ]]$
    \item GB$(1 + X, 1 + X^{2r - 1}, 2r^2)$ of parameters $[[ 4r^2, 2, 2r ]]$
    \item  GB$(1 + X, 1 + X^{2t + 1}, t^2 + (t + 1)^2 \, )$ of parameters $ [[(2t + 1)^2 + 1, 2, 2t + 1 ] \! ]$
\end{itemize}

It is worth noting that family with parameters $[[(2t + 1)^2 + 1, 2, 2t + 1 ] \! ]$ was previously identified by Kovalev and Pryadko as (2,2)-GB codes, rivalling the best 2D surface codes for odd distances \cite{KP13}. Their construction, $GB(1 + X^{2t^2 + 1}, X + X^{2t^2 - 1}, t^2 + (t + 1)^2)$, was also inspired by periodic lattices. Crucially, our construction of the $ [[ 4r^2, 2, 2r ]] $ family directly addresses a long-standing question: that no GB construction could yield optimal even-distance 2D codes with such parameters \cite{PW22}. This is particularly significant as the best previously known GB family for even distances, $ [[  4r^2 + 4, 2, 2r ]]$ existed only for even $r$. \\

Beyond merely achieving good parameters, the practical utility of quantum codes hinges on their decoding properties and feasibility of implementation. To effectively categorize and study these codes, appropriate notions of equivalence are essential, much like identifying equivalent problems in complexity theory. Here, all codes under study are CSS codes derived from Cayley graphs. However, the standard equivalence relation on quantum error-correcting codes, where $ \mathcal{Q} $ and $ \mathcal{Q}'$ are equivalent if $ \mathcal{Q}' = \bU\mathcal{Q}$ for $ \bU$ being a unitary operator is ill-suited for these codes rigorous comparison. 

This is because it fails to preserve either the CSS structure or the underlying graph structure, both of which are crucial given that established decoding algorithms for (2,2)-GB codes, such as the renormalization algorithm \cite{RZ23}, rely on these very properties.

To rigorously assess their novelty, we defined a tailored CSS code equivalence relation that preserves both the CSS and the underlying graph structures. Under this relation, we show that the first two families are inequivalent to any known Kitaev or even-distance surface codes (with initial simulations indicating distinct weight generator polynomials for their associated codes), while our third family offers an alternative realization of the optimal odd-distance 2D surface code.

While Pryadko and Wang had initiated a study on the performance of GB codes based on their weights, including  $GB(1 + X, 1 + X^a, n)$ with length $2n$ where $n$ is a integer for which $2$ is a primitive root \cite{PWsimu}, our work significantly advances prior research. This paper presents a complete study of all (2,2)-GB codes, providing a comprehensive comparison between our newly constructed (2,2)-GB codes and known local 2D surface codes built using periodic lattices.

Our study concludes with a complete classification of top-performing, non-equivalent (2,2)-GB codes with lengths under 200 and a comparison with notable 2D weight-4 surface codes. This culminating effort provides a rich dataset for future practical quantum code implementations, offering a valuable resource for identifying highly structured codes with optimal decoding properties. 

\subsection{High-Level Overview}
This section details our approach to developing new, optimal (2,2)-Generalized Bicycle (GB) codes and definitively proving their structural distinctness from existing 2D weight-4 high-performing surface codes.

\subsubsection{Construction}
Our method for constructing (2,2)-GB codes harnesses the unique interplay between topological features of Cayley graphs with two generators and their geometric structures:

\begin{itemize}
    \item \textbf{Graph Representation:} We interpret (2,2)-GB codes as quantum codes derived from unoriented Cayley graphs with two generators.
    \item \textbf{Lattice Association:} We define a corresponding 2D lattice, allowing us to map graph cycles to lattice elements.
    \item \textbf{Minimum Distance Guarantee:} We demonstrate that the code's minimum distance is lower bounded by $\lambda(\LL)$, representing the shortest Manhattan norm of a non-zero vector in the associated lattice.
    \item \textbf{Targeted Design:} We identify specific GB codes whose associated lattices $\LL$ feature a large $\lambda(\LL)$ that is also straightforward to compute. This strategy enables us to achieve desired minimum distances.
\end{itemize}

\subsubsection{Comparison}
To confirm the novelty of our newly constructed codes, we developed a robust comparison framework:

\begin{itemize}
    \item \textbf{Unified Perspective:} We provide a unified view of our GB codes and 2D surface codes with periodicity vectors by interpreting both as CSS codes derived from unoriented Cayley graphs on abelian groups.
    \item \textbf{Specialized Equivalence Relation:} We establish a new, specific equivalence relation for CSS codes that preserves the structure of their underlying graphs.
    \item \textbf{Graph Isomorphism Link:} For the codes under investigation, we show that code equivalence directly implies isomorphism of their underlying graphs.
    \item \textbf{Group Isomorphism Implication:} We further prove that this graph isomorphism necessitates the isomorphism of their associated abelian groups.
    \item \textbf{Distinctness Proof:} By conclusively demonstrating that the underlying abelian groups of our novel codes are not isomorphic to those of established optimal 2D weight-4 surface codes (like Kitaev codes), we rigorously prove their distinctness.
\end{itemize}

%%%% Organisation of the paper
\subsection{Organization of the paper }

This paper has eight sections. 

We start in Section~\ref{sec: notations and relevant facts} by covering the core concepts of quantum codes, including stabilizer, CSS, GB, and graph-based codes. Section \ref{section: Proof of the main theorem} then provides the proof for our minimum distance bound on (2,2)-GB codes. 

Next, Section \ref{section: application of the main theorem} details the explicit construction of (2,2)-GB codes that achieve parameters comparable to the best 2D weight-four surface codes. 
Section \ref{sec: equivalence rel for CSS}--\ref{sec: comparison 2D and GB} introduce the methodological framework used to compare our codes against these optimal surface codes. 

Finally, we conclude by presenting the results of our comprehensive classification of (2,2)-GB codes in Section \ref{sec: classification of (2,2)-GB codes} and a summarizing table of all top-performing 2D weight-4 surface codes in Section \ref{sec: summary of TP 2D weight-4 surface codes}.

%%% Reminders on quantum codes
\section{Notations and relevant facts}
\label{sec: notations and relevant facts}

Quantum error-correcting codes (QECCs) are defined as sub-vector spaces within the $n$-qubit complex vector space $(\mathbb{C}^2)^{\otimes n}$.

\subsection{Representing Qubits and Operators}

We typically use the standard basis for a single qubit:
$|0\rangle = \begin{pmatrix} 1 \\ 0 \end{pmatrix}$ and $|1\rangle = \begin{pmatrix} 0 \\ 1 \end{pmatrix}$.
For an $n$-qubit system, where each $t_i \in \{0, 1\}$, the computational basis vectors of the complex vector space $(\mathbb{C}^2)^{\otimes n}$ are commonly written as $|t_1\rangle \otimes\cdots \otimes |t_n\rangle$, or more simply as $|t_1 \dots t_n\rangle$ for conciseness.

Any unitary operator $\mathbf{U}$ acting on an $n$-qubit state is expressed as a combination of Pauli operators: \[i^m \bO_1 \otimes \bO_2 \otimes \dots \otimes \bO_n\] where $\bO_j \in \{\bI, \bX, \bY, \bZ\}$ are the identity and standard Pauli matrices, and $m \in \{0, 1, 2, 3\}$ accounts for a phase multiplier.

It is common practice to represent these Pauli operators using two binary strings, $ \bu, \bv \in \{0, 1\}^n$, by mapping $\bU = i^m \bX^\bu \bZ^\bv$ to the pair $(\bu, \bv)$. Here, the operator $\bX^\bu$ denotes $\bX_1^{u_1}  \otimes \dots \otimes \bX_n^{u_n}$ and $\bZ^\bv$ denotes $\bZ_1^{v_1}  \otimes  \dots \otimes \bZ_n^{v_n}$. When two quantum operators are multiplied, their corresponding binary pairs $(\bu, \bv)$ are simply added modulo 2.

\subsection{Stabilizer Codes}

An $[[n, k, d]]$ stabilizer code $\mathcal{Q}$ is a $2^k$-dimensional subspace of $(\mathbb{C}^2)^{\otimes n}$ defined by an Abelian stabilizer group  $\mathbb{S} = \langle \bG_1, \dots, \bG_{n-k} \rangle$, generated by $n-k$ commuting Pauli operators, and that does not contain the negative identity operator ($-\bI$). The associated code $\mathcal{Q}$ consists of all states $|\psi\rangle$ that remain unchanged when acted upon by any operator $S \in \mathbb{S}$: 
$$ \mathcal{Q} = \{|\psi\rangle \in (\C^2)^{\otimes n} : \bS |\psi\rangle = |\psi\rangle, \forall \bS \in \mathbb{S}\} $$
Each stabilizer code can also be represented by a binary generator matrix $ \bH = (\bA_X | \bA_Z)$. In this matrix, each row corresponds to a generator $\bG_i$, where the $\bu$ vector forms a row in $\bA_X$ and the $\bv$ vector forms a row in $\bA_Z$. For completeness, $\bH$ may include linearly dependent rows. The crucial condition for the stabilizer generators to commute is expressed in terms of these binary matrices as:
$$ \bA_X \bA_Z^\intercal + \bA_Z \bA_X^\intercal = 0 \pmod 2 $$

\subsection{CSS codes}

Calderbank-Shor-Steane (CSS) codes \cite{CS96, S96} are a crucial subclass of stabilizer codes.
They are characterized by stabilizer generators that are products of only $\bX$ or only $\bZ$ operators.
This structure allows them to be represented by a generator matrix:
\[ \bH = \begin{pmatrix}
    \bH_X & 0 \\ 0 & \bH_Z 
\end{pmatrix}\]
where $\bH_X$ and $\bH_Z$ are two binary matrices whose row spaces are orthogonal, i.e., \mbox{$\bH_X \bH_Z^\intercal = \mathbf{0}$}. 

The code length $n$, representing the number of physical qubits, is the number of columns of $\bH_X$ (or of $\bH_Z$). The code dimension $k$, corresponding to the number of logical qubits, is given by $k = n - \mathrm{rank} \, \bH_X - \mathrm{rank} \, \bH_Z$. The minimum distance $d$ of the code is defined as the minimum of $d_X$ and $ d_Z$, where $d_X$ (resp. $d_Z$) denotes the smallest Hamming weight of a codeword of $\ker \bH_X$ (resp. $\ker \bH_Z$) that is not in $ rs(\bH_Z) $, the row space of $\bH_Z$ (resp. not in $rs(\bH_X$) the row space of $ \bH_X$).

\subsection{\texorpdfstring{CSS associated to an unoriented Cayley graph }{CSS associated to an unoriented Cayley graph }}
\label{sec: css associated to a graph (G, g_1, g_2)}

\subsubsection{\texorpdfstring{Graph construction $(\gG, g_1, g_2)$}{Graph construction (G, g1, g2)}}

\begin{definition}
    \label{def: graph (G,g_1, g_2)} Let $ \gG$ be an abelian group and $g_1, g_2 \in \gG$ be two group elements. 
We construct a graph, denoted by $ (\gG, g_1, g_2)$, where the vertices correspond to the elements of $\gG$. In this graph, each vertex $ g \in \gG $ is connected to exactly four neighbours:  $ g + g_1$ , $ g - g_1$, $ g + g_2$ and $ g - g_2$. This count of four distinct connections holds even if some of these operations result in the same group element.
\end{definition}

\begin{figure}[ht!] % [h!] tente de placer la figure "ici"
    \centering % Centre l'ensemble des sous-figures
        \includegraphics[scale=0.35]{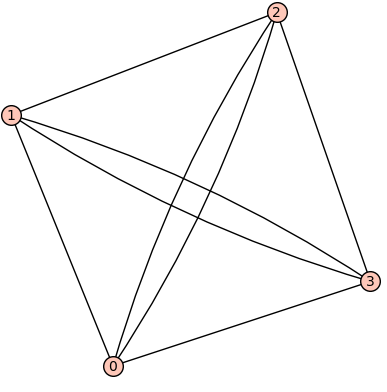}
        \caption{$(\Z/4\Z, \overline{1}, \overline{2})$ graph} % Légende spécifique à cette image
        \label{fig:graphZ4} % Label pour référencer cette sous-figure
\end{figure}

\begin{definition}[Cocycles and faces of $(\gG,g_1, g_2)$]
    Let $ (\gG,g_1, g_2) $ be a graph constructed as in Definition \ref{def: graph (G,g_1, g_2)} 
    \begin{itemize}
        \item  For any vertex $ g \in \gG$, we define its associated cocycle (in the context of this graph's construction) as the multiset consisting in the edges connecting $g$ to its four neighbours: ${ \{g, g + g_1\}, \{g, g - g_1\}, \{g, g + g_2\}, \{g,  g - g_2\} }$. \\

Note: This definition treats edges as distinct even if they connect to the same element, ensuring the multiset has four elements. \\

        \item For any vertex $ g \in \gG$, we define its associated face (in the context of this graph's construction)  as the multiset composed of the edges forming the $4$-cycle connecting the vertices $ g$, $ g + g_1$, $ g + g_1 + g_2 $ and $ g + g_2$: $ \{g, g+g_1\},  \{g+g_1, g+g_1+g_2\}, \{g+g_1+g_2, g+g_2 \}$, $ \{g+g_2, g\} $
    \end{itemize}
\end{definition}

\subsubsection{\texorpdfstring{Incidence matrices associated to $(\gG, g_1, g_2)$}{Incidence matrices associated to (G, g1, g2)}}

\label{sec: incidence matrix of (G, g_1, g_2)}

Consider a graph $ (\gG,g_1, g_2) $ as defined in Definition \ref{def: graph (G,g_1, g_2)}.
For the purposes of constructing incidence matrices, we fix an ordering of its vertices, edges, and faces.

\begin{definition}[Incidence matrices associated with $ (\gG,g_1, g_2)$]
\label{def: incidence_m of (G,g_1, g_2)}
We define the following incidence matrices over $ \F_2$:
\begin{itemize}
    \item $ \bH_X(\gG_,g_1, g_2) $: the vertex-edge incidence matrix of the graph $ (\gG, g_1, g_2) $. 
    \item $ \bH_Z (\gG, g_1, g_2) $: the face-edge incidence matrix of the graph  $ (\gG, g_1, g_2) $.  
\end{itemize}
\end{definition}

\subsubsection{\texorpdfstring{Construction of $CSS(\gG, g_1, g_2)$}{Construction of CSS(G, g1, g2)}}

A key property of the graph $ (\gG,g_1, g_2) $ is that every $ F$ face forms a cycle. Indeed, each vertex of the graph has an even number of edges of $F$ incident to it.  Consequently, this implies a fundamental relationship between the two incidence matrices: $ \bH_X(\gG,g_1, g_2) \cdot \bH_Z(\gG,g_1, g_2)^\intercal = \mathbf{0} $.

\begin{definition}
    \label{def: css(G,g_1,g_2) }
CSS$(\gG, g_1, g_2$) is defined as the CSS code whose stabilizer generator matrix is \[ \begin{pmatrix}
     \bH_X(\gG,g_1, g_2) & \mathbf{0} \\
     \mathbf{0} &  \bH_Z(\gG,g_1, g_2)
\end{pmatrix} \]
\end{definition}

\begin{remarks}
\label{remarks: different orderings in (G,g_1, g_2)}
Different orderings  of the vertices, edges and faces of the graph $ (\gG,g_1, g_2) $  will yield pairs of incidence matrices $ (\bH_X, \, \bH_Z)  $ and  $(\bH_X', \, \bH_Z')$. These pairs are related by permutation matrices $ \bP, \bQ, \bR$ such that: $ \bH_X' = \bP\bH_X\bQ$ and $ \bH_Z' = \bR\bH_Z\bQ$. 
In particular, the two CSS codes $ CSS(\bH_X, \bH_Z) $ and $ CSS(\bH_X', \bH_Z') $ have the same parameters.
\end{remarks}

% Lattices
\subsection{Background on lattices}
\label{sec: background on lattices}

\subsubsection{Lattices} A 2D lattice is a set
$\LL = \Z e_1 \oplus \Z e_2 \subseteq \R^2 $, where $ \{ e_1, \, e_2\} $ is an $\R$-basis of $ \R^2$. Its determinant $ det(\LL) =  |det(e_1, \, e_2)| $ only depends on $\LL$ and not on the choice of basis $\{e_1, e_2\}$. When $ \LL $ is a 2D sublattice of $\Z^2 $, its determinant is equal to the cardinality of the quotient $ \Z^2 / \LL $.

\subsubsection{Optimal surface codes}
\label{sec: optimal surface codes}
In their 2008 study of homological codes, Bombin and Delgado \cite{BD07} introduced an optimal family that significantly outperforms the original toric codes. This construction provides a specific code for each given distance:

\begin{itemize}
    \item Odd-distance optimal 2D surface code, with parameters $ \left[[ (2t + 1)^2 + 1, 2, 2t + 1 ] \!\right]$, achieved by a family of square lattice surface codes with periodicity vectors $\begin{pmatrix} t + 1 \\ t \end{pmatrix}, \begin{pmatrix} -t \\ t + 1 \end{pmatrix}$ where $t \geq 1$. \\

    \item  Even-distance optimal $ 2D$ surface code, of parameters $ \left[[  4r^2, 2, 2r |\right]$ achieved by $45$-degree-rotated surface codes with periodicity vectors $\begin{pmatrix}
        r  \\ r
    \end{pmatrix}, \begin{pmatrix}
        - r \\ r  
    \end{pmatrix}$. \\
\end{itemize}

Under the formalism introduced earlier, these codes correspond to Cayley-graph based CSS codes constructions:
\begin{itemize}
    \item Odd distance: $ CSS( \Z^2/\LL_t, \begin{pmatrix}
    1 \\ 0
\end{pmatrix} \mod \LL_t,\begin{pmatrix}
    0 \\ 1
\end{pmatrix} \mod \LL_t) $ 
where $ \LL_t = \Z \begin{pmatrix}
    t + 1 \\ t
\end{pmatrix}\bigoplus\Z \begin{pmatrix}
    -t \\ t   + 1
\end{pmatrix}$. \\

\item Even distance:
$ CSS( \Z^2/\LL_r, \begin{pmatrix}
    1 \\ 0
\end{pmatrix} \mod \LL_r,\begin{pmatrix}
    0 \\ 1
\end{pmatrix} \mod \LL_r) $ where $ \LL_r = \Z \begin{pmatrix}
 r  \\ r
\end{pmatrix}\bigoplus\Z \begin{pmatrix}
    r \\ - r
\end{pmatrix}$.

\end{itemize}

\subsection{Introduction to GB codes}

\subsubsection{Circulant matrices}

For an integer $ n \geq 1 $ and a polynomial $A(X) = \sum_{i=0}^{n-1} a_i X^i \in \F_2[X]_{\leq n - 1} $, the circulant matrix $Circ(A(X), \, n)$ is given by:
\begin{equation*}
Circ(A(X), \, n)  = \begin{pmatrix} 
a_0     & a_{n-1} & \dots  & a_{1}  \\
a_{1}   & a_0     & \ddots & \vdots \\
\vdots  & \ddots  & \ddots & a_{n - 1} \\
a_{n-1} & \dots   & a_{1}  & a_0
\end{pmatrix}.
\end{equation*}

The algebra of $ n \times n$ circulant matrices
over $ \F_2$ is isomorphic to that of polynomials
in $ \F_2[X]/(X^n - 1)$. This isomorphism arises because any circulant matrix can be expressed as a polynomial in the companion matrix of $X$. A direct consequence is that the product of any two circulant matrices is commutative. 

Throughout this article, we consider vectors in $ \F_2^n$
as columns. Under this convention, the
product $Circ(A(X),n)\bb$, where $ \bb = \begin{pmatrix}
    b_0 \\ \vdots \\ b_{n-1}
\end{pmatrix}$, directly
corresponds to the polynomial product:
\[  (\sum_{i = 0}^{n-1}a_iX^i) \cdot (\sum_{i = 0}^{n-1}b_iX^i) \mod  X^n  - 1 \]

\subsubsection{Generalized bicycle (GB) codes}
\label{subsec: intro GB codes}

The concept of generalized bicycle (GB) codes, introduced by Pryadko and Kovalev in 2013 \cite{KP13}, extends the Bicycle Codes previously studied by MacKay et al. \cite{MMM04}. While traditional Bicycle codes are defined by a single circulant matrix $\bA$ in their stabilizer generator matrices, specifically $\bH_X = \bH_Z = [\bA, \bA^\intercal]$, GB codes utilize two distinct circulant matrices.

For binary polynomials $A(X), B(X) \in \F_2[X]_{\leq n-1}$, the GB code $GB(A(X),B(X),n)$ is constructed with stabilizer generator matrices $\bH_X = [\bA, \bB]$ and $\bH_Z = [\bB^{\intercal}, \bA^{\intercal}]$, where $\bA = Circ(A(X), n)$ and $\bB = Circ(B(X), n)$. The code has a length of $2n$ and a dimension given by  $ 2 \deg (\gcd(A(X), B(X), X^n - 1))$. Since $\bH_Z$ can be obtained from $\bH_X$ through row and column permutations, the minimum distance of the code satisfies $ d = d_X = d_Z$ \cite{PW22}. \\

The isomorphic relationship between circulant matrices and polynomials over $\F_2[X]/(X^n - 1)$ allows us to express the code conditions of $ GB(A(X), B(X),n)$ in the polynomial domain. For a column vector $ \bc = \begin{pmatrix}
    u \\ v
\end{pmatrix} \in \F_2^{2n}$ with  $U(X)$ and $V(X) \in \F_2[X]_{\leq n-1}$ the polynomials associated with $\bu$ and $\bv$, respectively, the following holds true: 
\begin{itemize}
    \item $\bc$ belongs to the kernel of $\bH_X$ if and only if its associated polynomials satisfy the equality 
    \[  A(X)U(X) + B(X)V(X) \equiv  0 \mod X^n - 1 \]

    \item $\bc$ belongs to the row space of $\bH_Z$ if and only if there exists a polynomial $H(X) \in \F_2[X]$ such that:
    \begin{equation*} 
\left\{
\begin{aligned}
    U(X) &\equiv B(X)H(X) &\mod X^n - 1\\
    V(X) &\equiv A(X)H(X)  &\mod X^n - 1
\end{aligned}
\right.
\end{equation*}
\end{itemize}

In this work, we aim to maximize the minimum distance of (2,2)-GB codes generated by two binary polynomials $A(X)$ and $ B(X)$ with exactly two non-zero terms. 

As the next proposition summarizes, in our study of large-distance (2,2)-GB codes, we only need to investigate cases where $ A(X), B(X) \in \F_2[X]_{\leq n - 1}$ have non-zero constant terms.

%%% CODES GB EQUIVALENTS
\begin{proposition}  
\label{prop:equivalent_GB_codes}
Let $ n > 0 $ and $ A(X), B(X) \in \F_2 \left[X\right]_{\leq n - 1}$ be two polynomials. 

\begin{itemize}
    \item If $ k$ is an integer coprime with $n$, then $ GB(A(X),B(X),n)$ and 
    $GB( A(X^k) \mod  X^n - 1, B(X^k) \mod X^n - 1, n)$ have the same parameters \cite{PW22}. In particular, for distinct integers $r , s\geq 1 $, if $ r $ is coprime with $n$, then
 \begin{equation*}
      d(GB(1 + X^r, 1 + X^{s}, \,  n)) = d(GB(1 + X, 1 + X^\alpha, \, n))
 \end{equation*}
 where $ \alpha =   st \mod  n  $ with $ t = ( r \mod \, n)^{-1} $.

    \item For any integers $ i, j  \in \N $ the codes $GB(A(X),B(X),n)$ and
    $GB(A(X)X^i \mod X^n - 1, B(X)X^j \mod X^n - 1,n )$  have the same parameters.
\end{itemize}

\end{proposition} 

%%% Proof of proposition 3.1:
\textbf{Proof of the second point:}
Let $ R_i(X) $ and $S_j(X)$ be the residue of $A(X)X^i$ and $ B(X)X^j$ modulo $ X^n - 1$. We set 
$ \bA = Circ(A(X),n), \, \bB = Circ(B(X),n)$ and $ \bC =  Circ(X,n)$. The map $ h$ defined by
\begin{align*}
h: \F_2^{2n} &\to \F_2^{2n}\\
  \begin{pmatrix} \bu \\ \bv \end{pmatrix} &\mapsto \begin{pmatrix} \bC^i \bu \\ \bC^j \bv \end{pmatrix}
\end{align*}

 induces  isometries (for the Hamming weight) between:

\begin{itemize}
    \item the right kernels of $  \left[ \bA  , \, \bB \right] $ and $  \left[ \bC^i \bA, \,  \bC^j \bB \right] $ 
    \item the row spaces of $  \left[ \bB^\intercal , \,  \bA^\intercal \right] $ and the row space of $ \left[ (\bC^j \bB)^\intercal , \,  (\bC^i \bA)^\intercal \right]$ 

\end{itemize}

Since $ Circ(R_i(X), n )= \bC^i \bA$ and $ Circ(S_j(X), n ) = \bC^j \bB$ then the two codes $ GB(A(X), B(X), n)$ and $GB(R_i(X), S_j(X), n)$ have the same parameters.  \qed \\

The next section presents the theoretical foundation for our lower bound on the minimum distance of (2,2)-GB codes

% Lattice associated to a (2,2)-GB code
\subsection{Study of (2,2)-GB code}

 Let $ A(X) =  1 + X^{a} \in \F_2[X]_{\leq n - 1}$ and $ B(X) =  1 + X^{b} \in \F_2[X]_{\leq n - 1}$ be two polynomials over $ \F_2$. We define their associated circulant matrices as $ \bA = Circ(A(X), n)$ and $ \bB = Circ(B(X), n)$. 

\subsubsection{\texorpdfstring{Abstract graph associated with $GB(A(X), B(X), n)$}{Abstract graph associated with GB(A(X), B(X), n)}}
\label{sec: abstract graph associated to $GB(A(X), B(X), n)$}

The stabilizer generator matrices of $GB(A(X), B(X), n)$ can be directly interpreted in the context of the graph 
$ ( \Z/n\Z, \overline{a}, \, \overline{b}) $  (as defined in Definition \ref{def: graph (G,g_1, g_2)}). 

Here, the matrix $ \bH_X  = \left[ \bA , \, \bB \right]$ corresponds to the vertex-edge incidence matrix of the graph, while the matrix $ \bH_Z  = \left[ \bB^\intercal , \, \bA^\intercal \right]$ corresponds to the face-edge incidence matrix of the graph. From this perspective, $ GB(A(X), B(X), n)$ is precisely, the CSS code $ CSS( \Z/n\Z, \overline{a}, \, \overline{b})$. \\

While this abstract perspective will be crucial for distinguishing equivalent codes, representing $ GB(A(X), B(X), n)$ on concrete  2D surface lattice proved vital in the research of (2,2)-GB codes that outperform Kitaev codes. In the following sections, we explain how to build such representation.

\subsubsection{\texorpdfstring{Lattice associated with $GB(A(X), B(X), n)$}{Lattice associated with GB(A(X), B(X), n)}}

 The 2D lattice associated with $ GB(1 + X^a, 1 + X^b, n) $ is given by the kernel of the $ \Z$-linear application  
\begin{equation*}
   \begin{array}{ccccl}
\Phi  & : & \Z^2 & \to & \Z / n\Z\\
 & & (u, v)  & \mapsto & au  + bv \mod  n  
\end{array} 
\label{eq:Phi}
\end{equation*}

In the special case where $A(X) = 1 + X$ and $B(X) = 1 + X^\alpha$, the corresponding lattice is  $ \Z \begin{pmatrix}
    n \\ 0 
\end{pmatrix}\bigoplus \Z \begin{pmatrix}
    \alpha \\ - 1
\end{pmatrix}$.  With this mapping, the code $GB(A(X), B(X), n)$ can thus be interpreted as a 2D surface code defined on a torus with periodicity vectors $ \begin{pmatrix} n \\ 0 \end{pmatrix} $ and $ \begin{pmatrix} \alpha \\ -1 \end{pmatrix} $. \\

This next proposition shows that such code's minimum distance is at least the smallest Manhattan norm of any non-zero vector in its corresponding lattice.

\subsubsection{Lower bound on the minimum distance of (2,2)-GB codes}

% Theoreme 2.2 : d(GB(1 + x, 1 + x^alpha,n)) >= norme(L) 
\begin{theorem}%
\label{THM:MAIN_THM}%
Let $ n \geq 6 $ and $ 1  \leq \alpha  \leq n -1$. The minimum distance $ d_{min} $ of the Generalized Bicycle code $ GB(1 + X, \, 1 + X^\alpha,\, n)$ satisfies the following lower bound:
    \begin{equation*}
        d_{min} \geq \lambda ( \LL) = \min \{ \, ||\bu||_1 \, \, \vert \, \, \bu \in \LL \setminus \{ 0 \} \} 
    \end{equation*}
where $ \LL = \Z \begin{pmatrix}
    n \\ 0 
\end{pmatrix}\bigoplus \Z \begin{pmatrix}
    \alpha \\ - 1
\end{pmatrix} $ is a $ \Z^2$-sublattice and $ || \cdot ||_1 $ denotes the Manhattan norm.
\end{theorem}

This connection can be generalized to a broader class of GB codes, as shown in the following corollary.

 % Corollary of the main theorem
\begin{corollary} 
\label{corollary: corollary of main thm}

Let $ A(X) =  1 + X^u $ and $ B(X) =  1 + X^{v} $ be two polynomials of weight two. If $n$ is an integer such that $ n > max(6, u, v) $ and $u$ is coprime with $n$, then the minimum distance of $ GB(A(X), B(X), n)$ satisfies: 
\begin{equation*}
    d(GB(A(X), B(X), n)) \geq \lambda (\Z \begin{pmatrix}
    n \\ 0 
\end{pmatrix}\bigoplus \Z \begin{pmatrix}
    \alpha \\ - 1
\end{pmatrix}) 
\end{equation*}
where $\alpha =  (  v  \mod  n)\cdot (u  \mod  n)^{-1} $
\end{corollary}

%%%% Bicycle codes and graph theory
\section{Proof of Theorem~\ref*{THM:MAIN_THM}}
\label{section: Proof of the main theorem}

Throughout this section, we adopt the following notations, consistent with Theorem~\ref{THM:MAIN_THM}:
\begin{itemize}
    \item Let $ n \geq 6$ be an integer, and $ 1 \leq \alpha \leq n - 1   $.
    \item  Define $ A(X) = 1 + X $, and its associated circulant matrix $ \, \bA = Circ(A(X),n)$.
    \item  Let $ B(X) = 1 + X^\alpha$, and its associated circulant matrix $\, \bB  = Circ(B(X),n) $.
    \item Let $\LL = \Z \begin{pmatrix}
    n \\ 0 
\end{pmatrix}\bigoplus \Z \begin{pmatrix}
    \alpha \\ - 1
\end{pmatrix} $ be the lattice associated to $ GB(A(X), B(X), n)$.
\end{itemize}

As stated earlier in Proposition \ref{prop:equivalent_GB_codes}, the minimum distance of $ GB(A(X), B(X), n)$ is invariant under cyclic permutation of the columns of $ \bA$  or $\bB$. This allows us to simplify our analysis by assuming that $ 1 \leq \alpha \leq \frac{n}{2}$.  While Theorem \ref{THM:MAIN_THM} holds straightforwardly for $ \alpha = 1 $ or $\alpha = \frac{n}{2}$, our focus for the non-trivial cases, $ 1 < \alpha < \frac{n}{2}$, will involve a detailed graph-theoretic analysis. 

Our proof strategy consists of two main steps: 
\begin{itemize}
    \item \textbf{Graph reinterpretation:} We construct a graph on the vertices of $ \LL$. This construction ensures that each vector $ \bv $ of $ \ker [\bA, \bB]$  corresponds to a cycle $ \tilde{\bv} $ within this graph, of length equal to $  \wt(\bv) $ the Hamming weight of $\bv $. 
    
    \item \textbf{Lower bound on the cycle length:} We then prove that if $ \bv $ does not belong to the row-space of $[\bB^{\intercal}, \bA^{\intercal}]$, then the length of its corresponding cycle $ \tilde{\bv}$ is lower bounded by $ \lambda(\LL) $, the shortest Manhattan norm of a non-zero vector in the lattice associated to $GB(A(X), B(X), n)$.
\end{itemize}

% GB codes and graphs theory

\subsection{Reinterpretation of the minimum distance}
\label{sec: Reinterpretation of the minimum distance}

As previously stated in Section \ref{sec: abstract graph associated to $GB(A(X), B(X), n)$}, $ \bH_X$ and $ \bH_Z$, the stabilizer generator matrices of $GB(A(X), B(X) ,n)$ can be directly interpreted as the vertex-edge and face-edge incidence matrices of the graph
$ ( \Z/n\Z, \overline{1}, \, \overline{\alpha}) $.  The $  2n$ edges of this graph are defined as follows: for $ 0 \leq k < n$
 $ h_k $, the $k$-th edge, connects vertices $ \overline{k}$ and $ \overline{k} + \overline{1}$, and $ h_{n +k }$,  the $(n + k)$-th edge, connects vertices  $  \overline{k}$ and $ \overline{k} + \overline{\alpha} $.
 
 Because $ 1 < \alpha < \frac{n}{2}  $, this construction ensures a simple graph where $ h_k$ and $ h_{n + k}$ are always unique, as they connect to different vertices. \\

The set $ \F_2^{2n}$ is identified with the power set of edges from our graph. Under this identification, any vector $\bv  $ corresponds to the subset of edges $   e_{\bv} = \{ h_i \, \vert \, 0 \leq i < 2n, \, v_i = 1 \}$, which contains  $ \wt(\bv)$ elements. In the subsequent sections, we will refer to $ \bv$ as the characteristic vector of $ e_\bv$ and  any mention of the sum of edges will refer to the sum of their corresponding characteristic vectors.

% REINTERPRETATION DE rs(H_x) && Ker(H_X) 
\subsubsection{\texorpdfstring{Reinterpretation of $rs(\mathbf{H}_X)$ and $\ker \mathbf{H}_X$:}{Reinterpretation of rs(HX) and ker(HX):}}

\label{sec: reinterpretation of rs(H_X) and Ker(H_X)}

For any integer $ p \in \{0, \dots, n -1\} $, the $ p^{\text{th}}$ row of $\bH_X$ is defined as cocycle($p$). This row corresponds to the set of edges connecting $\op$ to its four neighbours:
\begin{equation*}
\{ \, \{ \op, \, \, \op + \oun \},  \, \{\op, \, \, \op - \oun\}, \, \{\op, \, \,  \op + \oalpha\},  \, \{\op, \, \,  \op - \oalpha\} \, \}
    \label{eq:cocycle(p)}
\end{equation*}

The cycles, the elements of $\ker \bH_X$, are orthogonal to all cocycles. Each cycle $\calC$ is characterized by the property that each vertex $ \overline{v} \in \Z / n\Z$ belongs to an even number of edges of $\calC$.

% Définition des faces
\subsubsection{\texorpdfstring{Reinterpretation of $ rs(\bH_Z)$ and $ \ker \bH_Z$:}{Reinterpretation of rs(HZ) and ker(HZ):}}

\label{sec: reinterpretation of rs(H_Z) and Ker(H_Z)}

For any integer $ p \in \{0, \dots, n -1\} $, the $ p^{\text{th}}$ row of $\bH_Z$  is defined as face($p$). This row corresponds to the set of edges of the $ 4$-cycle connecting $\op$, $\op + \oun$, $ \op + \oalpha + \oun $ and $ \op + \oalpha$:
\begin{equation*}
    \, \{ \, \{\op, \, \, \op + \oun \}, \,  \{\op + \oun, \, \, \op + \oun + \oalpha \}, \, \{\op + \oun + \oalpha , \, \, \op + \oalpha \}, \,  \{\op + \oalpha, \, \, \op\} \, \} 
    \label{eq:face(p)}
\end{equation*}

\subsubsection{Reinterpretation of the minimum distance}

The minimum distance of $GB(A(X), B(X), n)$ corresponds to the shortest length of a cycle of $ \G$ that cannot be expressed as a sum of faces. This implies that such a cycle must be simple (containing no proper sub-cycles) and, by extension, connected.

\subsection{Heart of the Proof:}

We decompose the proof of the minimum distance lower bound into three main steps:
\begin{itemize}
\item We first establish a correspondence between walks on the graph $ (\Z/n\Z, \, \,  \oun , \, \, \oalpha)$ and walks in the lattice $ \Z^2 $. This mapping allows us to associate any cycle $ \calC$ of the graph to a lattice element $ \bc \in \LL$ such that the Manhattan norm of $ \bc$ is less than or equal to the length of $ \calC$.

    \item We then demonstrate that if a cycle $ \calC$ is not a sum of faces of the graph, its corresponding lattice element $ \bc $ must be non-zero, i.e., $c\neq \begin{pmatrix}
        0 \\ 0
    \end{pmatrix} $

    \item From these steps, we can directly deduce that the minimum distance of $GB(A(X), B(X), n)$ is greater than or equal to the shortest Manhattan norm of a non-zero vector in $ \LL$, i.e., $ d(GB(A(X), B(X), n)) \geq \lambda(\LL)$.
\end{itemize}

% Link between the graph and the lattice
\subsection{\texorpdfstring{From $ \G$-walks to $ \Z^2$-walks}{From G-walks to Z²-walks}}
\label{sec: from G-walks to Z^2-walks}

\subsubsection{Walks and Edge Orientation}
A walk $ \calC$ of length $ r \geq 1$ in $ \G $ is a sequence of edges  $ e_0, \dots,  e_{r-1} $ such that consecutive edges share a common vertex. When $ e_i = \{ C_i,  C_{i +1} \} $ for $ 0 \leq i \leq r - 1$, we can specify the orientation of $ \calC$ by writing it as a sequence of vertices:
\begin{equation*}
 \calC  :  C_0 \xrightarrow{} \dots \xrightarrow{}  C_{r}
\end{equation*}

Along any given walk in $ \G$ we can encounter exactly four different types of directed steps (or oriented edges), corresponding to the connections between vertices: 
\begin{gather*}
    \ok  \, \, \xrightarrow[]{+ \oun} \, \, \ok + \oun \quad \text{or }\quad  \ok  + \oun \, \, \xrightarrow{- \oun } \, \, \ok \\ 
\ok  \, \, \xrightarrow{ + \oalpha} \, \, \ok  + \oalpha, \quad \text{or }\quad \ok + \oalpha  \, \, \xrightarrow{- \oalpha} \, \,  \oalpha 
\end{gather*}

\subsubsection{\texorpdfstring{Construction of associated $\Z^2$-walks}{Construction of associated Z2-walks}}
\begin{figure}[ht!] % [h!] tente de placer la figure "ici"
    \centering % Centre l'ensemble des sous-figures
        \includegraphics[scale=0.72]{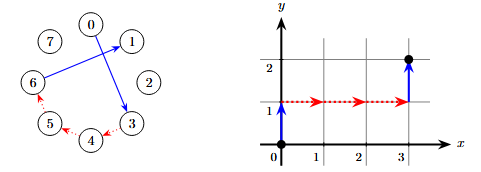}
        \caption{Mapping a walk in the Cayley graph \((\mathbb{Z}/8\mathbb{Z}, \oun, \overline{3})\) (left) to a path in the grid \(\mathbb{Z}^2\) (right).} % Légende spécifique à cette image
        \label{fig: from G-walk to Z^2 walk} % Label pour référencer cette sous-figure
\end{figure}

From any walk $ \calC : \, C_0 \xrightarrow{} \dots \xrightarrow{}  C_{r}$ in our graph $\G$, we define an associated walk in $ \Z^2$, denoted $ \gamma_\calC$. This walk precisely mimics the behaviour of $ \calC$ by tracking the net displacement caused by each step. (see Figure \ref{fig: from G-walk to Z^2 walk} )

The walk $\gamma_\calC$ is defined by a sequence of points  $ \bP_0 \xrightarrow{} \bP_1 \xrightarrow{} \dots \xrightarrow{} \bP_r $ in $ \Z^2$, starting from $ \bP_0 =  \begin{pmatrix}
    0 \\ 0
\end{pmatrix} $. For each step from $ C_i$ to $C_{i+1}$
  in $ \calC$ the corresponding point $\bP _{i+1}$ in $ \gamma_\calC$ is determined by adding a vector $\bv$ to $\bP_i$, which depends on the type of edge traversed in $ \calC$. 
\begin{itemize}
    \item If the step is $ C_i \to C_{i +1} = C_i + \oun $, we increment the $x$-coordinate: $ \bv= \begin{pmatrix}
        1 \\ 0
    \end{pmatrix}$.

    \item If the step is $ C_i \to C_{i +1}= C_i - \oun $, we decrement the $x$-coordinate: $ \bv = - \begin{pmatrix}
        1 \\ 0
    \end{pmatrix}$.

        \item If the step is $ C_i \to C_{i +1} = C_i + \oalpha $, we increment the $y$-coordinate: $ \bv =  \begin{pmatrix}
        0 \\ 1
    \end{pmatrix}$.

    \item If the step is $ C_i \to C_{i +1} = C_i - \oalpha $, we decrement the $y$-coordinate: $ \bv = - \begin{pmatrix}
        0 \\ 1
    \end{pmatrix}$.
\end{itemize}

\subsubsection{\texorpdfstring{Endpoint of $ \Z^2$ walks associated with $ \G$ cycles}{Endpoint of Z² walks associated with G-cycles}}

When the original walk $ \calC $ forms a connected cycle in the graph $ \G$, its associated $ \Z^2$-walk $ \gamma_\calC$ exhibits a specific and unique significant property.

% Proposition 
\begin{proposition}
     \label{prop:link_between__connected_cycle_and_L}

Let $ \calC \, : \, C_0 \xrightarrow{} \dots \xrightarrow{}  C_{r}  $ be a walk in $ \G$. The end point of $ \gamma_\calC$, $\bP_r$, represents the net displacement vector accumulated during the walk. Its Manhattan norm, $ \|\bP_r\|_1$, is less than or equal to the length of $ \calC$.

The coordinates of $\bP_r $ are given by:
$\bP_r = \begin{pmatrix} n_{\oun}(\calC) - n_{- \oun}(\calC) \\
    n_{\oalpha}(\calC) - n_{-\oalpha}(\calC)
\end{pmatrix}$ where for $ \epsilon \in \{ \pm \, \,  \oun , \, \, \pm \, \, \oalpha  \} $, $ n_{\epsilon} (\calC)$ denotes the number of times the walk $\calC$ traverses an edge in the direction corresponding to a step of $ \epsilon$:
\begin{equation*}
       n_{\epsilon} (\calC) =  |\{\, \,   i \in [[0, r  ]] \, \,  | \,  \, C_{i+1}  - C_i = \epsilon \, \, \}| 
   \end{equation*}

Furthermore, if $ \calC $ is a connected cycle within  $ \G$, then the endpoint $ \bP_r$ is guaranteed to be an element of the associated lattice $\LL = \Z\begin{pmatrix}
    n \\ 0 
\end{pmatrix} \bigoplus \Z\begin{pmatrix}
    \alpha \\ - 1
\end{pmatrix}$.
\end{proposition}

% Proof of Proposition 
\textbf{Proof:} Recall that  $ \LL  = \Z\begin{pmatrix}
    n \\ 0 
\end{pmatrix} \bigoplus \Z\begin{pmatrix}
    \alpha \\ - 1
\end{pmatrix}$ consists of all vectors $ \begin{pmatrix}
    x \\ y
\end{pmatrix} \in \Z^2$ satisfying $ x + \alpha y \equiv 0 [n]$.

The endpoint $ \bP_r  = \begin{pmatrix}
    n_{\oun}(\calC) - n_{-\oun}(\calC) \\
     n_{\oalpha}(\calC) - n_{-\oalpha}(\calC)
\end{pmatrix}$ of the associated $ \Z^2$ walk $ \gamma_\calC$ has a Manhattan norm $ \| \bP_r \|_1$ that is always less than or equal to the length of  $ \calC$, which is $ n_{\oun}(\calC) +  n_{-\oun }(\calC) +  n_{\oalpha}(\calC) +  n_{-\oalpha}(\calC)$. 

If $ \calC : \, C_0 \xrightarrow{}  \dots  \xrightarrow{}   C_{r} = C_0$ is a connected cycle of $\G$, then  $ S =  \sum_{i =  0}^{r - 1} C_{i +1} - C_i = \overline{0} $. This sum can also be expressed in terms of the net counts of each step type:
$S = (n_{\oun}(\calC) - n_{-\oun}(\calC)) \cdot \oun+ (n_{\oalpha}(\calC) - n_{-\oalpha}(\calC))\cdot \oalpha $.

Thus $ \bP_r  = \begin{pmatrix}
    n_{\oun}(\calC) - n_{-\oun}(\calC) \\
     n_{\oalpha}(\calC) - n_{-\oalpha}(\calC)
\end{pmatrix} \in \LL $. \qed \\
 
\subsubsection{\texorpdfstring{Simplification of the proof of Theorem \ref{THM:MAIN_THM}:}{Simplification of the proof of the main Theorem}}

Proposition \ref{prop:link_between__connected_cycle_and_L} offers a substantial simplification for proving Theorem \ref{THM:MAIN_THM}. Recall that the minimum distance of $GB(A(X), B(X), n)$ is defined by the length of the shortest simple cycle in $ \G $ that is not a sum of faces. 

Given this, to prove that $d(GB(A(X), B(X), n))$ is lower bounded by the shortest Manhattan norm of a non-zero lattice vector (i.e $ d(GB(A(X),B(X), n)) \geq \lambda (\LL)$ ), we only need to show one thing:
for any simple cycle of $ \G$, $ \calC$, that is not sum of faces, $ \bP_r$, the endpoint of its associated $ \Z^2$ walk, is non-zero: $ \bP_r \neq \begin{pmatrix}
    0 \\ 0
\end{pmatrix}$. This is because Proposition \ref{prop:link_between__connected_cycle_and_L} already confirms that $ \bP_r$ is an element of $ \LL$ whose Manhattan norm is less than or equal to $|\calC| $. \\

In the previous section, we constructed the $ \Z^2$-walk $\gamma_\calC$ that mimics the $ \G$ walk $\calC$. The next section reverses that perspective: constructing a $ \G $ walk from a specified $ \Z^2$ walk. This inverse mapping is crucial as it enables us to demonstrate that for any simple cycle $\calC$ that is not a sum of faces, the endpoint $ \bP_r$ of its associated $ \Z^2$-walk $ \gamma_\calC$
must have at least one non-zero coordinate. 

% FROM $ \Z^2$-WALKS TO $\G$-WALKS
\subsection{\texorpdfstring{From $ \Z^2$-walks to $\G$-walks}{From Z²-walks to G-walks}}
\label{sec: from Z^2 walks to G walks}

\subsubsection{\texorpdfstring{Construction of associated $\G$ walks}{Construction of associated G walks}}

For any given $ \Z^2$-walk $  \gamma \, : \,  \gamma_0  \xrightarrow{} \dots \xrightarrow{} \gamma_{r} $ we can create a collection $  \{ (C_{\gamma} )_{t_0} \, \vert \,  t_0 \in \Z / n\Z \} $ of $ \G$ walks that replicate the sequence of steps taken by $ \gamma$. Specifically, the walk $ (C_{\gamma})_{t_0}$ is defined by a sequence of vertices $(C_{\gamma})_{t_0} : t_0  \xrightarrow{} \dots   \xrightarrow{} t_r $ in $ \Z/n\Z $ starting at $ t_0$. For each step from $ \gamma_i $ to $ \gamma_{i + 1}$, the corresponding vector $ t_{i + 1} $ in $(C_{\gamma})_{t_0}$ is determined by adding an element modulo $n$ to $t_i$. This element is defined based on the type of step taken in $ \gamma$ as follows:

\begin{itemize}
    \item If the step is $ \gamma_i \to \gamma_{i  +1}$ where $ \gamma_{i  +1} = \gamma_i + \begin{pmatrix}
        1 \\ 0
    \end{pmatrix}$, we add $ \oun$: $ t_{i + 1} = t_i + \oun $.

    \item If the step is $ \gamma_i \to \gamma_{i  +1}$ where $ \gamma_{i  +1} = \gamma_i - \begin{pmatrix}
        1 \\ 0
    \end{pmatrix}$, we subtract $ \oun$: $ t_{i + 1} = t_i - \oun $.

    \item If the step is $ \gamma_i \to \gamma_{i  +1}$ where $ \gamma_{i  +1} = \gamma_i + \begin{pmatrix}
        0 \\ 1
    \end{pmatrix}$, we add $ \oalpha$: $ t_{i + 1} = t_i + \oalpha$.

    \item If the step is $ \gamma_i \to \gamma_{i  +1}$ where $ \gamma_{i  +1} = \gamma_i - \begin{pmatrix}
        0 \\ 1
    \end{pmatrix}$, we subtract $ \oalpha$: $ t_{i + 1} = t_i - \oalpha $.
\end{itemize}

\subsubsection{\texorpdfstring{End of the proof of Theorem \ref{THM:MAIN_THM}}{End of the proof of the main Theorem}}

The formalism described above allows us to express any $\G$-walk $\calC$ as $(\gamma_\calC)_{C_0}$, where $C_0$ is the starting vertex of $ \calC$. This interpretation enables us to identify conditions on $\gamma_\calC$ that guarantee that $\calC$ is a sum of faces of $ \G$.

%%%% KEY LEMMA
\begin{lemma}
\label{lemma: Z^2 cycles are sum of faces}
If $  \gamma \, : \,   \gamma_0 \to  \dots \to  \gamma_{r} $  is a simple $ \Z^2$-cycle starting at the origin, then for all $  t \in \Z / n\Z, \,(C_\gamma)_t $, the associated walk in $\G$ starting at $t$ is a sum of faces of $ \G$.
\label{lemma:key_lemma}
\end{lemma}

%%%% PROOF OF LEMMA 5.1
\textbf{Proof of Lemma \ref{lemma:key_lemma}:} We prove Lemma \ref{lemma:key_lemma} by induction. If $ \gamma $ is a simple $\Z^2$-cycle starting at the origin that surrounds exactly 1 square of $ \Z^2 $, then $ \gamma $ is equal to that square. So, $ \forall t \in \Z / n\Z, (C_\gamma)_t  $ is one of the faces of $ \G$ starting at $t$:
\begin{itemize}
    \item (DL) : $ t \xrightarrow{} t + \oun \xrightarrow{} t + \oun + \oalpha  \xrightarrow{} t + \oalpha \xrightarrow{} t  $

    \item (DR) : $ t \xrightarrow{} t + \oalpha \xrightarrow{} t - \oun + \oalpha  \xrightarrow{} t - \oun \xrightarrow{} t  $

    \item (UL) : $ t \xrightarrow{} t - \oalpha  \xrightarrow{} t  + \oun - \oalpha \xrightarrow{} t + \oun \xrightarrow{} t $

    \item (UR) : $ t \xrightarrow{} t - \oun \xrightarrow{} t - \oun - \oalpha  \xrightarrow{} t - \oalpha \xrightarrow{} t  $
\end{itemize}
 
%%% ---------  RECURENCY HYPOTHESIS ----------------
Now, assume there exists $q \geq 1$, such that for any simple $\Z^2$-cycle $ \gamma$ starting at the origin and surrounding exactly $ 1 \leq k \leq q$ $\, \Z^2$-elementary squares, all associated $\G$-walks $ (C_\gamma)_t$ are face sums. \\

%%% --------- P(q+1) IS TRUE ----------------
Consider a simple $ \Z^2$-cycle $ \Gamma $ going through the origin, which surrounds exactly $q + 1 $ $ \Z^2$-elementary grid squares. We will show that for any $ t \in \Z / n\Z $, $ \Gamma_t $ is a sum of faces of $ \G$. \\
 
\begin{figure}[ht!] % [h!] tente de placer la figure "ici"
    \centering % Centre l'ensemble des sous-figures
        \includegraphics[scale=0.78]{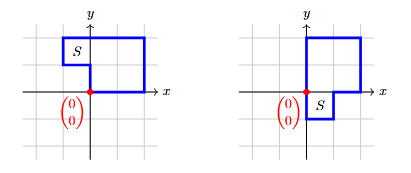}
        \caption{Two possible configurations for the boundary of \( \Gamma \): Either, there is a face non-incident to the origin (left). Either, there are multiple faces incident to the origin (right).} % Légende spécifique à cette image
        \label{fig: carre S dans bord Gamma}
\end{figure}

Let $ Int(\Gamma) $ denote the bounded connected region enclosed by $ \Gamma $.
Within $ Int(\Gamma) $, at least one of these two scenarios is true: either there is at least one grid square that shares edges with $ \Gamma$ and is not adjacent to the origin, or there are multiple such grid squares that are adjacent to the origin. We choose one of these squares, denoted  $S$, and let $\begin{pmatrix}
    u \\ v
\end{pmatrix}$ denote the coordinates of its lower-left corner. (see Figure \ref{fig: carre S dans bord Gamma})

Let $ \gamma $ be the simple $ \Z^2 $-cycle  that surrounds the same grid squares as $ \Gamma $ but $S$. Since $ \gamma$ goes through the origin and surrounds exactly $ q $ grid squares, then by induction, $ \gamma_t $ is a sum of faces of $\G$.  Let $ P = \overline{u + \alpha v}$, then $ \Gamma_t = \gamma_t + S(t+ P) $ where $ S(t + P) $ is one of the faces of $\G$ adjacent to $ t + P$: 
\begin{itemize}
    \item (DL):    $ t + P \xrightarrow{} t + P + \oun    \xrightarrow{} t + P + \oun  + \oalpha \xrightarrow{} t + P+ \oalpha \xrightarrow{}  t + P $
    \item (DR): $     t + P \xrightarrow{} t + P + \oalpha  \xrightarrow{} t + P + \oalpha  - \oun  \xrightarrow{}  t + P  - \oun   \xrightarrow{} t + P $
    \item (UL): $     t + P \xrightarrow{} t + P - \oalpha \xrightarrow{}  t + P - \oalpha + \oun \xrightarrow{} t + P + \oun  \xrightarrow{} t + P $
    \item (UR): $    t + P \xrightarrow{} t + P - \oun \xrightarrow{} t + P - \oun - \oalpha  \xrightarrow{}  t + P - \oalpha  \xrightarrow{} t + P$
\end{itemize}

Hence, $ \Gamma_t $ is a sum of faces of $ \G$. \qed

We can now present the final corollary, which concludes the proof of Theorem \ref{THM:MAIN_THM}.

% Corollary of the key Lemma
\begin{corollary}
     Let $ \calC \, : \,  C_0 \xrightarrow{} \dots \xrightarrow{} C_{r-1} \xrightarrow{} C_0 $ be a simple cycle of $\G$. If $ \calC$ is not a sum of faces of $\G$, then at least one of the following conditions must hold: $ n_{\oun}(\calC) - n_{-\oun} (\calC)\neq 0$ or $ n_{\oalpha}(\calC) - n_{-\oalpha}  (\calC) \neq 0$.
\label{corollary:corolloraly_of_key_lemma}
\end{corollary}

% Proof of the corollary of the key lemma
\textbf{Proof of Corollary \ref{corollary:corolloraly_of_key_lemma}:} We prove the contrapositive of the corollary. If $ \calC $ is a simple cycle of $ \G$ such that $ n_1(\calC) = n_{-1} (\calC) $ and $ n_{\alpha}(\calC) = n_{-\alpha}(\calC) $, then $ \gamma_\calC $, the associated $ \Z^2$-walk starting at the origin, is also a simple cycle.  
Hence, by Lemma \ref{lemma: Z^2 cycles are sum of faces}, $ \calC= (\gamma_\calC)_{C_0}$ is a sum of faces of $\G$. \qed

%%%% APPLICATION OF THE Theorem
\section{Application of Theorem \ref{THM:MAIN_THM}}
\label{section: application of the main theorem}

In quantum coding theory, a core objective is to design quantum codes with the largest possible minimum distance, as this directly enhances fault tolerance. For (2,2)-GB codes, their minimum distance is fundamentally constrained to scale with the square root of the code length. 

While lattice-based toric codes have historically represented the best known 2D surface codes with weight-four generators in terms of minimum distance (see Section \ref{sec: optimal surface codes}), our work makes a significant leap. We leverage Theorem \ref{THM:MAIN_THM}, demonstrating that constructing GB codes with large minimum distances effectively reduces to identifying associated lattices whose non-zero vectors possess large Manhattan norms.This insight allows us to construct explicit families of GB codes that attain optimal parameters, matching the performance of the best-known 2D weight-4 surface codes \cite{BD07}. Our method hinges on carefully choosing lattices where we can control the invariant $\lambda(\LL)$, which directly bounds the minimum distance. \\

While Kovalev and Pryadko \cite{KP13} demonstrated how to construct optimal odd-distance surface codes using the GB formalism (e.g., $GB(1 + X^{2t^2 + 1}, X + X^{2t^2}, t^2 + (t + 1)^2)$ with parameters $[(2t + 1)^2 + 1, 2, 2t + 1]$), we introduce an alternative family that achieves the same optimal odd-distance scaling through a distinct construction.

More notably, Pryadko and Wang \cite{PW22} claimed that optimal even-distance 2D surface codes—characterized by periodicity vectors $ \begin{pmatrix} r \\ r \end{pmatrix}, \begin{pmatrix} r \\-r \end{pmatrix} $ and parameters $[ \! [ 4r^2, 2, 2r ]]$—could not be realized using GB codes. According to their results, the next closest GB-compatible constructions only reached parameters $[ \! [ 4r^2 + 4, 2, 2r ]]$ for even values of $r$. Contrary to that belief, our approach yields a new family of (2,2)-GB codes that meets these optimal even-distance parameters, providing the first known GB realization of such codes. 

These constructions are summarized in the following proposition:

\begin{proposition}
\label{prop: GB optimaux}
Let $n \geq 2 $, $ r \geq 1$, and $ t \geq 1$ be three positive integers. 
\begin{itemize}
    \item GB$(1 + X, 1 + X^{n}, n^2)$ has parameters $[[ 2n^2, 2, n ]]$ 
    \item GB$(1 + X, 1 + X^{2r - 1}, 2r^2)$ has parameters $[[ 4r^2, 2, 2r ]]$  
    \item GB$(1 + X, 1 + X^{2t + 1}, t^2 + (t + 1)^2 \, ) $ has parameters $ [[(2t + 1)^2 + 1, 2, 2t + 1 ] \! ]$  
\end{itemize}
\end{proposition}

We will break down the proof into several parts, dedicating each to a specific code construction.

\subsection{\texorpdfstring{Study of $GB(1 + X, 1 + X^{n}, n^2)$ Minimum Distance}{Study of GB(1 + X, 1 + X**n, n²) minimum distance}}
\label{sec: Study of GB(1 + X, 1 + X^{n}, n^2) minimum distance}

\begin{lemma}
\label{lemma: borne inf d_min kitaev}
    For $ n \geq 2$, the minimum distance of GB$(1 + X, 1 + X^n, n^2)$ is equal to $ n$.
\end{lemma}

\textbf{Proof of Lemma \ref{lemma: borne inf d_min kitaev}:} For $ n = 2$, we confirm this equality using computer simulations.  For all $ n \geq 3$, we establish the equality by demonstrating that the minimum distance is both lower-bounded and upper-bounded by $n$. 

\subsubsection*{Upper Bounding the Minimum Distance} 

Under Section \ref{subsec: intro GB codes} formalism, the minimum distance of GB$(1 + X, 1 + X^n, n^2)$ corresponds to the smallest Hamming weight of a pair of polynomials $(U(X), V(X)) \in (\F_2[X]_{\leq n^2 - 1})^2$ that satisfy the two following congruences:
\begin{itemize}
    \item $ (1+X)U(X) + (1 + X^n)V(X) \equiv 0 \mod X^{n^2} - 1$
    \item There is no polynomial $ H(X) \in \F_2[X] $ such that:
\begin{equation*} 
\left\{
\begin{aligned}
    U(X) & \equiv (1 +X^n) H(X) & \mod X^{n^2} - 1 \\
    V(X) & \equiv (1 + X)H(X) &\mod X^{n^2} - 1
\end{aligned}
\right.
\end{equation*}
\end{itemize}

In $ \F_2[X], $ we observe the equality: $ (1 + X^n)\sum_{k = 0}^{n-1}X^{nk} = 1 + X^{n^2}$. 
This implies $  (1 + X)\times 0 +  (1 + X^n)\sum_{k = 0}^{n-1}X^{nk} \equiv 0 \mod X^{n^2} - 1$. Furthermore, suppose, for contradiction, that there exists a polynomial $ H(X) \in \F_2[X] $ such that: 
\begin{equation*} 
\left\{
\begin{aligned}
    (1 +X^n) H(X) & \equiv  0 & \mod X^{n^2} - 1 \\
   (1 + X)H(X)  & \equiv \sum_{k = 0}^{n-1}X^{nk} &  \mod X^{n^2} - 1
\end{aligned}
\right.
\end{equation*}

Hence, we would have: \[ (\sum_{q = 0}^{n-1} X^q)(1 + X)H(X) \equiv (1 + X^n)H(X) \equiv 0 \mod     X^{n^2} - 1\] However, since $  (1 + X)H(X) \equiv \sum_{k = 0}^{n-1}X^{nk} \mod X^{n^2} - 1$, it follows: $(\sum_{k = 0}^{n-1}X^{nk}) \cdot (\sum_{q = 0}^{n-1} X^q) \equiv 0 \mod X^{n^2} - 1$.
This leads to a contradiction because the polynomial $(\sum_{k = 0}^{n-1}X^{nk}) \cdot (\sum_{q = 0}^{n-1} X^q)$ is non-zero and its degree is equal to $ n(n-1) + n - 1 = n^2 - 1 < n^2$. Consequently, the minimum distance of $GB(1 +X, 1 + X^n, n^2)$ is upper bounded by the Hamming weight of the pair $ (0, \sum_{k = 0}^{n-1}X^{nk}) $ which is $n$.

\subsubsection*{Lower Bounding the Minimum Distance} 

According to Theorem \ref{THM:MAIN_THM}, for $ n \geq 3$, the minimum distance of $  GB(1 + X, 1 + X^n, n^2)$  is lower bounded by the smallest Manhattan norm of any non-zero elements of the lattice $\Z \begin{pmatrix}
    n ^2 \\ 0 
\end{pmatrix}\bigoplus \Z\begin{pmatrix}
    n \\ - 1
\end{pmatrix}$. 

Consider an arbitrary non-zero lattice element $ \bT = x\begin{pmatrix} n^2 \\ 0 \end{pmatrix} + y\begin{pmatrix} n \\  -1\end{pmatrix} $ where $ (x, y) \in \Z^2 \backslash \{ (0, 0) \}$. Its Manhattan norm is $ \| \bT\|_1 = \vert n( xn + y) \vert + \vert y \vert $. \\

\underline{If $ xn + y = 0$, then:} $ x \neq 0$ since $(x, y) \in \Z^2 \backslash \{(0,0)\}$.
Thus $\| \bT\|_1 = n \vert x \vert \geq n$.

\underline{If $ xn + y \neq 0$, then:} $\vert xn + y \vert$ is a non-zero integer. Thus, $\bT$ satisfies: 
\[ \| \bT\|_1 = \vert n( xn + y) \vert + \vert y \vert \geq n \vert xn + y \vert  \geq n    \]

In conclusion, $d( GB(1 + X, 1 + X^n, n^2) )  \geq n$.
\qed.

\subsection{\texorpdfstring{Study of $GB(1 + X, 1 + X^{2r - 1}, 2r^2)$ Minimum Distance}{Study of GB(1 + X, 1 + X**(2r - 1), 2r\texttwosuperior)}}

\label{sec: Study of GB(1 + X, 1 + X^{2r - 1}, 2r^2) minimum distance}

\begin{lemma}
\label{lemma: borne inf d_min paire}
    For $ r \geq 1$ the minimum distance of  $GB(1 + X, 1 + X^{2r - 1}, 2r^2)$ is equal to $ 2r$.
\end{lemma}

\textbf{Proof of Lemma \ref{lemma: borne inf d_min paire}:}
     For $ r = 1 $, the equality is trivially true. For all $ r \geq 2$, we establish the equality by demonstrating that the minimum distance is both lower-bounded and upper-bounded by $2r$.

\subsubsection*{Upper Bounding the Minimum Distance} 

Let $ \begin{pmatrix}
    \bu \\ \bv
\end{pmatrix} $ be the vector of weight $2r$ where $ \bu \in \F_2^{2r^2}$ has $1$s  on its first $ 2r - 1$ coordinates and $ \bv \in  \F_2^{2r^2} $ has a single $1$ at its first coordinate. The vector $ \begin{pmatrix}
    \bu \\ \bv
\end{pmatrix} $ belongs to $\ker \bH_X\backslash rs(\bH_Z)$ where the matrices $ \bH_X$ and $ \bH_Z$ are given by: $ \bH_X = \left[Circ( 1 + X, 2r^2) \, | \, Circ(1 +X^{2r -1}, 2r^2) \right] $ and $\bH_Z = [ Circ(1 +X^{2r -1}, 2r^2)^{\intercal} \, | \, Circ( 1 + X, 2r^2)^{\intercal} ]$. Hence, the minimum distance of $ GB(1 + X, 1 + X^{2r - 1}, 2r^2) $ is less than $ 2r$. 

\subsubsection*{Lower Bounding the minimum distance :} 

According to Theorem \ref{THM:MAIN_THM}, for $ r \geq 2$, the minimum distance of $  GB(1 + X, 1 + X^{2r - 1}, 2r^2) $  is lower bounded by the smallest Manhattan norm of any non-zero elements of the lattice $\Z \begin{pmatrix}
    2r^2 \\ 0 
\end{pmatrix}\bigoplus \Z\begin{pmatrix}
    2r - 1 \\ - 1
\end{pmatrix}$. Consider an arbitrary non-zero lattice element $ \bT = x\begin{pmatrix} 2r^2 \\  0\end{pmatrix} + y\begin{pmatrix} 2r - 1 \\  - 1 \end{pmatrix}  $ with $(x, y) \in \Z^2 \backslash \{ (0,0) \}$. \\

\underline{If $ y + rx = 0$, then:} $ x \neq 0$ since $ (x,y)  \in \Z^2 \backslash \{(0,0) \}$ and $ r \geq 2$. Thus: 
\[\| \bT \|_1 = \| \begin{pmatrix} rx \\  rx\end{pmatrix} 
\|_1 = 2r \vert x\vert \geq 2r  \]

\underline{If $ y + rx \neq 0$, then:} $ \vert y + rx  \vert \geq 1 $ since $ r, x, y \in \Z$. 
Thus:
\begin{align*}
    \| \bT \|_1  = \vert\,  2r^2x + (2r - 1)y \,\vert + \vert y \vert  &= \vert 2r (rx + y ) - y \vert + \vert y \vert \\
              & \geq   2r \vert rx + y  \vert   -    \vert y \vert  \, \, \vert + \vert y \vert \\
              & \geq 2r  
\end{align*}
which proves, $ d(GB(1 +X , 1 + X^{2r - 1}, 2r^2)) \geq 2r$. \qed 

\subsection{\texorpdfstring{Study of $GB(1 + X, 1 + X^{2t + 1}, t^2 + (t + 1)^2)$ minimum distance}{Study of GB(1 + X, 1 + X**{2t + 1}, t**2 + (t + 1)**2) minimum distance}}

\label{sec: Study of GB(1 + X, 1 + X^{2t + 1}, t^2 + (t + 1)^2) minimum distance}

\begin{lemma}
\label{lemma: borne inf d_impaire}
    For $ t \geq 1$ the minimum distance of $GB(1 + X, 1 + X^{2t + 1}, t^2 + (t + 1)^2 )$  is equal to $ 2t + 1$.
\end{lemma}

\textbf{Proof of Lemma \ref{lemma: borne inf d_impaire}:}
For $ t = 1 $,  we confirm the equality using computer simulations. For all $ t \geq 2$, we establish the equality by demonstrating that the minimum distance is both lower-bounded and upper-bounded by $2t + 1$. 

\subsubsection*{Upper bounding the Minimum Distance :} 

As detailed in Section \ref{subsec: intro GB codes}, the minimum distance of the code $GB(1 + X, 1 + X^{2t + 1}, t^2 + (t + 1)^2)$ corresponds to the smallest Hamming weight of a pair of polynomials $(U(X), V(X)) \in (\F_2[X]_{\leq t^2 + (t + 1)^2 - 1})^2$ that satisfy the two following conditions: 

\begin{itemize}
    \item $ (1+X)U(X) + (1 + X^{2t + 1})V(X) \equiv 0 \mod X^{t^2 + (t + 1)^2} - 1$
    \item There is no polynomial $ H(X) \in \F_2[X] $ such that 
\begin{equation*}
\left\{
\begin{aligned}
    U(X) & \equiv (1 +X^{2t + 1})H(X) & \mod{X^{t^2 + (t + 1)^2} - 1} \\
    V(X) & \equiv (1 + X)H(X)       & \mod{X^{t^2 + (t + 1)^2} - 1}
\end{aligned}
\right.
\end{equation*}

\end{itemize}

In the quotient $ \F_2[X]/(X^{t^2 + (t +1)^2} - 1)$, we have the following equalities:
\begin{align*}
     ( 1 +X^{2t + 1})( 1 + \sum_{k = 0}^{t - 1} X^{2t^2 - k(2t +1)}) & = X^{2t + 1} + X^{2t^2 - (t -1)(2t + 1)}  \\
     & = X^{2t +1} + X^{t +1} \\
     & = X^{t +1}( 1 + X^{t})  \\
     & = X^{t + 1}(\sum_{q= 0}^{t-1} X^q)(1 + X)
\end{align*}

Consequently, setting $ U(X) = X^{t + 1}(\sum_{q= 0}^{t-1} X^q)  $ and 
$ V(X) = ( 1 + \sum_{k = 0}^{t - 1} X^{2t^2 - k(2t +1)})$, yields the following congruence modulo $ X^{t^2 + (t+1)^2} - 1$:
\[  (1 + X)U(X) + ( 1 +X^{2t + 1})V(X) \equiv 0 \mod X^{t^2 + (t+1)^2} - 1 \]

The Hamming weight of $ V(X)$ is $ t + 1$ and the Hamming weight of $ U(X)$ is $t$ since $ deg(U(X)) = 2t < t^2 + (t + 1)^2 $. Since one polynomial must have odd weight, it is impossible to find a  polynomial $ H(X) \in \F_2[X] $ that simultaneously  satisfies the congruences:
\begin{equation*}
    \left\{ 
    \begin{aligned}
        U(X) &\equiv (1 +X^{2t + 1}) H(X) &\mod X^{t^2 + (t + 1)^2} - 1 \\
        V(X) &\equiv (1 + X)H(X) &\mod X^{t^2 + (t + 1)^2} - 1
    \end{aligned}
    \right.
\end{equation*}

Thus $ d(GB(1 + X, 1 + X^{2t + 1}, t^2 + (t +1)^2)$ is upper bounded by the Hamming weight of the pair $(U(X), V(X))$ which is $2t + 1$. 

\subsubsection*{Lower Bounding the Minimum Distance :}

According to Theorem \ref{THM:MAIN_THM}, for $ t \geq 2$, the minimum distance of $ GB(1 + X, 1 + X^{2t + 1}, t^2 + (t + 1)^2)$  is lower bounded by the smallest Manhattan norm of any non-zero elements of the lattice $\Z \begin{pmatrix}
    t^2 + ( t+1)^2 \\ 0 
\end{pmatrix} \bigoplus \Z \begin{pmatrix}
     2t + 1 \\  -1
\end{pmatrix} $.  \\

Let $ \bT = x \begin{pmatrix}
    t^2 + ( t+1)^2 \\ 0
\end{pmatrix}   + y \begin{pmatrix}
    2t + 1 \\  -1
\end{pmatrix} $ with $(x, y) \in \Z^2 \backslash \{ (0,0) \}$. \\

If $ x = 0$, then: $y$ is a non-zero integer. Therefore we have, $ \| \bT\|_1 \geq \vert y \vert  (2t + 1) \geq 2t + 1$.  We can now assume $ x \neq 0$. 

Having established this base case, we proceed by distinguishing two main scenarios for the proof, based on
whether $ \vert y \vert \leq t \vert x \vert $ or not. \\

\underline{Case 1. $ \vert y \vert \leq t \vert x \vert $:}
\begin{align*}
   \|\bT\|_1 & = \vert x(t^2 + (t + 1)^2) + y(2t + 1) \vert   + \vert y \vert \\
    & \geq \vert x \vert (t^2 + (t + 1)^2) -  \vert y \vert  (2t + 1)   + \vert y \vert \\
    & = \vert x \vert(2t^2 + 2t + 1) - \vert y \vert \cdot 2t \\
    & = \vert x \vert(2t + 1) + 2t( t\vert x \vert - \vert y \vert) \\
    & \geq \vert x \vert(2t +1) \\
    & \geq 2t + 1 
\end{align*}

\underline{Case 2. $ \vert y \vert > t \vert x \vert $:} $ \bT $ can be written:
\[ \bT   = \begin{pmatrix}
        (2t + 1)((t  + 1)x + y\, ) -tx
        \\ -y
    \end{pmatrix} \] 

If $ \vert (t +1)x + y \vert = 0 $, then $ \| \bT\|_1 = (2t + 1) \vert x\vert \geq 2t  + 1$.

Else, the Manhattan norm of $ \bT$ satisfies:
\begin{align*}
    \| \bT \|_1 & = \vert \, (2t + 1)( (t  + 1)x + y) -t \vert x \vert\, \vert  + \vert y\vert \\
                & \geq \vert \, (2t + 1)\vert(t + 1)x +  y \vert    - t\vert x \vert \,  \vert + \vert y\vert \\
                & \geq (2t + 1)\vert(t + 1)x +  y \vert    - t\vert x \vert + \vert y\vert \\
                & > 2t + 1
\end{align*}

Thus, $ d(GB(1  + X, 1 +X^{2t + 1}, t^2 + (t +1)^2) \geq 2t + 1$. \qed

\section{Equivalence Relations for Quantum Codes}
\label{sec: equivalence rel for CSS}

In the previous section, we introduced three families of Generalized Bicycle (GB) codes that achieve optimal performance, matching the best known 2D toric codes with weight-four generators. Now, to rigorously establish the originality of our contribution, it is essential to determine whether these newly constructed codes represent genuinely new structures or are simply equivalent to known optimal codes. 

To address this, we first delve into the concept of equivalent quantum codes. In quantum coding theory, codes are generally deemed "the same" if they are interconvertible via transformations that preserve their essential error correction capabilities (such as error detection, correction, and decoding properties). However, as will be demonstrated, various types of equivalence relations exist, each possessing distinct behaviours and implications.

\subsection{General Quantum Equivalence relation}

\subsubsection{Definition}
\label{def: general qu equiv relation}
Two quantum codes, $\mathcal{Q}_1$ and $\cQ_2$, are equivalent if one can be transformed into the other by a unitary operation $\bU$, such that $\cQ_2 = \bU\cQ_1$. For stabilizer codes, this $\bU$ must be a sequence of qubit permutations and local Clifford transformations which are defined as tensor products of single-qubit Clifford gates, meaning $\bU = \bU_1 \otimes \dots \otimes \bU_n$, where each $\bU_i$ is composed of Hadamard, phase, and CNOT gates.

\subsubsection{Limitations}
When a code $\cQ_1$ is stabilized by a set of operators $\mathcal{S}$, the equivalent code $\cQ_2$ is stabilized by the transformed set $\bU\mathcal{S}\bU^\dagger$. However, this general equivalence relation does not always preserve the specific structure of CSS codes. This is because local Clifford gates like the Hadamard ($\bH$) transform Pauli $\bX$ operators into $\bZ$ operators ($\bH \bX \bH^\dagger = \bZ$) and $\bZ$ operators into $\bX$ operators ($\bH \bZ \bH^\dagger = \bX$). Similarly, the Phase ($\bS$) gate transforms $\bX$ into $\bY$ ($\bS \bX \bS^\dagger = \bY = i\bX\bZ$). Since CSS codes require their stabilizer generators to be composed \textit{only} of $\bX$ or $\bZ$ operators (and identity), these transformations can change the fundamental form of the generators, thus breaking the CSS structure. \\

To compare the GB optimal codes (identified in Section \ref{section: application of the main theorem}) with the optimal weight-4 2D surface codes, which are both derived from Cayley graphs on abelian groups, our focus will be on equivalence relations that preserve both their CSS structure and their underlying graph structure.

\subsection{CSS Graph-preserving  (CGP) Equivalence}

\subsubsection{Definition:}
\label{def: equiv qu codes}

Let $\mathcal{Q}$ and $\mathcal{Q}'$ be two quantum codes of length $n$. We say that $\mathcal{Q}$ is CGP-equivalent to $\mathcal{Q}'$, denoted $\mathcal{Q} \sim \mathcal{Q}'$, if and only if there exists a permutation $\phi$ of the qubit indices $\{1, \dots, n\}$ such that $\mathcal{Q}' = \mathbf{P}_\phi(\mathcal{Q})$ where $ \mathbf{P}_\phi \in GL_{2^n}(\mathbb{C})$ is a permutation matrix acting on basis states as follows: 
$\mathbf{P}_\phi |t_1\dots t_n\rangle = |t_{\phi(1)}\dots t_{\phi(n)}\rangle$ for any computational basis state $|t_1\dots t_n\rangle$.
This implies that $\mathcal{Q}'$ consists of all states formed by permuting the qubits of states in $\mathcal{Q}$:
$$\mathcal{Q}' = \left\{ \sum_{t \in \mathbb{F}_2^n} \alpha_t |t_{\phi(1)}\dots t_{\phi(n)}\rangle \quad \middle| \quad \sum_{t \in \mathbb{F}_2^n} \alpha_t |t_1\dots t_n\rangle \in \mathcal{Q} \right\}$$

\subsubsection{CSS Structure Compatibility:}
Two CGP-equivalent stabilizer codes, $\mathcal{Q}$ and $\mathcal{Q}'$ (i.e., $\mathcal{Q}' = \mathbf{P}_\phi(\mathcal{Q})$), have the same parameters and their stabilizer groups are directly related. A Pauli operator $ \bO_1 \otimes \dots \otimes \bO_n$ stabilizes $\mathcal{Q}$ if and only if the permuted operator $\bO_{\phi(1)} \otimes \dots \otimes \bO_{\phi(n)}$ stabilizes $\mathcal{Q}'$. Crucially, this implies that the CSS code structure is preserved under such equivalence: $\mathcal{Q}$ is a CSS code if and only if $\mathcal{Q}'$ is also a CSS code.

\subsubsection{Relation between Stabilizer Generator Matrices}
For two stabilizer codes $\mathcal{Q}$ and $\mathcal{Q}'$ that are CGP-equivalent (i.e., $\mathcal{Q}' = \mathbf{P}_\phi(\mathcal{Q})$), their respective generator matrices, $\mathbf{H} = \begin{pmatrix} \mathbf{H}_X & 0 \\ 0 & \mathbf{H}_Z \end{pmatrix}$ and $\mathbf{H}' = \begin{pmatrix} \mathbf{H}_X' & 0 \\ 0 & \mathbf{H}_Z' \end{pmatrix}$, are directly related. The row spaces of their submatrices are connected by the following equalities:
\begin{equation*}
    \left\{ \begin{aligned}
         rs(\mathbf{H}_X') = rs(\mathbf{H}_X \mathbf{Q})     \\
        rs(\bH_Z')  = rs(\bH_Z\bQ) 
    \end{aligned} \right.
\end{equation*}
where $ \bQ $ denote the $n \times n$ permutation matrix corresponding to $\phi$.

\subsubsection{Graph Structure Compatibility}
\label{sec: graph compatibily}
 For the class of codes under consideration—where the generator submatrices $\mathbf{H}_X, \bH_X'$ and $\mathbf{H}_Z, \bH_Z'$ correspond respectively to the vertex-edge incidence and face-edge incidence (i.e., dual vertex-edge incidence) matrices of Cayley graphs—the conditions $rs(\mathbf{H}_X') = rs(\mathbf{H}_X\mathbf{Q})$ and $rs(\mathbf{H}_Z') = rs(\mathbf{H}_Z\mathbf{Q})$ play a central role. 

These constraints induced by the CGP-equivalence between codes, guarantee a bijective correspondence between the edges of the underlying graphs, as well as their duals. Consequently, cycles in one graph (or its dual) are mapped directly to cycles in the other, a property known in graph theory as 2-isomorphism. 

While more restrictive than general stabilizer code equivalence, the CGP-equivalence enables a comparison between codes that faithfully preserves the graph-theoretic structure at the heart of their construction. \\

As we will show in the next section, in the specific case of the GB and optimal weight-4 surface codes examined here,  having 2-isomorphism of the associated graphs is, in fact, equivalent to full graph isomorphism. This means the incidence relations between edges are fully preserved, providing an even stronger form of structural equivalence.

\begin{remarks}
\label{rem: simple cycles corresponds to simple cycles}
Consider two CGP-equivalent CSS codes, $\mathcal{Q}$ and $\mathcal{Q}'$, based on Cayley graphs ($\mathcal{G}$ and $\mathcal{G}'$), whose stabilizer generator matrices satisfy $rs(\mathbf{H}_X') = rs(\mathbf{H}_X\mathbf{Q})$ and $rs(\mathbf{H}_Z') = rs(\mathbf{H}_Z\mathbf{Q})$ where $\mathbf{Q}$ is the permutation matrix.

This directly implies a one-to-one correspondence, induced by $ \bQ$, between the edges of $\mathcal{G}$ and $\mathcal{G}'$. As a result, every simple cycle in $\mathcal{G}$ maps directly to an equivalent \textbf{simple cycle} in $\mathcal{G}'$ through this correspondence.
\end{remarks}

\textbf{Proof of Remarks \ref{rem: simple cycles corresponds to simple cycles}:} We can represent the power set of edges for both graph $\mathcal{G}$ and $\mathcal{G}'$ as $\mathbb{F}_2^N$. The mapping from $\mathbf{v} \in \mathbb{F}_2^N$ to $\mathbf{Q}^{-1}\mathbf{v} \in \mathbb{F}_2^N$ is a bijective transformation. This transformation establishes a one-to-one correspondence between edge subsets of $\mathcal{G}$ and $\mathcal{G}'$, ensuring that any simple cycle in $\mathcal{G}$ maps to a simple cycle in $\mathcal{G}'$. \\

Consider a simple cycle $ \calC_\bv$ in the graph $ \mathcal{G}$.  Since, $ \bH_X$ is a vertex-edge incidence matrix of $ \mathcal{G}$, this means that, the cycle  $ \calC_\bv$ corresponds to a vector $ \bv  \in \ker (\bH_X) $. Applying the transformation, we find that $\mathbf{Q}^{-1}\mathbf{v} \in \ker(\mathbf{H}_X\mathbf{Q})$. Because $rs(\mathbf{H}_X\mathbf{Q})^\perp = rs(\mathbf{H}_X')^\perp = \ker(\mathbf{H}_X')$, it follows that $\mathbf{Q}^{-1}\mathbf{v} \in \ker(\mathbf{H}_X')$. This directly means that $\mathcal{C}'_{\mathbf{Q}^{-1}\mathbf{v}}$, the corresponding edge subset in $\mathcal{G}'$, is indeed a cycle within $\mathcal{G}'$. 

Now, let's assume for contradiction that $ \calC'_{\bQ^{-1}\bv} $ is not a simple cycle in $ \mathcal{G'}$. This implies there must exist a proper sub-cycle $\emptyset \subsetneq \mathcal{C}'_{\mathbf{h}} \subsetneq \mathcal{C}_{\mathbf{Q}^{-1}\mathbf{v}}$ in $\mathcal{G}'$. This sub-cycle is represented by a non-zero vector $\mathbf{h} = \mathbf{Q}^{-1}\mathbf{w} \in \ker(\mathbf{H}_X') = \ker(\mathbf{H}_X\mathbf{Q})$, where the support of $\mathbf{h}$ is a proper subset of the support of $\mathbf{Q}^{-1}\mathbf{v}$. Since $\mathbf{Q}$ is a permutation matrix, this implies that $\mathbf{w} \in \ker(\mathbf{H}_X)$ and its support is a proper subset of the support of $\mathbf{v}$. Consequently, the cycle $\mathcal{C}_{\mathbf{w}}$ in graph $\mathcal{G}$ corresponding to $\mathbf{w}$, is a proper subset of $\mathcal{C}_{\mathbf{v}}$, contradicting the simplicity of $ \calC_\bv$. \qed.

\section{Comparing GB codes to notable 2D Surface Codes}
\label{sec: comparison 2D and GB}

Both our novel Generalized Bicycle (GB) codes introduced in Section \ref{section: application of the main theorem} and the 2D surface codes detailed in Section \ref{sec: optimal surface codes} are founded upon degree-4 regular Cayley graphs that exhibit strong connectivity. This shared structural basis allows for a rigorous comparison between them.

A core contribution of our work is proving the distinctness of our two new GB code families from Kitaev toric codes and optimal 2D weight-4 surface codes. This proof hinges on a critical graph-theoretic insight: for graphs possessing the specific structure of those underlying our codes, a 2-isomorphism (which preserves cycles) implies a full graph isomorphism (preserving incidence relations). 

\begin{definition}[Graphs isomorphism]
    \label{def: graph isomorphism}
Two graphs $ \mathcal{G} $ and $ \mathcal{H}$ are isomorphic if and only if there is a bijective map $ \tau : \mathcal{G} \to \mathcal{H} $ such that: $u$ and $v$ are neighbours in $ \mathcal{G}$ if and only if $ \tau(u)$ and $ \tau(v)$ are neighbours in $ \mathcal{H}$.
\end{definition}

Precisely, our argument leverages the following established graph-theoretic results, particularly Whitney's Theorem. \cite{Whitney1932}

\subsection{Whitney's Theorem}
\label{sec: whitney theorem}

\begin{definition}[3-connectedness]
    A graph $\mathcal{G}$ is triply-connected if it is connected and  removing any two distinct $(u,v)$ vertices, leaves the remaining graph $ \mathcal{G} -\{ u, v\}$ still connected.
\end{definition}

\begin{theorem}[Whitney's Theorem \cite{Whitney1932}]
\label{thm: whitney thm}
Let $ \mathcal{G}_1, \,  \mathcal{G}_2 $ be two triply connected graphs. If there is a one-to-one correspondence between their edges such that simple cycles in $ \mathcal{G}_1$ corresponds to simple cycles in $ \mathcal{G}_2$ then $ \mathcal{G}_1$ and $ \mathcal{G}_2$ are isomorphic. 
\end{theorem}

\subsection{Overview of the Distinctness Proof}

Our proof strategy is to reduce the question of code CGP-equivalence to the existence of group isomorphisms, for which we already know the answer. \\

Both proofs demonstrating the distinctness between optimal 2D surface codes and our Generalized Bicycle codes will follow a consistent pattern. Since these are CSS codes derived from unoriented Cayley graphs ($\mathcal{G}$ and $\mathcal{G}'$) based on abelian groups ($\gG$ and $\gG'$), we will proceed as follows:

\begin{enumerate}
    \item First, we will establish that the underlying graphs, $\mathcal{G}$ and $\mathcal{G}'$, are triply-connected.

    \item Next, we will prove by contradiction that the codes are not CGP-equivalent. If we assume they are CGP-equivalent, their associated graphs, $\mathcal{G}$ and $\mathcal{G}'$, must be 2-isomorphic. By Whitney's Theorem, this 2-isomorphism would then imply that the graphs themselves are isomorphic in the traditional sense.

    \item Finally, we will show that this graph isomorphism must induce a group isomorphism between their associated groups $\gG$ and $\gG'$. However, these groups will not be isomorphic, thus leading to a contradiction and proving the distinctness of the codes.
\end{enumerate}

\subsection{Even-distance 2D Surface Code Comparison}
\label{sec: comparison between GB and Kitaev}

\subsubsection{Triple connectivity of the underlying graphs}

\begin{lemma}
    \label{lemma: Z/2r^2Z 3-connected}
    For $ r \geq 3$, the Cayley graph $(\Z/2r^2\Z, \oun , \overline{2r - 1}) $, which underlies $GB(1 + X, 1 + X^{2r - 1}, 2r^2)$ is triply-connected.
\end{lemma}

\textbf{Proof of Lemma \ref{lemma: Z/2r^2Z 3-connected}:} The goal is to prove that if we delete any two distinct vertices then the remaining graph is still connected. \\

\underline{\textbf{Case 1: We delete $\overline{a}$ and $ \overline{a+ 1} $}:} \\

The original graph contains the Hamiltonian cycle encompassing all vertices $ \overline{a} \to \overline{a + 1}  \to \dots \to \overline{a - 1} \to \overline{a}$. Even after the deletions, all remaining vertices are still connected to one  another via the path $ \overline{a + 2} \xrightarrow{}\overline{a + 3} \xrightarrow{} \dots \xrightarrow{} \overline{a - 1} $. \\

\underline{\textbf{Case 2: We delete $\overline{a} $ and $ \overline{b} $ with $ 0 \leq a < b \leq 2r^2 - 1$}} \\
\underline{\textbf{ and $ \overline{a} \notin \{ \overline{b + 1}, \overline{b - 1} \} $}} \\

In the graph $(\Z/2r^2\Z - \{ \overline{a}, \overline{b}\} \, , \oun , \, \overline{2r - 1})$ the original Hamiltonian cycle, which traverses all vertices using only the $ +\oun  =  1 \mod 2r^2 $ edges, is now split into two distinct sections due to the removal of $ \oa$ and  $ \ob$.

\begin{itemize}
    \item One section extends from $ \overline{b + 1 } $ to $ \overline{a - 1}$ 
    \[ \overline{b + 1 } \to \overline{b + 2} \to  \dots \to \ozero \to \oun \to \dots \to \overline{a - 1} \]
    \item The other section runs from $ \overline{a + 1}$ to $ \overline{b - 1} $
    \[ \overline{a + 1} \to \overline{a + 2} \to \dots \to \overline{b  - 1}\]
\end{itemize}

If $  \overline{a} = \overline{b + 1 - (2r - 1)} = \overline{b - 2r + 2}$, then the edge $ \overline{a - 2} \to \overline{a - 2 + (2r - 1) } $ allows for the reconnection of the two sections. 

\begin{figure}[htbp] % [h!] tente de placer la figure "ici"
    \centering % Centre l'ensemble des sous-figures
        \includegraphics[scale=0.78]{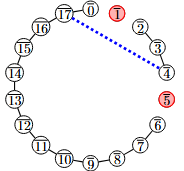}
        \caption{Connectivity in the Cayley graph $(\mathbb{Z}/2r^2\mathbb{Z}, \oun, \overline{2r - 1})$ with $r = 3$ is preserved after removing vertices $a = \overline{1}$ and $\overline{a + 1 - (2r - 1)} = \overline{5}$ thanks to the edge between $\overline{a - 2} = \overline{17}$ and $\overline{a - 2 + (2r - 1)} = \overline{4}$, represented with the dotted blue line.}
\end{figure}

The edge $ \overline{a - 2} \to \overline{a - 2 + (2r - 1) } $ is a valid edge in the original graph and its two incident vertices are, indeed, distinct from the removed vertices $ \oa$ and $ \ob$:
\begin{itemize}
    \item $ \overline{a - 2} \neq \oa $ since $ \overline{2} \neq \ozero $ in $ \Z/2r^2\Z$ because $ r \geq 3$.
    
\item $ \overline{a - 2} \neq \ob $, otherwise we would have $ \overline{b  - 2r} = \overline{b} $ thus $ \overline{2r} = \ozero $ in $ \Z/2r^2\Z $ which is not possible because  $ 0 < 2r < 2r^2 $.

    \item $\overline{a - 2 + (2r - 1) } \neq \oa $ in $ \Z/2r^2\Z$ else, $ \overline{2r - 3} = \ozero $ in $ \Z/2r^2\Z$, which is not possible since for $ r \geq 3$, $ 0 < 2r - 3 < 2r^2$.
    
    \item $\overline{a - 2 + (2r - 1) } \neq \ob  $ otherwise, this would lead to $ \overline{b - 1} = \ob$ in $ \Z/2r^2\Z$, implying $ \oun = \ozero$ which is false. \\
\end{itemize}

If $ \oa \neq \overline{b + 1 - (2r - 1) } $, then the edge $ \overline{b + 1} \to \overline{b + 1 - (2r - 1)} $ allows the two branches to form one connected component in the graph $(\Z/2r^2\Z - \{ \overline{a}, \overline{b}\} \, , \oun , \, \overline{2r - 1})$. \\

\begin{figure}[htbp] % [h!] tente de placer la figure "ici"
    \centering % Centre l'ensemble des sous-figures
        \includegraphics[scale=0.78]{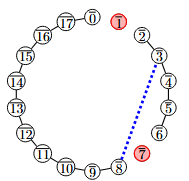}
        \caption{ Connectivity in the Cayley graph $(\mathbb{Z}/2r^2\mathbb{Z}, \oun, \overline{2r - 1})$ with $r = 3$ is preserved after removing vertices $a = \overline{1}$ and $b = \overline{7} \neq\overline{a + 1 - (2r - 1)}$ thanks to the edge between $\overline{b  + 1} = \overline{8}$ and $\overline{b + 1  - (2r - 1)} = \overline{3}$, represented with the dotted blue line.}
\end{figure}

The edge $ \overline{b + 1} \to \overline{b + 1 - (2r - 1)} $is a valid edge in the original graph and its two incident vertices are distinct from the removed vertices $\oa$ and $\ob$:
\begin{itemize}
    \item $ \overline{b + 1} \neq \oa $ by hypothesis
    \item $ \overline{b + 1} \neq \ob $ since $ \overline{1} \neq \ozero $ in $ \Z/2r^2\Z$

    \item $\overline{b + 1 - (2r -1) } \neq \oa $ by hypothesis
    \item $\overline{b + 1 - (2r - 1)} \neq \ob  $ in $ \Z/2r^2\Z$ since $ 0 < 2r  - 2 < 2r^2$,
    for $ r \geq 3$.
\end{itemize}

Consequently, regardless of the specific pair of vertices removed, the remaining graph$(\Z/2r^2\Z - \{ \overline{a}, \overline{b}\} \, , \oun , \, \overline{2r - 1})$ remains connected. Therefore, the graph $(\Z/2r^2\Z,  \, \oun , \, \overline{2r - 1})$ is triply-connected. \qed

\begin{lemma}
\label{lemma: 2D surface pair 3-connexe}
    For an integer $r \geq 3$, let $\LL = \mathbb{Z}\begin{pmatrix} r \\ r \end{pmatrix} \oplus \mathbb{Z}\begin{pmatrix} 2r \\ 0 \end{pmatrix}$.
    Then the Cayley graph $(\mathbb{Z}^2/\LL, \begin{pmatrix} 1 \\ 0 \end{pmatrix} \mod \LL, \begin{pmatrix} 0 \\ 1 \end{pmatrix} \mod \LL)$ that underlies the optimal even-distance 2D surface code with parameters $[[ 4r^2, 2, 2r ]]$ (as introduced in Section \ref{sec: optimal surface codes}) is triply-connected.
\end{lemma}

\textbf{Proof of Lemma \ref{lemma: 2D surface pair 3-connexe}:} Our proof proceeds into several steps. 

\begin{enumerate}
    \item First, we will demonstrate that the graph $ \Z^2/\LL - \{ P, Q \}$ remains connected for any two distinct non-zero points $ P, Q \neq \zeroL$. Our strategy here involves constructing three independent paths (paths that share no common vertices except their start and end points) between any non-zero point in the quotient group  $ \Z^2/ \LL$ and $ \zeroL$. This construction ensures that any pair distinct non-zero points , neither of which is $P$ or $Q$, will have a path connecting them that avoids crossing $P$ or $Q$.

    \item Second, we will establish the connectivity of the graph $ \Z^2/ \LL - \{ \zeroL, Q\}$  for any non-zero point $ Q  \neq \zeroL$. We will achieve this by exhibiting an isomorphism between this graph and a known connected graph. \\ 
\end{enumerate}

\underline{First Step: Building independent paths:} \\

The vertices of our graph reside in the quotient space $ \Z^2/\LL$ where the lattice $ \LL$ is defined by $  \LL = \mathbb{Z}\begin{pmatrix} r \\ r \end{pmatrix} \oplus \mathbb{Z}\begin{pmatrix} 2r \\ 0 \end{pmatrix}$. 

To facilitate the construction of paths that share no common intermediate vertices, we will work with a canonical representative for each equivalence class. Every element of $ \Z^2$ can be uniquely expressed as the sum of an element from $ \LL$ and an element of $ \Z^2 \cap \mathcal{P} $ where $ \mathcal{P}$, the fundamental domain of $ \LL$ associated to the basis $ \begin{pmatrix}
    2r \\ 0 
\end{pmatrix} \begin{pmatrix}
    r \\ r 
\end{pmatrix} $, is given by: 
\[ \mathcal{P} = \{ \lambda \begin{pmatrix}
    2r \\ 0 
\end{pmatrix}  + \mu 
\begin{pmatrix}
    r \\ r 
\end{pmatrix} \quad \vert \quad 0 \leq \lambda, \mu < 1 \} \]
    
The elements of $\mathbb{Z}^2 \cap \mathcal{P} $ are integer points $\begin{pmatrix}
        x \\ y 
    \end{pmatrix}$ satisfying the conditions $0 \leq y < r$ and $y \leq x < y + 2r$. Since, by definition, any two distinct points within the fundamental domain $\mathcal{P}$ are inequivalent modulo $\LL$, these integer points precisely constitute a unique set of representatives for the equivalence classes of $\mathbb{Z}^2 / \LL$. Consequently, building independent paths simplifies to constructing paths that traverse different representatives within this fundamental domain. \\

Throughout the remainder of this proof, we will refer to an equivalence class $\begin{pmatrix} x \\ y \end{pmatrix} \mod{\LL}$ simply by its integer representative $ \begin{pmatrix}
    x \\ y 
\end{pmatrix}$ where $0 \leq y < r$ and $y \leq x < y + 2r$. \\

Let $ R =  \begin{pmatrix}
    a \\ b 
\end{pmatrix} $ be a non-zero integer representative with $ 0 \leq b < r$ and $ b \leq a < b  + 2r$. Our goal is to construct three independent paths connecting $\begin{pmatrix}
    0 \\ 0
\end{pmatrix}$ to $ R$. \\

We start by noticing that, since $ R \neq \begin{pmatrix}
    0 \\ 0
\end{pmatrix}$ and $ a \geq b \geq 0$ then $ a > 0$. \\

\underline{Case 1: $ 0 < a \leq r $} (see Figure \ref{fig:cas-a-between-0-and-r})

\begin{itemize}
    \item \textbf{Path 1 (green path in Figure \ref{fig:cas-a-between-0-and-r}):} This path proceeds rightwards along the x-axis from the origin to the vertex $\begin{pmatrix} a \\ 0 \end{pmatrix}$. Subsequently, it moves upwards from $\begin{pmatrix} a \\ 0 \end{pmatrix}$ along the vertical axis to reach the destination $R = \begin{pmatrix} a \\ b \end{pmatrix}$.

    \item  \textbf{Path 2 (red solid and dashed line path in Figure \ref{fig:cas-a-between-0-and-r}):} This path leverages the equivalence of the origin $\begin{pmatrix} 0 \\ 0 \end{pmatrix}$ to $\begin{pmatrix} r \\ r \end{pmatrix}$ modulo $\LL$. It begins at $\begin{pmatrix} 0 \\ 0 \end{pmatrix}$, moves downwards to $\begin{pmatrix} r \\ b \end{pmatrix}$. Then proceeds leftwards from $\begin{pmatrix} r \\ b \end{pmatrix}$ to connect to $R= \begin{pmatrix} a \\ b \end{pmatrix}$.

    \item \textbf{Path 3: (dotted blue path in Figure \ref{fig:cas-a-between-0-and-r})} This path initiates from $\begin{pmatrix} 0 \\ 0 \end{pmatrix}$, which is equivalent to $\begin{pmatrix} 2r \\ 0 \end{pmatrix}$ modulo $\LL$. It then ascends to $\begin{pmatrix} 2r \\ b + 1 \end{pmatrix}$, proceeds rightwards to $\begin{pmatrix} a \\ b + 1 \end{pmatrix}$, and finally descends one step to reach $R = \begin{pmatrix} a \\ b \end{pmatrix}$. \\
\end{itemize}

\tikzset{
    path2_custom_dash/.style={
        line width=2pt,
        red!70!black,
        % Motif: long tiret, petit espace, point, petit espace, long tiret...
        dash pattern=on 4mm off 2mm on 0.5mm off 2mm,
    }
}
\begin{figure}[htbp]
\centering
\begin{tikzpicture}[scale=0.53]

    \def\r{6}      % Parameter r
    \def\a{4}      % x-coordinate of the endpoint
    \def\b{2}      % y-coordinate of the endpoint

    % Axes with arrows (no ticks/labels)
    \draw[->] (0,0) -- ({3*\r + 1},0) node[right] {$x$};
    \draw[->] (0,0) -- (0,{\r + 1}) node[above] {$y$};

    % Z^2 grid from 0 to 3r horizontally, 0 to r vertically
    \draw[very thin, gray!40] (0,0) grid ({3*\r},\r);

    % Fundamental parallelogram in thick black lines
    \coordinate (P0) at (0,0);
    \coordinate (P1) at (\r,\r);
    \coordinate (P2) at ({2*\r},0);
    \coordinate (P3) at ({3*\r},\r);
    \draw[thick, black] (P0) -- (P1) -- (P3) -- (P2) -- cycle;

\foreach \pt/\labeltext/\pos in {%
    (2*\r,{\b + 1})/$\begin{pmatrix} 2r \\ b+1 \end{pmatrix}$/above left,%
    (\r,\r)/$\begin{pmatrix} r \\ r \end{pmatrix}$/above left,%
    (\r,\b)/$\begin{pmatrix} r \\ b \end{pmatrix}$/below right,%
    (0,0)/$\begin{pmatrix} 0 \\ 0 \end{pmatrix}$/below left,%
    (2*\r,0)/$\begin{pmatrix} 2r \\ 0 \end{pmatrix}$/below,%
    (3*\r,\r)/$\begin{pmatrix} 3r \\ r \end{pmatrix}$/above right,%
    ({\b + 1},{\b + 1})/$\begin{pmatrix} b+1 \\ b+1 \end{pmatrix}$/above left,%
    (\a,0)/$\begin{pmatrix} a \\ 0 \end{pmatrix}$/below%
}
{
    \filldraw[black] \pt circle (3pt);
    \node[\pos, font=\small] at \pt {\labeltext};
}

    % Path 1 (green)
    \draw[line width=2pt, green!80!black, -] (0,0) -- (\a,0) -- (\a,\b);

    % Path 2 (red)
    \draw[path2_custom_dash, -] (\r,\r) -- (\r,\b) -- (\a,\b);

    % Path 3 (blue)
    \draw[line width=2pt, blue!80!black, -, dotted] 
        ({2*\r},0) -- ({2*\r},{\b + 1}) -- 
        ({2*\r + \b + 1},{\b + 1});
    \draw[line width=2pt, blue!80!black, -, dotted] 
        ({\b + 1}, {\b + 1}) -- ({\a},{\b + 1}) -- 
        ({\a},{\b});

    % Legend - shifted to the right outside the drawing area
    \node[green!80!black,right, font=\small] at ({3*\r + 2}, \r) {Path 1};
    \node[red!80!black,right, font=\small] at ({3*\r + 2}, {\r - 0.7}) {Path 2};
    \node[blue!80!black,right, font=\small] at ({3*\r + 2}, {\r - 1.4}) {Path 3};

        % Mark endpoint R
    \filldraw[black] (\a,\b) circle (3pt);
    \node[above right, font=\small] at (\a,\b) {$R$};

\end{tikzpicture}
\caption{Three independent paths from $\begin{pmatrix}0 \\ 0\end{pmatrix}$ to $R=\begin{pmatrix}a \\ b\end{pmatrix}$ on the torus $\mathbb{Z}^2 / \LL$ where $\LL = \Z\begin{pmatrix}
    2r \\ 0
\end{pmatrix} \bigoplus \Z \begin{pmatrix}
    r \\ r 
\end{pmatrix}  $ and parameters satisfying $ 0 < a \leq r $ and $ 0 \leq b < r - 1$. Here  $r=6$, $a=3$ and $b=2$.}
\label{fig:cas-a-between-0-and-r}
\end{figure}
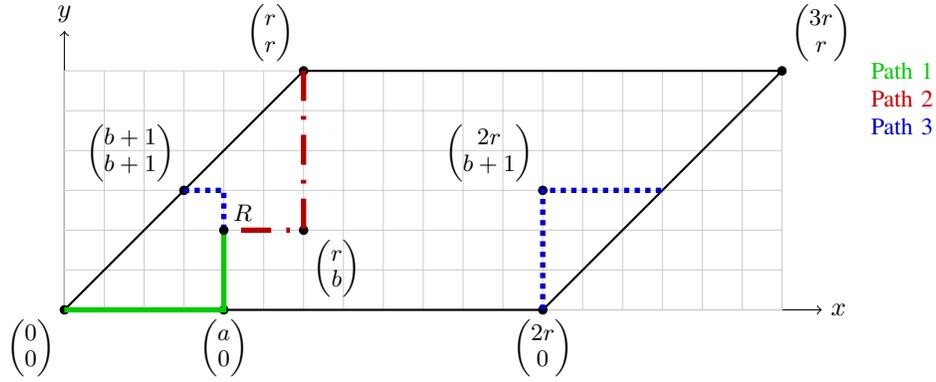

\begin{figure}[htbp]
\centering
\begin{tikzpicture}[scale=0.53]
    \def\r{6}      % Parameter r
    \def\a{\r - 1 } % a = r 
    \def\b{\r - 1} % b = r - 1
    
    % Axes with arrows (no ticks/labels)
    \draw[->] (0,0) -- ({3*\r + 1},0) node[right] {$x$};
    \draw[->] (0,0) -- (0,{\r + 1}) node[above] {$y$};
    
    % Z^2 grid from 0 to 3r horizontally, 0 to r vertically
    \draw[very thin, gray!40] (0,0) grid ({3*\r},\r);
    
    % Fundamental parallelogram in thick black lines
    \coordinate (P0) at (0,0);
    \coordinate (P1) at (\r,\r);
    \coordinate (P2) at ({2*\r},0);
    \coordinate (P3) at ({3*\r},\r);
    \draw[thick, black] (P0) -- (P1) -- (P3) -- (P2) -- cycle;
    
    % Points with labels - only the specified ones
    \foreach \pt/\labeltext/\pos in {%
        (0,0)/$\begin{pmatrix} 0 \\ 0 \end{pmatrix}$/below left,%
        ({\r - 1},0)/$\begin{pmatrix} r - 1 \\ 0 \end{pmatrix}$/below,%
        (\b,\b)/$\begin{pmatrix} r-1 \\ r-1 \end{pmatrix}$/left,%
        (2*\r,\b)/$\begin{pmatrix} 2r \\ r-1 \end{pmatrix}$/left,%
        (2*\r,0)/$\begin{pmatrix} 2r \\ 0 \end{pmatrix}$/below,%
        (3*\r,\r)/$\begin{pmatrix} 3r \\ r \end{pmatrix}$/above right,%
        (\r,\r)/$\begin{pmatrix} r \\ r \end{pmatrix}$/above left%
    }
    {
        \filldraw[black] \pt circle (3pt);
        \node[\pos, font=\small] at \pt {\labeltext};
    }
    
    % Mark endpoint R
    \filldraw[black] (\a,\b) circle (3pt);
    \node[above right, font=\small] at (\a,\b) {$R$};
    
    % Path 1 (green): (0,0) -> (a,0) -> (a,b)
    \draw[line width=2pt, green!80!black, -] (0,0) -- (\a,0) -- (\a,\b);
    
    % Path 2 (red): (0,0) -> (r,r) -> (r,b) -> (a,b)
    % Since a = r-1, we go from (r,r-1) to (r-1,r-1)
    \draw[path2_custom_dash, -] (\r,\r) -- (\r,\b) -- (\a,\b);
    
    % Path 3 (blue): (0,0) -> (2r,0) -> (2r,b) -> (a,b)
    % From (2r,r-1) we need to go to (r-1,r-1)
    % This wraps around: (2r,r-1) -> ... -> (3r-2,r-1) -> (r-1,r-1)
    \draw[line width=2pt, blue!80!black, -, dotted] 
        ({2*\r},0) -- ({2*\r},\b);
    \draw[line width=2pt, blue!80!black, -, dotted] 
        ({2*\r},\b) -- ({3*\r - 1},\b);
    \draw[line width=2pt, blue!80!black, -, dotted] 
        (\b,\b) -- (\a,\b);
    
    % Legend - shifted to the right outside the drawing area
    \node[green!80!black,right, font=\small] at ({3*\r + 2}, \r) {Path 1};
    \node[red!80!black,right, font=\small] at ({3*\r + 2}, {\r - 0.7}) {Path 2};
    \node[blue!80!black,right, font=\small] at ({3*\r + 2}, {\r - 1.4}) {Path 3};
\end{tikzpicture}
\caption{Three independent paths from $\begin{pmatrix}0 \\ 0\end{pmatrix}$ to $R =\begin{pmatrix}a \\ b\end{pmatrix}$ on the torus $\mathbb{Z}^2 / \LL$ where $\LL = \Z\begin{pmatrix}
    2r \\ 0
\end{pmatrix} \bigoplus \Z \begin{pmatrix}
    r \\ r 
\end{pmatrix}  $ and parameters satisfying $  r - 1 \leq a \leq   r $ and $ b  = r - 1$. 
Here $ r = 6$ and $ a = b = r - 1$.}
\label{fig:cas-a-between-0-and-r et b=r-1}
\end{figure}
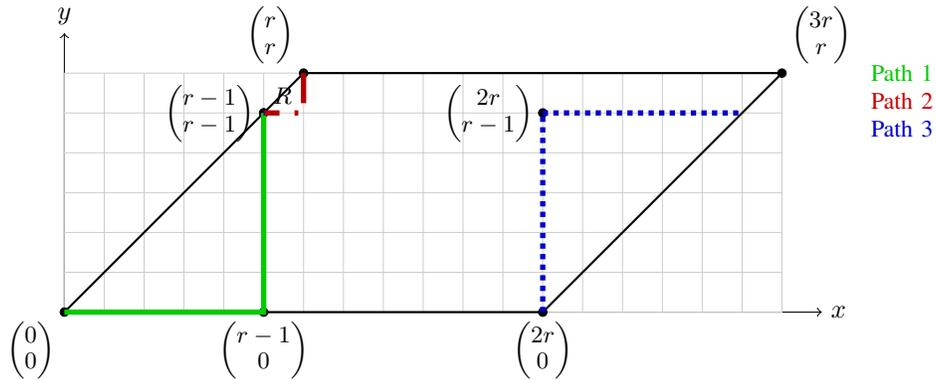

\FloatBarrier
\underline{Case 1.2: $ b = r - 1 $} (see Figure \ref{fig:cas-a-between-0-and-r et b=r-1})

% Cas b = r - 1
\begin{itemize}
    \item \textbf{Path 1 (green path in Figure \ref{fig:cas-a-between-0-and-r et b=r-1}):} This path starts at the origin $\begin{pmatrix} 0 \\ 0 \end{pmatrix}$, moves right along the x-axis to $\begin{pmatrix} a \\ 0 \end{pmatrix}$, and then ascends vertically to reach $R = \begin{pmatrix} a \\ r - 1 \end{pmatrix}$.

\item \textbf{Path 2 (red solid and dashed line in Figure \ref{fig:cas-a-between-0-and-r et b=r-1}):} This path starts from $\begin{pmatrix} 0 \\ 0 \end{pmatrix}$, which is equivalent to $\begin{pmatrix} r \\ r \end{pmatrix}$ modulo $\LL$. It then move downwards from $\begin{pmatrix} 0\\ 0\end{pmatrix}$ to $\begin{pmatrix} r \\ r - 1 \end{pmatrix}$, before proceeding leftward to arrive at $R = \begin{pmatrix} a \\ r - 1 \end{pmatrix}$. 

\item \textbf{Path 3 (dotted blue line path in Figure \ref{fig:cas-a-between-0-and-r et b=r-1}):} This path begins at $\begin{pmatrix} 0 \\ 0 \end{pmatrix}$, which is equivalent to $\begin{pmatrix} 2r \\ 0 \end{pmatrix}$ modulo $\LL$. It then ascends to $\begin{pmatrix} 2r \\ r - 1 \end{pmatrix}$. From this point, it moves rightward to $\begin{pmatrix} a \\ r - 1 \end{pmatrix}$. 
\end{itemize}

\underline{Case 2: $r < a < 2r $}

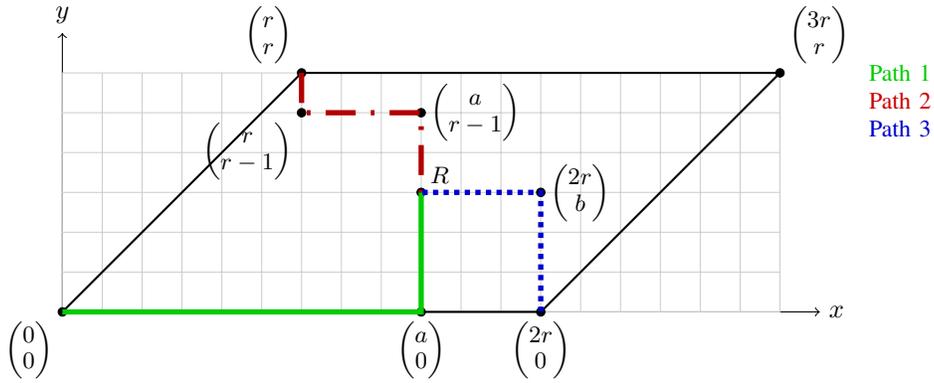
\begin{figure}[htbp]
\centering
\begin{tikzpicture}[scale=0.53]
    \def\r{6}      % Parameter r
    \def\a{9}      
    \def\b{3}      
    
    % Axes with arrows (no ticks/labels)
    \draw[->] (0,0) -- ({3*\r + 1},0) node[right] {$x$};
    \draw[->] (0,0) -- (0,{\r + 1}) node[above] {$y$};
    
    % Z^2 grid from 0 to 3r horizontally, 0 to r vertically
    \draw[very thin, gray!40] (0,0) grid ({3*\r},\r);
    
    % Fundamental parallelogram in thick black lines
    \coordinate (P0) at (0,0);
    \coordinate (P1) at (\r,\r);
    \coordinate (P2) at ({2*\r},0);
    \coordinate (P3) at ({3*\r},\r);
    \draw[thick, black] (P0) -- (P1) -- (P3) -- (P2) -- cycle;
    
    % Points with labels
    \foreach \pt/\labeltext/\pos in {%
        (0,0)/$\begin{pmatrix} 0 \\ 0 \end{pmatrix}$/below left,%
        (\a,0)/$\begin{pmatrix} a \\ 0 \end{pmatrix}$/below,%
        (2*\r,0)/$\begin{pmatrix} 2r \\ 0 \end{pmatrix}$/below,%
        (3*\r,\r)/$\begin{pmatrix} 3r \\ r \end{pmatrix}$/above right,%
        (\r,\r)/$\begin{pmatrix} r \\ r \end{pmatrix}$/above left,%
        (\r,{\r-1})/$\begin{pmatrix} r \\ r-1 \end{pmatrix}$/below left,%
        (\a,{\r-1})/$\begin{pmatrix} a \\ r-1 \end{pmatrix}$/right,%
        (2*\r,\b)/$\begin{pmatrix} 2r \\ b \end{pmatrix}$/right%
    }
    {
        \filldraw[black] \pt circle (3pt);
        \node[\pos, font=\small] at \pt {\labeltext};
    }
    
    % Mark endpoint R
    \filldraw[black] (\a,\b) circle (3pt);
    \node[above right, font=\small] at (\a,\b) {$R$};
    
    % Path 1 (green): (0,0) -> (a,0) -> (a,b)
    \draw[line width=2pt, green!80!black, -] (0,0) -- (\a,0) -- (\a,\b);
    
    % Path 2 (red): (0,0) -> (r,r) -> (r,r-1) -> (a,r-1) -> (a,b)
    \draw[path2_custom_dash, -] (\r,\r) -- (\r,{\r-1}) -- (\a,{\r-1}) -- (\a,\b);
    
    % Path 3 (blue): (0,0) -> (2r,0) -> (2r,b) -> (a,b)
    \draw[line width=2pt, blue!80!black, -, dotted] 
        ({2*\r},0) -- ({2*\r},\b) -- (\a,\b);
    
    % Legend - shifted to the right outside the drawing area
    \node[green!80!black,right, font=\small] at ({3*\r + 2}, \r) {Path 1};
    \node[red!80!black,right, font=\small] at ({3*\r + 2}, {\r - 0.7}) {Path 2};
    \node[blue!80!black,right, font=\small] at ({3*\r + 2}, {\r - 1.4}) {Path 3};
\end{tikzpicture}
\caption{Three independent paths from $\begin{pmatrix}0 \\ 0\end{pmatrix}$ to $R=\begin{pmatrix}a \\ b\end{pmatrix}$ on the torus $\mathbb{Z}^2 / \LL$ where $\LL = \Z\begin{pmatrix}
    2r \\ 0
\end{pmatrix} \bigoplus \Z \begin{pmatrix}
    r \\ r 
\end{pmatrix}  $ and parameters satisfying $ r < a < 2r$ and $0 \leq  b < r$. Here $ r = 6$, $ a = 9$ and $ b = 3$.}
\label{fig: r<a<2r et b<r}
\end{figure}

\begin{itemize}
    \item \textbf{Path 1 (green path in Figure \ref{fig: r<a<2r et b<r}):} This path begins at the origin $\begin{pmatrix} 0 \\ 0 \end{pmatrix}$ and proceeds rightward along the x-axis to reach $\begin{pmatrix} a \\ 0 \end{pmatrix}$. From there, it ascends vertically to reach $R=\begin{pmatrix} a \\ b \end{pmatrix}$.

    \item \textbf{Path 2 (red solid and dashed line in Figure \ref{fig: r<a<2r et b<r}):} This path starts from $\begin{pmatrix} 0 \\ 0 \end{pmatrix}$, which is equivalent to $\begin{pmatrix} r \\ r \end{pmatrix}$ modulo $\LL$. It first moves downwards to $\begin{pmatrix} r \\ r - 1 \end{pmatrix}$, then proceeds rightward  to $\begin{pmatrix} a \\ r - 1 \end{pmatrix}$. Finally, it descends  to reach the point $R=\begin{pmatrix} a \\ b \end{pmatrix}$. 

    \item \textbf{Path 3 (blue dotted line path in Figure \ref{fig: r<a<2r et b<r}):} This path begins at $\begin{pmatrix} 0 \\ 0 \end{pmatrix}$, which is equivalent to $\begin{pmatrix} 2r \\ 0 \end{pmatrix}$ modulo $\LL$. It then ascends vertically to $\begin{pmatrix} 2r \\ b \end{pmatrix}$. From this point, it proceeds leftward to connect to the point $R=\begin{pmatrix} a \\ b \end{pmatrix}$. 
\end{itemize}

\underline{Case 3: $ 2r \leq a \leq 3r - 2 $} \\

If $ a$ can be equal to $2r$, we have to take different shape of paths for the first and third path. Additionally, notice that since $ a $ is greater or equal to $ 2r$ and must be strictly less than $ b + 2r$, then $ b $ must be positive. \\

\underline{Case 3.1: $ b < r - 1  $} \\

\begin{figure}[htbp]
\centering
\begin{tikzpicture}[scale=0.53]
    \def\r{6}      % Parameter r
    \def\a{13}     % a = 13
    \def\b{3}      % b = 3
    
    % Axes with arrows (no ticks/labels)
    \draw[->] (0,0) -- ({3*\r + 1},0) node[right] {$x$};
    \draw[->] (0,0) -- (0,{\r + 1}) node[above] {$y$};
    
    % Z^2 grid from 0 to 3r horizontally, 0 to r vertically
    \draw[very thin, gray!40] (0,0) grid ({3*\r},\r);
    
    % Fundamental parallelogram in thick black lines
    \coordinate (P0) at (0,0);
    \coordinate (P1) at (\r,\r);
    \coordinate (P2) at ({2*\r},0);
    \coordinate (P3) at ({3*\r},\r);
    \draw[thick, black] (P0) -- (P1) -- (P3) -- (P2) -- cycle;
    
    % Points with labels
    \foreach \pt/\labeltext/\pos in {%
        (0,0)/$\begin{pmatrix} 0 \\ 0 \end{pmatrix}$/below left,%
        ({2*\r-1},0)/$\begin{pmatrix} 2r-1 \\ 0 \end{pmatrix}$/below left,%
        ({2*\r-1},\b)/$\begin{pmatrix} 2r-1 \\ b \end{pmatrix}$/left,%
        (2*\r,0)/$\begin{pmatrix} 2r \\ 0 \end{pmatrix}$/below right,%
        (\r,\r)/$\begin{pmatrix} r \\ r \end{pmatrix}$/above left,%
        (\r,{\r-1})/$\begin{pmatrix} r \\ r-1 \end{pmatrix}$/below,%
        (\a,{\r-1})/$\begin{pmatrix} a \\ r-1 \end{pmatrix}$/right,%
        (3*\r,\r)/$\begin{pmatrix} 3r \\ r \end{pmatrix}$/above right,%
        (\a,{\b-1})/$\begin{pmatrix} a \\ b-1 \end{pmatrix}$/below right%
    }
    {
        \filldraw[black] \pt circle (3pt);
        \node[\pos, font=\small] at \pt {\labeltext};
    }

    % Path 1 (green): (0,0) -> (2r-1,0) -> (2r-1,b) -> (a,b)
    \draw[line width=2pt, green!80!black, -] (0,0) -- ({2*\r-1},0) -- ({2*\r-1},\b) -- (\a,\b);
    
    % Path 2 (red): (0,0) -> (r,r) -> (r,r-1) -> (a,r-1) -> (a,b)
    \draw[path2_custom_dash, -] (\r,\r) -- (\r,{\r-1}) -- (\a,{\r-1}) -- (\a,\b);
    
    % Path 3 (blue): (0,0) -> (2r,0) -> (2r,b-1) -> (a,b-1) -> (a,b)
    \draw[line width=2pt, blue!80!black, -, dotted] 
        ({2*\r},0) -- ({2*\r},{\b-1}) -- (\a,{\b-1}) -- (\a,\b);
    
    % Legend - shifted to the right outside the drawing area
    \node[green!80!black,right, font=\small] at ({3*\r + 2}, \r) {Path 1};
    \node[red!80!black,right, font=\small] at ({3*\r + 2}, {\r - 0.7}) {Path 2};
    \node[blue!80!black,right, font=\small] at ({3*\r + 2}, {\r - 1.4}) {Path 3};

        % Mark endpoint R
    \filldraw[black] (\a,\b) circle (3pt);
    \node[right, font=\small] at (\a,\b) {$R$};
\end{tikzpicture}
\caption{Three independent paths from $\begin{pmatrix}0 \\ 0\end{pmatrix}$ to $R=\begin{pmatrix}a \\ b\end{pmatrix}$ on the torus $\mathbb{Z}^2 / \LL$ where $\LL = \Z\begin{pmatrix}
    2r \\ 0
\end{pmatrix} \bigoplus \Z \begin{pmatrix}
    r \\ r 
\end{pmatrix}  $ and parameters satisfying $ 2r \leq a  \leq 3r - 2$ and $0 <  b < r - 1$. Here $ r = 6$, $ a = 13$ and $ b = 3$.}

\label{fig: 2r <= a <= 3r - 2 et b < r-1}
\end{figure}
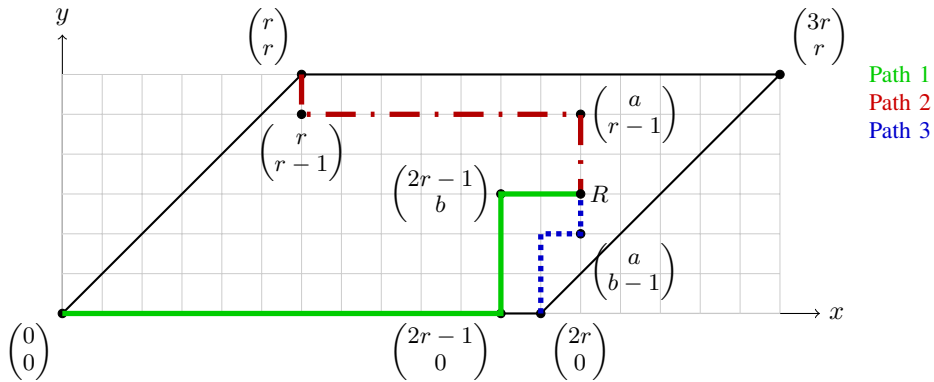

\begin{itemize}
    \item \textbf{Path 1 (green path in Figure \ref{fig: 2r <= a <= 3r - 2 et b < r-1}):} This path begins at the origin $\begin{pmatrix} 0 \\ 0 \end{pmatrix}$ and moves rightward along the x-axis to  $\begin{pmatrix} 2r - 1 \\ 0 \end{pmatrix}$. It then ascends vertically to $\begin{pmatrix} 2r - 1 \\ b \end{pmatrix}$ before continuing rightward again to reach the point $R=\begin{pmatrix} a \\ b \end{pmatrix}$.
\end{itemize}

\textbf{Path 2 (red and solid dashed line in Figure \ref{fig: 2r <= a <= 3r - 2 et b < r-1}) :} This path starts from $\begin{pmatrix} 0 \\ 0 \end{pmatrix}$. Utilizing its equivalence to $\begin{pmatrix} r \\ r \end{pmatrix}$ modulo $\LL$, It moves downwards from $\begin{pmatrix} 0 \\ 0 \end{pmatrix}$ to $\begin{pmatrix} r \\ r - 1 \end{pmatrix}$, then proceeds rightward to $\begin{pmatrix} a \\ r - 1 \end{pmatrix}$ and finally descends to $R=\begin{pmatrix} a \\ b \end{pmatrix}$.

\textbf{Path 3 (blue dotted line in Figure \ref{fig: 2r <= a <= 3r - 2 et b < r-1}):} This path initiates at $\begin{pmatrix} 0 \\ 0 \end{pmatrix}$, which is equivalent to $\begin{pmatrix} 2r \\ 0 \end{pmatrix}$ modulo $\LL$. It then ascends vertically to $\begin{pmatrix} 2r \\ b - 1 \end{pmatrix}$. From this point, it proceeds rightward to $\begin{pmatrix} a \\ b - 1 \end{pmatrix}$ and then makes a final upward step to $\begin{pmatrix} a \\ b \end{pmatrix}$. \\

\FloatBarrier

\underline{Case 3.2: $ b =  r - 1  \geq 2$}

\begin{figure}[htbp]
\centering
\begin{tikzpicture}[scale=0.53]
    \def\r{6}      % Parameter r
    \def\a{13}     % a = 13
    \def\b{\r - 1} % b = r - 1 = 5
    
    % Axes with arrows (no ticks/labels)
    \draw[->] (0,0) -- ({3*\r + 1},0) node[right] {$x$};
    \draw[->] (0,0) -- (0,{\r + 1}) node[above] {$y$};
    
    % Z^2 grid from 0 to 3r horizontally, 0 to r vertically
    \draw[very thin, gray!40] (0,0) grid ({3*\r},\r);
    
    % Fundamental parallelogram in thick black lines
    \coordinate (P0) at (0,0);
    \coordinate (P1) at (\r,\r);
    \coordinate (P2) at ({2*\r},0);
    \coordinate (P3) at ({3*\r},\r);
    \draw[thick, black] (P0) -- (P1) -- (P3) -- (P2) -- cycle;
    
    % Points with labels
    \foreach \pt/\labeltext/\pos in {%
        (0,0)/$\begin{pmatrix} 0 \\ 0 \end{pmatrix}$/below left,%
        (2*\r,0)/$\begin{pmatrix} 2r \\ 0 \end{pmatrix}$/below right,%
        ({\r - 1},0)/$\begin{pmatrix} r - 1 \\ 0 \end{pmatrix}$/below,%
        (\r,\r)/$\begin{pmatrix} r \\ r \end{pmatrix}$/above left,%
        ({\r - 1},{\r - 1})/$\begin{pmatrix} r - 1 \\ r - 1 \end{pmatrix}$/left,%
        (\r,{\r-1})/$\begin{pmatrix} r \\ r-1 \end{pmatrix}$/below right,%
        (3*\r,\r)/$\begin{pmatrix} 3r \\ r \end{pmatrix}$/above right,%
        (\a,{\b-1})/$\begin{pmatrix} a \\ b-1 \end{pmatrix}$/below right%
    }
    {
        \filldraw[black] \pt circle (3pt);
        \node[\pos, font=\small] at \pt {\labeltext};
    }

    % Path 1 (green): (0,0) -> (2r-1,0) -> (2r-1,b) -> (a,b)
    \draw[line width=2pt, green!80!black, -] (0,0) -- ({\r-1},0) -- ({\r-1},{\r - 1}) ;
    \draw[line width=2pt, green!80!black, -] ({3*\r-1}, {\r - 1}) -- ({\a},{\r - 1}) ;
    
    % Path 2 (red): (0,0) -> (r,r) -> (r,r-1) -> (a,r-1) -> (a,b)
    \draw[path2_custom_dash, -] (\r,\r) -- (\r,{\r-1}) -- (\a,{\r-1}) -- (\a,\b);
    
    % Path 3 (blue): (0,0) -> (2r,0) -> (2r,b-1) -> (a,b-1) -> (a,b)
    \draw[line width=2pt, blue!80!black, -, dotted] 
        ({2*\r},0) -- ({2*\r},{\b-1}) -- (\a,{\b-1}) -- (\a,\b);
    
    % Legend - shifted to the right outside the drawing area
    \node[green!80!black,right, font=\small] at ({3*\r + 2}, \r) {Path 1};
    \node[red!80!black,right, font=\small] at ({3*\r + 2}, {\r - 0.7}) {Path 2};
    \node[blue!80!black,right, font=\small] at ({3*\r + 2}, {\r - 1.4}) {Path 3};

        % Mark endpoint R
    \filldraw[black] (\a,\b) circle (3pt);
    \node[above right, font=\small] at (\a,\b) {$R$};
\end{tikzpicture}
\caption{Three independent paths from $\begin{pmatrix}0 \\ 0\end{pmatrix}$ to $R=\begin{pmatrix}a \\ b\end{pmatrix}$ on the torus $\mathbb{Z}^2 / \LL$ where $\LL = \Z\begin{pmatrix}
    2r \\ 0
\end{pmatrix} \bigoplus \Z \begin{pmatrix}
    r \\ r 
\end{pmatrix}  $ and parameters satisfying $ 2r \leq a \leq 3r - 2$ and $b = r - 1$. Here $ r = 6$, $ a = 13$ and $ b =  5$.}

\label{fig: 2r <= a <= 3r - 2 et b = r - 1}
\end{figure}
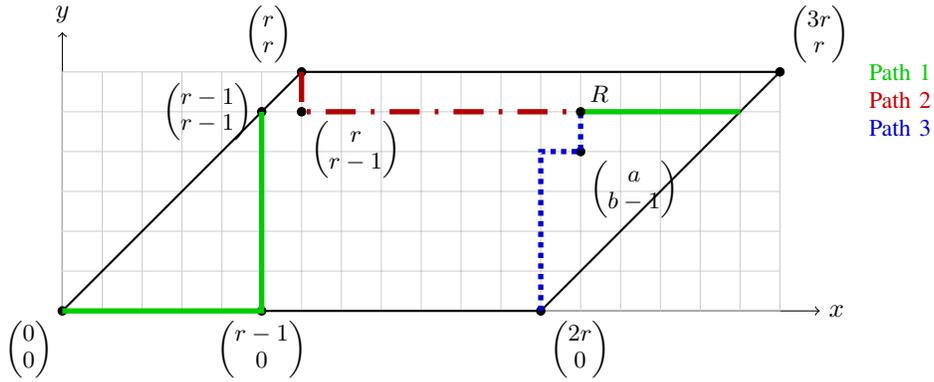

\begin{itemize}
    \item \textbf{Path 1 (green path in Figure \ref{fig: 2r <= a <= 3r - 2 et b = r - 1}:} This path starts at the origin $\begin{pmatrix} 0 \\ 0 \end{pmatrix}$ and moves rightward along the x-axis to $\begin{pmatrix} r - 1 \\ 0 \end{pmatrix}$. It then ascends vertically to $\begin{pmatrix} r - 1 \\ r - 1 \end{pmatrix}$. Finally, it proceeds leftward from $\begin{pmatrix} r - 1 \\ r - 1 \end{pmatrix}$ to reach the destination $R=\begin{pmatrix} a \\ r - 1 \end{pmatrix}$.

    \item \textbf{Path 2 (red solid and dashed lines in Figure \ref{fig: 2r <= a <= 3r - 2 et b = r - 1}):} This path begins at $\begin{pmatrix} 0 \\ 0 \end{pmatrix}$, which is equivalent to $\begin{pmatrix} r \\ r \end{pmatrix}$ modulo $\LL$. It first descends to $\begin{pmatrix} r \\ r - 1 \end{pmatrix}$, then continue rightward to  $R=\begin{pmatrix} a \\ r - 1 \end{pmatrix}$.

    \item \textbf{Path 3 (dotted blue line in Figure \ref{fig: 2r <= a <= 3r - 2 et b = r - 1}):} This path starts at $\begin{pmatrix} 0 \\ 0 \end{pmatrix}$, which is equivalent to $\begin{pmatrix} 2r \\ 0 \end{pmatrix}$ modulo $\LL$. It first ascends vertically to $\begin{pmatrix} 2r \\ r - 2 \end{pmatrix}$, then moves rightward to $\begin{pmatrix} a \\ r - 2 \end{pmatrix}$, and finally ascends one step to $\begin{pmatrix} a \\ r - 1 \end{pmatrix}$. 
\end{itemize}

In every case, using vertices of $\mathbb{Z}^2 \cap \mathcal{P}$, we have constructed 3 paths between $\begin{pmatrix}
    0 \\ 0
\end{pmatrix}$ and $R$ paths that share no common vertices except for their endpoints.  This finding is significant: for any four distinct non-zero points $(P, Q, R, S)$ in $\mathbb{Z}^2/\LL$, there is always at least one path between $R$ and $S$ that completely avoids $P$ and $Q$. 

Therefore, the graph $\mathbb{Z}^2/\LL - \{P, Q\}$ remains connected. \\

\underline{Second step: $ \Z^2/\LL - \{\zeroL, Q\} $ is connected for $ Q \neq \zeroL$} \\

Since the order of the group $|\mathbb{Z}^2/\LL|$ is $2r^2$ where $r \geq 3$, we can always select a vertex $P \in \mathbb{Z}^2/\LL$ such that $P \neq \zeroL$ and $P \neq -Q$. 

The map $ \psi$ : $ T \in \mathbb{Z}^2/\LL - \{ \zeroL, Q\} \to T + P \in  \mathbb{Z}^2/\LL - \{ P, P + Q\} $ induces a graph isomorphism. From our previous result (Step One), we know that the graph $ \Z^2/ \LL$ with two distinct non-zero points removed remains connected. Since $P \neq \zeroL$, $P + Q \neq \zeroL$, and $P + Q \neq P$, the graph $\mathbb{Z}^2/\LL - \{ P, P + Q\}$ is connected.
Therefore, its isomorphic counterpart, $\mathbb{Z}^2/\LL - \{ \zeroL, Q\}$, must also be connected. \\

In conclusion, we have successfully demonstrated that the graph $\mathbb{Z}^2/\LL$ remains connected after removing any two vertices, covering both scenarios:

\begin{itemize}
    \item Removing two distinct non-zero vertices $P, Q$ from $ \mathbb{Z}^2/\LL$.
    \item Removing $\zeroL$ and any other non-zero vertex $Q $ from $ \mathbb{Z}^2/\LL $.
\end{itemize}

Thus, we conclude that $\mathbb{Z}^2/\LL$ is triply-connected. \qed 

\subsubsection{Non-CGP Equivalence of the codes}

\begin{proposition}
\label{prop: kitaev pair pas eq a un GB}
    For $ r \geq 3 $, the optimal even-distance 2D surface code with periodicity vectors $ \begin{pmatrix}
    r \\ r
\end{pmatrix}, \, \begin{pmatrix}
    r \\ - r
\end{pmatrix}$  and $ GB(1 + X, 1 + X^{2r-1}, 2r^2)  $ are not CGP-equivalent. 
\end{proposition}

\textbf{Proof of Proposition \ref{prop: kitaev pair pas eq a un GB}:} Suppose, for contradiction, that these two codes are CGP-equivalent. 

These two codes come from the Cayley graphs:
\begin{itemize}
    \item $(\mathbb{Z}^2/\LL, \begin{pmatrix} 1 \\ 0 \end{pmatrix} \mod{\LL}, \begin{pmatrix} 0 \\ 1 \end{pmatrix} \mod{\LL})$, where $\LL = \mathbb{Z}\begin{pmatrix} r \\ r \end{pmatrix} \oplus \mathbb{Z}\begin{pmatrix} r \\ - r \end{pmatrix}$
        \item $(\mathbb{Z}/2r^2\mathbb{Z}, \overline{1}, \overline{2r - 1})$ 
\end{itemize}

As discussed in Section \ref{sec: graph compatibily} and Remarks \ref{rem: simple cycles corresponds to simple cycles}, their underlying graphs are 2-isomorphic, meaning there is a direct correspondence between their simple cycles. We have already shown in Lemma \ref{lemma: Z/2r^2Z 3-connected} and Lemma \ref{lemma: 2D surface pair 3-connexe} that both of these graphs are triply-connected. Thus, according to Whitney's Theorem (Theorem \ref{thm: whitney thm}), these two 2-isomorphic triply-connected graphs are, in fact, isomorphic in the usual sense. \\

Now, we will explain how this graph isomorphism leads to a contradiction by following this plan:
\begin{itemize}
    \item First, we will prove that the graph isomorphism between $(\mathbb{Z}/2r^2\mathbb{Z}, \overline{1}, \overline{2r - 1})$ and $(\mathbb{Z}^2/\LL, \begin{pmatrix} 1 \\ 0 \end{pmatrix} \mod{\LL}, \begin{pmatrix} 0 \\ 1 \end{pmatrix} \mod{\LL})$ induces a group isomorphism between $\mathbb{Z}/2r^2\mathbb{Z}$ and $\mathbb{Z}^2/\LL$.
    \item Then, we will prove that these two groups cannot be isomorphic because $\mathbb{Z}^2/\LL$ lacks an element of order $2r^2$.
\end{itemize}

\subsubsection*{First Step: From graph isomorphism to group isomorphism:} 

Let $ h : (\Z/2r^2 \Z, \, \overline{1}, \overline{2r-1}) \to( \Z^2/\LL, \begin{pmatrix}
        1 \\ 0
    \end{pmatrix} \mod \LL, \begin{pmatrix}
        0 \\ 1
    \end{pmatrix} \mod \LL) $ be the graph isomorphism and let $ \psi $ be the map defined as:
\[ \begin{array}{ccccl}
\psi& : &  \Z/2r^2 \Z & \xrightarrow{} & \Z^2/\LL \\
& & \overline{T} & \longmapsto & h(\overline{T}) - h(\overline{0})
\end{array} \]

\subsubsection*{$ \psi$ is a graph isomorphism} 

Its bijectivity is ensured by the bijectivity of $h$. Furthermore, $ \psi $ preserves neighbourhood relationships: $\overline{x}$ and $\overline{y}$ are neighbours in $ (\Z/2r^2 \Z, \, \overline{1}, \overline{2r-1}) $ if and only if $h(\overline{x}) $ and $ h(\overline{y}) $ are neighbours in 
    $( \Z^2/\LL, \begin{pmatrix}
        1 \\ 0
    \end{pmatrix} \mod \LL, \begin{pmatrix}
        0 \\ 1
    \end{pmatrix} \mod \LL)$ which is equivalent to asking $\psi(\overline{x})$  and  $\psi(\overline{y})$  are neighbours in 
    $( \Z^2/\LL, \begin{pmatrix}
        1 \\ 0
    \end{pmatrix} \mod \LL, \begin{pmatrix}
        0 \\ 1
    \end{pmatrix} \mod \LL) $.

\subsubsection*{ $\psi $ is a group morphism}

Let $ \psi(\oun) = U$.
We will demonstrate that $\forall k \in \Z, \psi(\overline{k}) = kU $. We break down the proof into two steps: first, we show by induction that this equality holds for all non-negative $ k$.
Second, we are going to extend this to all negative integers. \\

Let's demonstrate by induction that $\forall k \in \N , \, \, P(k): \psi(\overline{k}) = kU $ holds true. \\

$P(0)$ and $ P(1)$ hold true, since $ \psi(\overline{0}) = \zeroL $ and $ \psi(\oun) = U$. 
Now, assume for some positive integer $ k \geq 1 $ that the property holds for all $  0 \leq t \leq k$, i.e $ \psi(\overline{t}) = tU$. We aim to show that $ \psi(\overline{k + 1}) = (k +1)U $. 

Since $ \psi $ is a graph isomorphism between $ (\Z/2r^2\Z, \, \,  \oun, \, \, \overline{2r - 1})$
and $ (\Z^2/\LL, \begin{pmatrix}
    1 \\0 
\end{pmatrix}\mod \LL , \, \begin{pmatrix}
    0 \\1 
\end{pmatrix} \mod \LL)$ and knowing that $ \overline{0}$ and $ \overline{1}$ are neighbours in  $ \Z/2r^2\Z$ it follows that $ \psi(\overline{1})$ is a neighbour of $ \psi(\overline{0}) = \zeroL $ in  $ \Z^2/\LL$. Therefore, $ U =  \psi(\overline{1}) $ belongs to $ \{ \pm \begin{pmatrix}
        1 \\ 0
    \end{pmatrix}  \mod \LL, \, \begin{pmatrix}
        0 \\ 1
    \end{pmatrix} \mod \LL\}$. \\

Now let $ V \in \Z^2 / \LL $ be such that: $\{ \pm \begin{pmatrix}
        1 \\ 0
    \end{pmatrix}  \mod \LL, \, \begin{pmatrix}
        0 \\ 1
    \end{pmatrix} \mod \LL\} = \{ \pm U, \pm V \}$. For $ r \geq 3$, the elements $ \zeroL, U, V, -U, -V $ are pairwise distinct. This is because vectors such as $ \begin{pmatrix}
        \pm1 \\ 0
    \end{pmatrix}, \begin{pmatrix}
        0 \\ \pm 1
    \end{pmatrix}, 
\begin{pmatrix}
        \pm1 \\ \pm 1
    \end{pmatrix}, 
, \begin{pmatrix}
        \pm2 \\ 0
    \end{pmatrix}, \begin{pmatrix}
        0 \\ \pm 2
    \end{pmatrix}
      $ are not contained in  $  \LL = \Z\begin{pmatrix}
        r \\ r
    \end{pmatrix} \bigoplus \Z\begin{pmatrix}   r \\ - r
    \end{pmatrix}$ \\

Since $ \psi$ is graph isomorphism, and that  $ \overline{k}$ and $\overline{k  + 1}$ are neighbours in  $ (\Z/2r^2\Z, \overline{1}, \overline{2r-1}) $, it follows that $ \psi(\overline{k + 1})$ is a neighbour of $ \psi(\overline{k}) = kU$ in $ (\Z^2/ \LL,\, \,  U, \, \, V) $.  Therefore, $ \psi(\overline{k+1}) \in \{ kU \pm U, \, kU \pm V\}$. \\

We will now prove, that the only possible value for $ \psi(\overline{k+1}) $ is $ (k + 1)U$. \\

\underline{Case 1. $ \psi(\overline{k+1}) = (k - 1)U$:} \\

By induction, $ \psi(\overline{k+1}) = (k - 1)U =  \psi(\overline{k -1})$. Thus, due to $ \psi $ bijectivity we must have $ \overline{k +1} = \overline{k- 1}$ in $ \Z/2r^2\Z$. This implies, $ \overline{2} = \overline{0}$ in $ \Z/2r^2\Z$ which is impossible for $ r \geq 3$, since $ 0 < 2 < 2r^2$. \\

\underline{Case 2. $ \psi(\overline{k+1}) = kU \pm V$:} \\

$ \psi(\overline{k + 1})$ belongs to one of these two  $( \Z^2/\LL, U, V) $  $4$-cycles: 
\begin{itemize}
    \item $ kU \to kU + V \to (k-1)U + V \to (k - 1)U \to kU $
    \item $ kU \to kU - V \to (k -1)U - V \to (k-1)U \to kU$
\end{itemize}

Given that $ \psi$ is a bijective map from $  \Z/2r^2\Z$ to $  \Z^2/\LL$, for which
$ \psi(\overline{k - 1})  = (k-1)U$, and $ \psi(\overline{k })  = kU$, we can express these two cycles in a standardized form.
\[  \psi(\overline{k}) \to \psi(\overline{k + 1}) \to \psi(\overline{T}) \to \psi(\overline{k-1}) \to \psi(\overline{k})\]
where $ T \in \Z$ is an integer such that:
\begin{itemize}
     \item $ \psi(\overline{T}) \in \{  (k-1)U +  V,  (k-1)U -  V\}$ 
     \item $\overline{T}$ is a neighbour of $\overline{k-1}$ in the graph $(\Z/2r^2\Z,\, \, \overline{1}, \overline{2r-1})  $: 
     \[  \overline{T} \in \{ \overline{k - 2}, \, \,\overline{k},  \, \, \overline{k + 2r -2}, \, \, \overline{k  - 2r}\}\]
     
     \item $\overline{T}$ is a neighbour of $\overline{k + 1}$ in the graph $(\Z/2r^2\Z,\, \, \overline{1}, \overline{2r-1})  $: 
     \[ \overline{T} \in \{ \overline{k}, \, \, \overline{k + 2}, \, \, \overline{k + 2r}, \, \, \overline{k - 2r + 2} \} \] 
 \end{itemize}

Let's see that $\overline{T} \notin \{ \overline{k - 2}, \, \,\overline{k},  \, \, \overline{k + 2r -2}, \, \, \overline{k  - 2r}\} $. \\

We cannot have $\overline{T} = \overline{k}$. If this were true, then we would have $(k - 1)U \pm  V = \psi(\overline{T}) = \psi(\overline{k})  = kU$ which would imply $  U = \pm V$. However, we previously established that $ U \neq \pm V$. \\

We can also not have $ \overline{T} = \overline{k-2}$. If this were true, then considering that $\oT$ must also be one of $\{ \overline{k + 2}, \, \, \overline{k - 2r  + 2}, \, \, \overline{k + 2r} \}$, one of the following scenarios would occur in $ \Z/2r^2\Z$:
\begin{itemize}
    \item $\overline{k - 2}  = \overline{k + 2} \implies \overline{4} = \overline{0}$
        \item $\overline{k - 2}  = \overline{k - 2r + 2} \implies \overline{2r - 4} = \overline{0}$ 
    \item $\overline{k - 2}  = \overline{k + 2r} \implies \overline{2r + 2} = \overline{0}$ 
\end{itemize}
However, for $ r \geq 3$, we have $ 0 <  4 < 2r^2$ and, $ 0 < 2r - 4 < 2r + 2 < 2r^2$, which disproves our assumption.\\

We can also not have $  \overline{T} = \overline{k - 2r}$. If this were true,  then considering that $ \oT$ must also be one of $\{ \overline{k + 2}, \, \, \overline{k - 2r  + 2}, \, \, \overline{k + 2r} \}$,  one of the following scenarios would occur in $ \Z/2r^2\Z$:
\begin{itemize}
    \item $\overline{k  - 2r}  = \overline{k + 2} \implies \overline{2r +  2 } = \overline{0}$
        \item $\overline{k - 2r}  = \overline{k - 2r + 2} \implies \overline{2} = \overline{0}$ 
    \item $\overline{k  - 2r}  = \overline{k + 2r} \implies \overline{4r } = \overline{0}$ 
\end{itemize}
However, for $ r \geq 3$, we have $ 0 < 2 < 2r + 2 < 4r < 2r^2$. Thus $ \overline{T} \neq \overline{k  - 2r}$. \\

Finally, We can also not have $ \overline{T} = \overline{k+ 2r - 2} $. If this were true,  then considering that $ \oT$ must also be one of $\{ \overline{k + 2}, \, \, \overline{k - 2r  + 2}, \, \, \overline{k + 2r} \}$,  one of the following scenarios would occur in $ \Z/2r^2\Z$:
\begin{itemize}
    \item $\overline{k + 2r - 2}  = \overline{k + 2} \implies \overline{2r - 4} = \overline{0}$
        \item $\overline{k + 2r - 2}  = \overline{k  - 2r + 2} \implies \overline{4r - 4} = \overline{0}$
    \item $\overline{k+ 2r - 2}  = \overline{k + 2r} \implies \overline{2} = \overline{0}$ 

\end{itemize}
However, for $ r \geq 3$, we have $ 0 < 2 \leq  2r - 4 < \, 4r - 4 < 2r^2$, which disproves our assumption. Thus, $ \overline{T}  \neq \overline{k +  2r - 2}$. \\

Consequently, our assumption for Case 2 that $ \psi(\overline{k + 1}) =  kU \pm V $ cannot be true. In summary, we have proved that $ \psi(\overline{k + 1}) $ belongs to $ \{ (k + 1)U, \, \, (k -1)U, \, \, kU + V, \, \, kU - V \} $, but that it cannot be $ (k -1)U, \, \, kU + V$ or $ kU - V $. Thus, it must be that $ \psi(\overline{k + 1}) = (k +1)U$, proving P($k+1$) is true. \\

By induction, we therefore have $ \forall k \in \N, \, \, \psi(\overline{k}) = kU$. Now let's prove that for any $ q \in \N$ we have $ \psi(\overline{-q}) = -qU$. In $ \Z/2r^2\Z, \, \, \overline{-q} = \overline{-1}\cdot \overline{q} = \overline{(2r^2 - 1)}\cdot \overline{q}  = \overline{(2r^2 - 1)q} $. So by induction, $ \psi(\overline{-q}) = (2r^2 - 1) q U$. 

Given that $ U \in \{ \pm   \begin{pmatrix}
    1 \\ 0
\end{pmatrix} \mod \LL ,  \quad \pm \begin{pmatrix}
    0 \\ 1
\end{pmatrix} \mod \LL \} $, where $ \LL = \Z\begin{pmatrix}
    r \\ r
\end{pmatrix}  \bigoplus \Z\begin{pmatrix}
    r \\ -r
\end{pmatrix} $, it follows that $ 2rU = \zeroL$. Hence, $\psi(\overline{-q}) = (2r^2 - 1) q U = -q U$. \\

Consequently, $ \forall k \in \Z, \, \, \psi(\overline{k}) = kU$, proving $ \psi$ is group homomorphism, and thus an isomorphism.

\subsubsection*{Second step: proving that there is no group isomorphism:} 

We have shown that, for $ r \geq 3$, $ \psi : \Z/2r^2\Z \to \Z^2/\LL $ is a group isomorphism.
However, in $ \Z/2r^2\Z $, there are elements of order $ 2r^2$ (e.g. $ \oun)$, whereas in $ \Z^2/\LL$ with $\LL = \Z\begin{pmatrix} r \\ r \end{pmatrix} \oplus \Z\begin{pmatrix} r \\ -r \end{pmatrix}$ every element's order  is less than $ 2r < 2r^2$.
Indeed: 
\[ \begin{pmatrix}
    x \\ y 
\end{pmatrix} \in \Z^2 \implies  2r\begin{pmatrix}
    x \\ y 
\end{pmatrix} = x\begin{pmatrix}
    2r \\ 0
\end{pmatrix}  + y\begin{pmatrix}
    0 \\ 2r
\end{pmatrix} \in  \LL \] 

In conclusion, despite, sharing the same parameters $ [[ 4r^2, 2, 2r ] \! ]$, the $45$ degree-rotated surface codes with periodicity vectors $ \begin{pmatrix}
    r \\ r
\end{pmatrix}, \, \begin{pmatrix}
    r \\ -r
\end{pmatrix}$ and the GB code $ GB(1 + X, 1 + X^{2r- 1}, 2r^2)$ are not CGP-equivalent. \qed  

\subsection{Standard Kitaev Toric-code Comparison}

\subsubsection{Triple-connectivity of the underlying graphs}

\begin{lemma}
    \label{lemma: Z/n^2Z 3-connected}
    For $ n \geq 3$, the Cayley graph $(\Z/n^2\Z, \oun , \overline{n}) $, which underlies $GB(1 + X, 1 + X^{n}, n^2)$ is triply-connected.
\end{lemma}

\textbf{Proof:}Applying the same proof strategy as in Lemma~\ref{lemma: Z/2r^2Z 3-connected}, we show that the graph $(\mathbb{Z}/n^2\mathbb{Z}, \oun, \overline{n})$ remains connected even after the removal of any pair of distinct vertices. This establishes the 3-connectivity of the graph. \qed

\begin{remarks}
More generally, With this same strategy, we can show that for (2,2)-GB code of the form $ GB(1 + X, 1 + X^\alpha, N) $, with $ 2 \leq \alpha \leq N - 2 $ the underlying graph $ (\Z/N\Z, \oun, \oalpha) $ is triply-connected.
\end{remarks}

\begin{lemma}
\label{lemma: kita standard 3-connexe}
    For an integer $ n \geq 3$,  let  $ \LL = \Z\begin{pmatrix}
    n \\ 0
\end{pmatrix} \bigoplus \Z\begin{pmatrix}
    0 \\ n 
\end{pmatrix} $. Then the Cayley graph $ (\Z^2/\LL, \begin{pmatrix}
    1 \\ 0
\end{pmatrix} \mod \LL, \begin{pmatrix}
    0 \\ 1
\end{pmatrix}\mod \LL) $ that underlies the standard Kitaev toric code with parameters $ [[ 2n^2, 2, n]]$ is triply-connected.
\end{lemma}

\textbf{Proof of Lemma \ref{lemma: kita standard 3-connexe}:} Let  $ \LL = \Z\begin{pmatrix}
    n \\ 0
\end{pmatrix} \bigoplus \Z\begin{pmatrix}
    0 \\ n 
\end{pmatrix} $. 

Following the same strategy as in Lemma~\ref{lemma: 2D surface pair 3-connexe}:
\begin{enumerate}
    \item First, proving that: $ \Z^2/\LL - \{ P, Q \}$ remains connected for any two distinct non-zero points $ P, Q \neq \zeroL$ by 
    constructing three independent paths  between any non-zero point in the quotient $ \Z^2/ \LL$ and $ \zeroL$.
    
    \item Then, establishing the connectivity of the graph $ \Z^2/ \LL - \{ \zeroL, Q\}$  for any non-zero point $ Q  \neq \zeroL$, by exhibiting an isomorphism with a known connected graph.
\end{enumerate}
We show that the graph $(\Z^2/\LL, \begin{pmatrix}
    1 \\ 0
\end{pmatrix} \mod \LL, \begin{pmatrix}
    0 \\ 1
\end{pmatrix}\mod \LL) $ is triply-connected. \qed

\begin{proposition}
\label{prop: kitaev standard pas eq a un GB}
    For $ n \geq 4 $, the surface code with periodicity vectors $ \begin{pmatrix}
    n \\ 0
\end{pmatrix}, \, \begin{pmatrix}
    0 \\ n 
\end{pmatrix}$ vectors and $ GB(1 + X, 1 + X^{n}, n^2)$ are not CGP-equivalent. 
\end{proposition}

\textbf{Proof of Proposition \ref{prop: kitaev standard pas eq a un GB}: } Suppose, for contradiction, that these two codes are CGP-equivalent. These two codes come from the Cayley graphs:
\begin{itemize}
    \item $(\mathbb{Z}^2/\LL, \begin{pmatrix} 1 \\ 0 \end{pmatrix} \mod{\LL}, \begin{pmatrix} 0 \\ 1 \end{pmatrix} \mod{\LL})$, where $\LL = \mathbb{Z}\begin{pmatrix} n \\ 0 \end{pmatrix} \oplus \mathbb{Z}\begin{pmatrix} 0 \\ n \end{pmatrix}$
        \item $(\mathbb{Z}/n^2\mathbb{Z}, \, \overline{1}, \, \overline{n})$ 
\end{itemize}

 By applying the exact same proof strategy as in Proposition \ref{prop: kitaev pair pas eq a un GB}, we can show that if these codes were equivalent, then there would be a graph isomorphism: 
    \[ (\Z^2/\LL, \begin{pmatrix}
        1 \\ 0
    \end{pmatrix} \mod \LL, \begin{pmatrix}
        0 \\ 1
    \end{pmatrix} \mod \LL) \cong (\Z/n^2\Z, \, \, \overline{1}, \, \, \overline{n}) \]
    This graph isomorphism, in turn, would induce a group isomorphism between $\Z/n^2\Z $ and $  \Z^2/\LL$.
    However, these two groups are not isomorphic because $ \Z^2 / \LL $ does not contain any element of order $ n^2$. Indeed, every element of $ \Z^2/\LL $ has an order that divides $ n $ which is inferior to $ n^2$:
\[ \begin{pmatrix}
    x \\ y
\end{pmatrix} \in \Z^2 \implies n\begin{pmatrix}
    x \\ y
\end{pmatrix} = x\begin{pmatrix}
    n \\ 0
\end{pmatrix} + y \begin{pmatrix}
    0 \\ n
\end{pmatrix} \in \LL\]

In conclusion, the standard Kitaev toric code of length $ 2n^2$ and the GB code $ GB( 1 + X, 1 + X^n, n^2) $ are not CGP-equivalent. \qed

\subsection{Odd-distance 2D Surface Codes Comparison}

\begin{proposition}
\label{prop: kitaev_impair eq a un GB}
For $ t \geq 1 $, the optimal odd-distance 2D surface code with periodicity vectors $ \begin{pmatrix}
    t \\ t  +1 
\end{pmatrix}, \, \begin{pmatrix}
    t + 1 \\ - t
\end{pmatrix}$  and the GB code $ GB(1 + X, 1 + X^{2t + 1}, t^2 + (t + 1)^2)  $ are equivalent.
\end{proposition}

\textbf{Proof of prop \ref{prop: kitaev_impair eq a un GB}:} Let $ \LL = \Z\begin{pmatrix}
        t  \\ t + 1
    \end{pmatrix} \bigoplus \Z\begin{pmatrix}
        t + 1 \\ - t
    \end{pmatrix}$. \\

        The graph $ (\Z/\left(  t^2 + (t + 1)^2 \right)\Z, \, \, \overline{1}, \,\, \overline{2t + 1})$  serves as the underlying structure for the code  $ GB(1 + X, 1 + X^{2t + 1}, t^2 + ( t + 1)^2) $ (see Section \ref{sec: abstract graph associated to $GB(A(X), B(X), n)$}).  \\

        Similarly, the optimal odd-distance 2D surface code, characterized by periodicity vectors 
        $ \begin{pmatrix}
    t \\ t  +1 
\end{pmatrix}, \, \begin{pmatrix}
    t + 1 \\ - t
\end{pmatrix}$  is derived from the graph $ ( \Z^2/\LL, \begin{pmatrix}
        1 \\ 0
    \end{pmatrix} \mod \LL, \begin{pmatrix}
        0 \\ 1
    \end{pmatrix} \mod \LL)$ (see Section \ref{sec: optimal surface codes}). \\
    
Our objective is to demonstrate the equivalence of these two codes. We will prove this in three steps:
\begin{itemize}
    \item First, we will construct a group isomorphism between $ \Z/\left( t^2 + (t + 1)^2 \right)\Z$ and $\Z^2/\LL $. 

    \item Next, we will prove that this group isomorphism induces a graph isomorphism between the two underlying graphs.

    \item Finally, we will show that for specific orderings on the (edges, vertices and faces) the vertex-edge incidence matrices $ \bH_X, \bH_X'$ and face-edge incidence matrices $ \bH_Z, \bH_Z'$ of the two graphs satisfy the relations: 
    \begin{equation*}
       \left\{  \begin{aligned}
            \bH_X' & = \bH_X \\
            rs(\bH_Z') & = rs(\bH_Z)
        \end{aligned} \right.
    \end{equation*}
    from which the code CGP-equivalence, and thus the equivalence will follow. 
\end{itemize}

\subsubsection*{Step 1: Constructing the group isomorphism}

Let $ \eta $ be the map defined as:
\[ \begin{array}{ccccl}
\eta & : &   \Z/ \left(  t^2 + (t + 1)^2 \right) \Z& \xrightarrow{} & \Z^2/\LL \\
& &  T \mod (t^2 + (t +1)^2)& \longmapsto & T\begin{pmatrix}
    1 \\ 0
\end{pmatrix} \mod \LL
\end{array} \]

$ \eta $ is a group isomorphism that sends $ \overline{1} \mapsto \begin{pmatrix}
    1 \\ 0
\end{pmatrix}$ and $ \overline{2t + 1} \mapsto - \begin{pmatrix}
    0 \\ 1
\end{pmatrix} \mod \LL$ \\

\underline{$ \eta$ is well-defined:} Let $ k , q \in \Z $ be two integers such that $ \overline{k} =  \overline{q} $ in $ \Z/\left( t^2 + (t +1)^2  \right)  \Z$. By hypothesis, $ \exists \lambda \in \Z, \, \, k - q = \lambda( t^2 + (t +1)^2) $.  Thus: \[  k \begin{pmatrix}
    1 \\ 0
\end{pmatrix} - q \begin{pmatrix}
    1 \\ 0
\end{pmatrix} = \lambda  \left( (t + 1) \begin{pmatrix}
     t + 1 \\ -t
\end{pmatrix} + t \begin{pmatrix}
     t \\ t  +1
\end{pmatrix} \right) \]

So $  k \begin{pmatrix}
    1 \\ 0
\end{pmatrix} - q \begin{pmatrix}
    1 \\ 0
\end{pmatrix} \in  \LL = \Z\begin{pmatrix}
        t \\ t + 1
    \end{pmatrix} \bigoplus \Z\begin{pmatrix}
         t + 1 \\ -t
    \end{pmatrix} $. 
Thus,  $ \eta(\overline{k}) = \eta(\overline{q}) $. \\

Moreover, $ \eta$ is group homomorphism, sending $ \oun$ to $ \begin{pmatrix}
    1 \\ 0 
\end{pmatrix} \mod \LL $ and  $ \overline{2t + 1}$ to $- \begin{pmatrix}
    0 \\ 1
\end{pmatrix} \mod \LL$. \\

\underline{$ \eta$ is bijective:} Since $ \LL $ is a sub-lattice of $ \Z^2$, the order of $ \Z^2 / \LL $ is given by:\[ \vert \Z^2 / \LL \vert = det \begin{pmatrix}
    t + 1 & t \\
    -t    & t + 1
\end{pmatrix} = (t +1)^2 + t^2 \] 
which corresponds to the order of $\Z/ \left( t^2 + (t + 1)^2 \right) \Z$. \\

Moreover, $ \eta $ is an injective map. Indeed, if $ k \in \Z $ is such that $ \eta(\overline{k}) = \zeroL $, then:  \[  k \begin{pmatrix}
    1 \\ 0
\end{pmatrix} \in \LL =  \Z\begin{pmatrix}
        t \\ t + 1
    \end{pmatrix} \bigoplus \Z\begin{pmatrix}
         t + 1 \\ -t
    \end{pmatrix}.\] Therefore, there exist $ \lambda, \mu \in \Z$ such that $ \, k \begin{pmatrix}
    1 \\ 0
\end{pmatrix}  = \lambda \begin{pmatrix}
    t \\  t + 1
\end{pmatrix} + \mu \begin{pmatrix}
    t + 1\\ -t
\end{pmatrix} $. This yield two integer equations:
\[ k = \lambda t + \mu (t + 1) \quad \text{and} \quad \mu t  = \lambda ( t + 1) \]

Since, $ t $ and $ t  + 1$ are coprime it must be that $ t $ divides $ \lambda $, and $ t + 1$ divides $ \mu$. Thus, there exists integers $ p, q $ such that $ \lambda = p t $ and $ \mu = q(t  +1) $.
Substituting these into the two previous equations:
\[ k = pt^2 + q(t + 1)^2 \quad \text{and} \quad qt(t + 1)  = pt(t +1)\]

Since $ t \geq 1$ then,  $ p = q $ and thus $ k = p\left( t^2 + (t + 1)^2 \right)$. Thus, we have proved that $ \eta(\overline{k}) = \zeroL $ implies $\ok =  \overline{0} $ in $ \Z/\left(t^2 + (t + 1)^2 \right) \Z$. Hence, $ \eta $ is injective. \\

In conclusion, since $ \eta $ is an injective homomorphism between groups of the same finite size, it is an isomorphism.

\subsubsection*{Step 2: Proving $ \eta$ is graph isomorphism:}

We have shown that the map $\eta$ defined by
\[ \begin{array}{ccccl}
\eta & : &  \Z/ \left( t^2 + (t + 1)^2 \right) \Z & \xrightarrow{} & \Z^2/\LL \\
& & \overline{T} & \longmapsto & T\begin{pmatrix}
    1 \\ 0
\end{pmatrix} \mod \LL
\end{array} \]
is a group isomorphism sending $ \overline{1}$ to $ \begin{pmatrix}
    1 \\ 0
\end{pmatrix}  \mod \LL$  and $ \overline{2t + 1} $ to  $ - \begin{pmatrix}
    0 \\ 1
\end{pmatrix} \mod \LL $.

Two vertices $ \oT $ and $ \overline{S}$ are neighbours in the graph $ (\Z/\left( t^2 + (t +1)^2\right)\Z, \, \, \overline{1}, \, \,  \overline{2t + 1} )$ if and only if $\oT - \overline{S} \in \{ \pm \oun, \, \, \pm \, \overline{2t + 1}  \}$, which due to the isomorphism $ \eta$, is equivalent to:
\[ \eta(\oT - \overline{S}) =  \eta(\oT) - \eta(\overline{S}) \in \{ \pm \begin{pmatrix}
    1 \\ 0
\end{pmatrix}\mod \LL,  \pm \, \begin{pmatrix}
    0 \\ 1
\end{pmatrix}\mod \LL \}\]

Hence, $ \eta $ induces a graph isomorphism between the two graphs $ (\Z/\left( t^2 + (t +1)^2\right)\Z, \, \, \overline{1}, \, \,  \overline{2t + 1} )$ and $ (\Z^2/\LL, \, \, \begin{pmatrix}
    1 \\ 0
\end{pmatrix} \mod \LL, \, \, \begin{pmatrix}
    0 \\ 1
\end{pmatrix}\mod \LL )$. \\

\subsubsection*{Step 3: Proving the codes CGP-equivalence }

    Proving the CGP-equivalence of these two codes necessitates demonstrating that their respective incidence matrices, $ \bH_X, \bH_X'$ and $ \bH_Z, \bH_Z' $  are related by the following expressions $ rs(\bH_X') = rs(\bH_X\bQ)$ and $rs(\bH_Z') = rs(\bH_Z\bQ) $ where $ \bQ$ represents a permutation matrix. 

    Specifically, we will demonstrate that for specific, carefully chosen orderings, the vertex-edge and face-edge incidence matrices of  $ \Z/\left( t^2 + (t + 1)^2\right)\Z$ (namely $ \bH_X$ and $\bH_Z$) and   $ \Z^2/\LL$  (namely $ \bH_X'$ and $ \bH_Z'$) can be made to satisfy the relations: 
    \begin{equation*}
       \left\{  \begin{aligned}
            \bH_X' & = \bH_X \\
            rs(\bH_Z') & = rs(\bH_Z) 
        \end{aligned} \right.
    \end{equation*}
    
\subsubsection*{Step 3.1 Choosing the orderings:} 

Since $ \eta $ induces a graph isomorphism between the two graphs $  (\Z/ \left( t^2 + (t + 1)^2 \right)\Z, \, \, \overline{1}, \,\, \overline{2t + 1})$ and $ (\Z^2/\LL, \, \, \begin{pmatrix}
    1 \\0
\end{pmatrix} \mod \LL , \begin{pmatrix}
    0 \\ 1
\end{pmatrix} \mod \LL)$ then these two graphs have the same number of edges and vertices. \\

For the vertices, we establish a parallel ordering. For each $ i \in \left[[ 0, t^2 + (t + 1)^2 - 1 |\right], $ we define: the $i$-th vertex in $ \Z/ \left( t^2 + (t + 1)^2 \right) \Z $ as $ S_i = \overline{i}$ and the $i$-th vertex in $\Z^2/\LL$ as $ T_i = \eta(\overline{i}) $. 

On the edges, we define the ordering such that: in $  (\Z/ \left(  t^2 + (t + 1)^2\right)\Z, \, \, \overline{1}, \,\, \overline{2t + 1})  $ the $k$-th edge is $ \{ \overline{x}, \overline{y}\} $  if and only if in $ (\Z^2/\LL, \, \, \begin{pmatrix}
    1 \\0
\end{pmatrix} \mod \LL , \begin{pmatrix}
    0 \\ 1
\end{pmatrix} \mod \LL)$ the $k$-th edge is $\{ \eta(\overline{x}), \, \eta(\overline{y}) \} $. \\

With this specific ordering, the vertex-edge incidence matrices of these two graphs are identical, i.e $ \bH_X' = \bH_X$. 

\subsubsection*{Step 3.2 Proving $rs(\bH_Z') = rs(\bH_Z) $:} 

The rows of $ \bH_Z$ are the characteristic vectors of the faces of $ (\Z/ \left(  t^2 + (t + 1)^2\right)\Z, \, \, \overline{1}, \,\, \overline{2t + 1})$, which are of the form:
\[ \oT \to \oT + \overline{1} \to  \oT + \overline{1} + \overline{2t + 1} \to \oT + \overline{2t + 1} \to \oT \]

Similarly, the rows of $ \bH_Z'$ are the characteristic vectors of the faces of $ (\Z^2/\LL, \, \, \begin{pmatrix}
    1 \\0
\end{pmatrix} \mod \LL , \begin{pmatrix}
    0 \\ 1
\end{pmatrix} \mod \LL)$ which corresponds to 4-cycle of vertices within $ \Z^2 / \LL$, of the form:
\[  S \to   S + \begin{pmatrix}
     1\\ 0
\end{pmatrix}   \to S + \begin{pmatrix}
    1  \\  1
\end{pmatrix}  \to S + \begin{pmatrix}
    0   \\ 1
\end{pmatrix} \to S\]

We will now demonstrate that the rows of both matrices are the same. First, let's start by proving that $ \{ \text{rows of } \bH_Z \} \subseteq   \{ \text{rows of } \bH_Z' \}$. \\

Let $ \bv $ be a row of $ \bH_Z$. By definition of $ \bv \in  \F_2^{2(t^2 + (t + 1)^2)}$, $\bv$ is the characteristic vector of a face of $(\Z/ \left(  t^2 + (t + 1)^2\right)\Z, \, \, \overline{1}, \,\, \overline{2t + 1})$. Such a face $\mathcal{F}$ is defined by vertices:
\[  \oT \to \oT + \overline{1} \to  \oT + \overline{1} + \overline{2t + 1} \to \oT + \overline{2t + 1} \to \oT \]

By our chosen ordering, $\bv$ is also the characteristic vector of $ \eta(\mathcal{F})$, the image of this face under the graph isomorphism $\eta$: $ \eta(\oT) \to  \eta(\overline{T + 1}) \to   \eta(\overline{T + 1 + 2t + 1}) \to \eta(  \overline{T + 2t + 1}) \to \eta(\oT)$. 

Since $\forall q \in \Z, \, \, \eta(\overline{q}) = \begin{pmatrix}
    q \\ 0 
\end{pmatrix} \mod \LL $ and $ \begin{pmatrix}
    2t + 1 \\ 0 
\end{pmatrix} \mod \LL = - \begin{pmatrix}
    0 \\ 1
\end{pmatrix} \mod \LL $, then $ \eta(\mathcal{F}) $ corresponds to the following cycle within $ \Z^2/\LL$: 
\[ \begin{pmatrix}
    T \\ 0
\end{pmatrix}  \to \begin{pmatrix}
    T + 1\\ 0
\end{pmatrix} \to \begin{pmatrix}
    T + 1\\ - 1
\end{pmatrix} \to \begin{pmatrix}
    T \\ - 1
\end{pmatrix} \to \begin{pmatrix}
    T \\ 0
\end{pmatrix} \] 

Since the graphs are unoriented, this corresponds to a valid face in $ \Z^2/ \LL$:
\[ \begin{pmatrix}
    T \\ - 1
\end{pmatrix} \to \begin{pmatrix}
    T + 1 \\ -1
\end{pmatrix} \to  \begin{pmatrix}
    T + 1 \\ 0
\end{pmatrix}  \to \begin{pmatrix}
    T \\ 0
\end{pmatrix} \to \begin{pmatrix}
    T  \\ -1
\end{pmatrix} \] 

Therefore, $ \bv $ is the characteristic vector of a face of  $ \Z^2/\LL$, which means $ \bv $ is a row of $\bH_Z'$. \\

Now, let's demonstrate that $ \{ \text{rows of } \bH_Z' \} \subseteq  \{ \text{rows of } \bH_Z \}$.
Let $ \bw$ be a row of  $ \bH_Z'$. By definition of $ \bw \in  \F_2^{2(t^2 + (t + 1)^2)}$, $\bw$ is the characteristic vector of a face of $(\Z^2/\LL, \, \, \begin{pmatrix}
    1 \\0
\end{pmatrix} \mod \LL , \begin{pmatrix}
    0 \\ 1
\end{pmatrix} \mod \LL)$. Such a face $\mathcal{F_\LL}$ is defined a sequence of vertices of $ \Z^2/\LL$:
\[  \begin{pmatrix}
    x \\ y 
\end{pmatrix}  \to \begin{pmatrix}
    x + 1 \\ y 
\end{pmatrix}  \to \begin{pmatrix}
    x + 1\\ y + 1 
\end{pmatrix}  \to \begin{pmatrix}
    x \\ y + 1 
\end{pmatrix}  \to \begin{pmatrix}
    x \\ y 
\end{pmatrix} \]

Since the map $ \eta : \Z/\left( t^2 + (t + 1)^2 \right)\Z \to \Z^2/\LL $,  $ u \mapsto \begin{pmatrix}
    u \\ 0 
\end{pmatrix} \mod \LL$ is an isomorphism and $ \eta(\overline{2t  + 1}) = - \begin{pmatrix}
    0 \\ 1
\end{pmatrix} \mod \LL $, there exists a unique $ \ok \in \Z/\left( t^2 + (t + 1)^2 \right)\Z$ such that: $  \eta(\overline{k}) = \begin{pmatrix}
    x \\ y 
\end{pmatrix} \mod \LL  $. Therefore $ \mathcal{F}_\LL$ corresponds to: $ \eta(\overline{k}) \to \eta(\overline{k + 1}) \to \eta(\overline{k + 1 - (2t  + 1)}) \to \eta(\overline{k - (2t + 1)}) \to \eta(\overline{k}) $. \\

By our chosen ordering, $\bw$ is also the characteristic vector of $ \eta^{-1}(\mathcal{F}_\LL)$ the face whose vertices are the preimages under $ \eta^{-1}$ of the vertices of $ \mathcal{F}_\LL $:
\[ \overline{k} \to \overline{k + 1} \to \overline{k + 1 - (2t  + 1)} \to \overline{k - (2t + 1)} \to \overline{k} \]

This is a face of the graph $ (\Z/ \left( t^2 + (t + 1)^2 \right) \Z, \, \, \overline{1}, \,\, \overline{2t + 1})$. Consequently $\bw$ is a row of  $ \bH_Z$. \\

\underline{Conclusion:} The set of rows of $\bH_Z$ is identical to the set of rows of $\bH_Z'$. Hence, in particular, $ rs(\bH_Z) = rs(\bH_Z') $.  

In conclusion, the code $GB(1  + X, 1 + X^{2t + 1}, t^2 + (t +1)^2)$ is CGP-equivalent, and thus equivalent under the standard notion of quantum code equivalence introduced in Definition \ref{def: general qu equiv relation}, to the optimal odd-distance 2D surface code of parameters with periodicity vectors $ \begin{pmatrix}
    t \\ t + 1
\end{pmatrix}$ and $ \begin{pmatrix}
    t  + 1 \\ - t
\end{pmatrix}$.  \qed

\section{Classification of (2,2)-GB codes}
\label{sec: classification of (2,2)-GB codes}
This section details our classification of (2,2)-GB codes, a specialized subclass of CSS codes, intrinsically linked to Cayley graphs of the form $(\mathbb{Z}/n\mathbb{Z}, a, b)$. 

Our main goal is to systematically identify, for all lengths below 200, the best-performing (2,2)-GB codes up to CGP-equivalence. This classification is crucial, as the practical decoding performance of these codes directly depends on their underlying graph structures. In this context, the CSS Graph-preserving (CGP) equivalence relation, defined in Definition \ref{def: equiv qu codes}, corresponds precisely to 2-isomorphisms between these associated graphs, thus revealing a deep structural correspondence key to their performance. \\

Since any (2,2)-GB code can be decomposed into a direct sum of codes of the form $GB(1 + X^a, 1 + X^b, n)$ where $\gcd(a, b, n) = 1$, our classification efforts focused on this canonical form, which serves as the foundational component in the broader construction of GB codes.

\subsection{Method:}

To achieve this classification, we first identified all unique graph representatives under the isomorphism equivalence relation. Recognizing that some of these representatives might still be 2-isomorphic, we proceeded to compute the parameters of all associated GB codes. This step is critical because 2-isomorphic codes are guaranteed to share the same parameters, thus providing a strong initial filter.

Finally, for any set of codes that yielded identical parameters, we performed a manual verification to confirm their non-2-isomorphism. This verification was conducted using two primary methods:
\begin{itemize}
    \item \textbf{3-Connectivity Verification:} We checked if the underlying graphs of all listed GB codes were 3-connected. According to Whitney's Theorem, for 3-connected graphs, non-isomorphism directly implies non-2-isomorphism. Crucially, for our work, all codes of the form  $GB(1 + X^a, 1 + X^b, n) $  with $ 1 \leq a  < b \leq n - 2$ and $gcd(a,b,n) = 1$ have underlying graphs that exhibit this property for $n$ between 4 and 100.

    \item \textbf{Permutation Equivalence Analysis:} Alternatively, we ensured that the row spaces of the different vertex-edge incidence matrices were not permutation equivalent. This serves as another robust method to distinguish non-CGP-equivalent codes even if they share parameters.
\end{itemize}

The computational aspects of this classification were performed using SageMath and Magma. Specifically, SageMath was utilized for determining equivalence classes of isomorphic graphs and evaluating their k-connectivity, while Magma was employed for computing the minimum distance of the quantum codes.

\subsection{Classification of Extremal Non-Equivalent (2,2)-GB Codes}

We present in Table \ref{table: classification GB(2,2) extremaux}, in which we classified all extremal non-equivalent (2,2)-GB codes with lengths under 200. This table focuses exclusively on (2,2)-GB codes. For each length, we detail the best achievable parameters $[[2N,2,d]]$, the total number of non-equivalent codes, the parameters A(X) and B(X) for their representatives, and their sources if previously documented. (see Table \ref{table: classification GB(2,2) extremaux}).

\begin{longtable}{|c|c|c|c|c|}
\caption{Extremal Non-Equivalent (2,2)-GB Codes GB(A(X), B(X), N) for Lengths Under 200}
\label{tab:our_gb_codes_modified_v2} \\
\toprule 
\thead{{$N$} }& \thead{Parameters \\ $[ \![2N, k, d]]$} & \thead{Number of \\Non-CGP-Equivalent \\ GB Codes} & \thead{ Polynomial \\ A(X)} & \thead{Polynomial \\ B(X)} \\
\midrule
\endfirsthead
\toprule 
\thead{{$N$} }& \thead{Parameters \\ $[ \![2N, k, d]]$} & \thead{Number of \\Non-CGP-Equivalent \\ GB Codes} & \thead{ Polynomial \\ A(X)} & \thead{Polynomial \\ B(X)} \\
\midrule
\endhead
\bottomrule
\endfoot
\bottomrule
\endlastfoot

% -----------------------------------------------------------
% Data extracted from "Nos Codes" column and restructured
% -----------------------------------------------------------

% N = 2
{2} & {$[ \![4,2,2]]$} &{1} & {$1+X$} & $1+X$ (optimal surface code \cite{BD07}) \\[0.5ex]
\hline \noalign{\vskip 0.3ex}
% N = 3
3 & $[ \![6,2,2]]$ & 1 & $1+X$ & $1+X^2$ \\[0.5ex]
\hline \noalign{\vskip 0.3ex}
% N = 4
{4} & {$[ \![8,2,2]]$} & {2} & {$1+X$} & $1+X$ (Kitaev code \cite{K03}) $\quad 1+X^2$ \\[0.5ex]
\hline \noalign{\vskip 0.3ex}
% N = 5
{5} & {$[ \![10,2,3]]$} & {1} & {$1+X$} & $1+X^3$ (optimal surface code\cite{BD07}) \\[0.5ex]
\hline \noalign{\vskip 0.3ex}
% N = 6
6 & $[ \![12,2,3]]$ & 1 & $1+X$ & $1+X^2$ \\[0.5ex]
\hline \noalign{\vskip 0.3ex}
% N = 7
7 & $[ \![14,2,3]]$ & 1 & $1+X$ & $1+X^2$ \\[0.5ex]
\hline \noalign{\vskip 0.3ex}
% N = 8
8 & $[ \![16,2,4]]$ & 1 & $1+X$ & $1+X^3$ \\[0.5ex]
\hline \noalign{\vskip 0.3ex}
% N = 9
9 & $[ \![18,2,3]]$ & 2 & $1+X$ & $1+X^2, \quad 1+X^3$ \\[0.5ex]
\hline \noalign{\vskip 0.3ex}
% N = 10
10 & $[ \![20,2,4]]$ & 2 & $1+X$ & $1+X^3, \quad 1+X^4$ \\[0.5ex]
\hline \noalign{\vskip 0.3ex}
% N = 11
11 & $[ \![22,2,4]]$ & 1 & $1+X$ & $1+X^3$ \cite{PWsimu} \\[0.5ex]
\hline \noalign{\vskip 0.3ex}
% N = 12
\multirow{3}{*}{12} & \multirow{3}{*}{$[ \![24,2,4]]$} & \multirow{3}{*}{3} & $1+X$ & $1+X^3, \quad 1+X^5$ \\[2.5ex]
    & & & $1+X^2$ & $1+X^3$ \\[0.5ex]
\hline \noalign{\vskip 0.3ex}
% N = 13
{13} & {$[ \![26,2,5]]$} & {1} &{$1+X$} & $1+X^5$ (optimal surface code \cite{BD07}) \\[0.5ex]
\hline \noalign{\vskip 0.3ex}
% N = 14
14 & $[ \![28,2,5]]$ & 1 & $1+X$ & $1+X^4$ \\[0.5ex]
\hline \noalign{\vskip 0.3ex}
% N = 15
15 & $[ \![30,2,5]]$ & 2 & $1+X$ & $1+X^4, \quad 1+X^6$ \\[0.5ex]
\hline \noalign{\vskip 0.3ex}
% N = 16
16 & $[ \![32,2,5]]$ & 1 & $1+X$ & $1+X^6$ \\[0.5ex]
\hline \noalign{\vskip 0.3ex}
% N = 17
17 & $[ \![34,2,5]]$ & 2 & $1+X$ & $1+X^4, \quad 1+X^5$ \\[0.5ex]
\hline \noalign{\vskip 0.3ex}
% N = 18
18 & $[ \![36,2,6]]$ & 1 & $1+X$ & $1+X^5$ \\[0.5ex]
\hline \noalign{\vskip 0.3ex}
% N = 19
19 & $[ \![38,2,5]]$ & 2 & $1+X$ & $1+X^4 $ \cite{PWsimu}, $ \quad 1+X^7$ \\[0.5ex]
\hline \noalign{\vskip 0.3ex}
% N = 20
20 & $[ \![40,2,5]]$ & 3 & $1+X$ & $1+X^4, \quad 1+X^6, \quad 1+X^8$ \\[0.5ex]
\hline \noalign{\vskip 0.3ex}
% N = 21
21 & $[ \![42,2,6]]$ & 2 & $1+X$ & $1+X^6, \quad 1+X^8$ \\[0.5ex]
\hline \noalign{\vskip 0.3ex}
% N = 22
22 & $[ \![44,2,6]]$ & 2 & $1+X$ & $1+X^5, \quad 1+X^6$ \\[0.5ex]
\hline \noalign{\vskip 0.3ex}
% N = 23
23 & $[ \![46,2,6]]$ & 1 & $1+X$ & $1+X^5$ \\[0.5ex]
\hline \noalign{\vskip 0.3ex}
% N = 24
\multirow{3}{*}{24} & \multirow{3}{*}{$[ \![48,2,6]]$} & \multirow{3}{*}{5} & $1+X$ & $1+X^5, \quad 1+X^7, \quad 1+X^9, \quad 1+X^{10}$ \\[2.5ex]
    & & & $1+X^3$ & $1+X^4$ \\[0.5ex]
\hline \noalign{\vskip 0.3ex}
% N = 25
25 & {$[ \![50,2,7]]$} & {1} & {$1+X$} & $1+X^7$ (optimal surface code \cite{BD07}) \\[0.5ex]
\hline \noalign{\vskip 0.3ex}
% N = 26
26 & $[ \![52,2,7]]$ & 1 & $1+X$ & $1+X^{10}$ \\[0.5ex]
\hline \noalign{\vskip 0.3ex}
% N = 27
27 & $[ \![54,2,7]]$ & 1 & $1+X$ & $1+X^6$ \\[0.5ex]
\hline \noalign{\vskip 0.3ex}
% N = 28
28 & $[ \![56,2,7]]$ & 2 & $1+X$ & $1+X^6, \quad 1+X^8$ \\[0.5ex]
\hline \noalign{\vskip 0.3ex}
% N = 29
29 & $[ \![58,2,7]]$ & 2 & $1+X$ & $1+X^8$ \cite{PWsimu}, $\quad 1+X^{12}$ \\[0.5ex]
\hline \noalign{\vskip 0.3ex}
% N = 30
30 & $[ \![60,2,7]]$ & 1 & $1+X^2$ & $1+X^9$ \\[0.5ex]
\hline \noalign{\vskip 0.3ex}
% N = 31
31 & $[ \![62,2,7]]$ & 2 & $1+X$ & $1+X^7, \quad 1+X^{12}$ \\[0.5ex]
\hline \noalign{\vskip 0.3ex}
% N = 32
32 & $[ \![64,2,8]]$ & 1 & $1+X$ & $1+X^7$ \\[0.5ex]
\hline \noalign{\vskip 0.3ex}
% N = 33
33 & $[ \![66,2,7]]$ & 3 & $1+X$ & $1+X^6, \quad 1+X^7, \quad 1+X^9$ \\[0.5ex]
\hline \noalign{\vskip 0.3ex}
% N = 34
34 & $[ \![68,2,8]]$ & 1 & $1+X$ & $1+X^{13}$ \\[0.5ex]
\hline \noalign{\vskip 0.3ex}
% N = 35
{35} & {$[ \![70,2,7]]$} & {4} & {$1+X$} & $1+X^6, \quad 1+X^8, \quad 1+X^{10}, \quad 1+X^{15}$ \\[0.5ex]
\hline \noalign{\vskip 0.3ex}
% N = 36
36 & $[ \![72,2,8]]$ & 3 & $1+X$ & $1+X^8, \quad 1+X^{10}, \quad 1+X^{15}$ \\[0.5ex]
\hline \noalign{\vskip 0.3ex}
% N = 37
37 & $[ \![74,2,8]]$ & 1 & $1+X$ & $1+X^8$ \\[0.5ex]
\hline \noalign{\vskip 0.3ex}
% N = 38
38 & $[ \![76,2,8]]$ & 2 & $1+X$ & $1+X^7, \quad 1+X^{16}$ \\[0.5ex]
\hline \noalign{\vskip 0.3ex}
% N = 39
39 & $[ \![78,2,8]]$ & 2 & $1+X$ & $1+X^7, \quad 1+X^{15}$ \\[0.5ex]
\hline \noalign{\vskip 0.3ex}
% N = 40
\multirow{3}{*}{40} & \multirow{3}{*}{$[ \![80,2,8]]$} & \multirow{3}{*}{5} & $1+X$ & $1+X^7, \quad 1+X^9, \quad 1+X^{11}, \quad 1+X^{15}$ \\[2.5ex]
    & & & $1+X^4$ & $1+X^5$ \\[0.5ex]
\hline \noalign{\vskip 0.3ex}
% N = 41
{41} & {$[ \![82,2,9]]$} & {1} & {$1+X$} & $1+X^9$ (optimal surface code \cite{BD07}) \\[0.5ex]
\hline \noalign{\vskip 0.3ex}
% N = 42
42 & $[ \![84,2,9]]$ & 1 & $1+X$ & $1+X^{16}$ \\[0.5ex]
\hline \noalign{\vskip 0.3ex}
% N = 43
43 & $[ \![86,2,9]]$ & 1 & $1+X$ & $1+X^{12}$ \\[0.5ex]
\hline \noalign{\vskip 0.3ex}
% N = 44
44 & $[ \![88,2,9]]$ & 1 & $1+X$ & $1+X^8$ \\[0.5ex]
\hline \noalign{\vskip 0.3ex}
% N = 45
45 & $[ \![90,2,9]]$ & 3 & $1+X$ & $1+X^8, \quad 1+X^{10}, \quad 1+X^{19}$ \\[0.5ex]
\hline \noalign{\vskip 0.3ex}
% N = 46
46 & $[ \![92,2,9]]$ & 1 & $1+X$ & $1+X^{10}$ \\[0.5ex]
\hline \noalign{\vskip 0.3ex}
% N = 47
47 & $[ \![94,2,9]]$ & 1 & $1+X$ & $1+X^{13}$ \\[0.5ex]
\hline \noalign{\vskip 0.3ex}
% N = 48
\multirow{3}{*}{48} & \multirow{3}{*}{$[ \![96,2,9]]$} & \multirow{3}{*}{3} & $1+X$ & $1+X^{14}, \quad 1+X^{20}$ \\[2.5ex]
    & & & $1+X^2$ & $1+X^9$ \\[0.5ex]
\hline \noalign{\vskip 0.3ex}
% N = 49
49 & $[ \![98,2,9]]$ & 1 & $1+X$ & $1+X^9$ \\[0.5ex]
\hline \noalign{\vskip 0.3ex}
% N = 50
50 & $[ \![100,2,10]]$ & 2 & $1+X$ & $1+X^9, \quad 1+X^{19}$ \\[0.5ex]
\hline \noalign{\vskip 0.3ex}
% N = 51
51 & $[ \![102,2,9]]$ & 4 & $1+X$ & $1+X^8, \quad 1+X^9, \quad 1+X^{11}, \quad 1+X^{15}$ \\[0.5ex]
\hline \noalign{\vskip 0.3ex}
% N = 52
52 & $[ \![104,2,9]]$ & 3 & $1+X$ & $1+X^8, \quad 1+X^{20}, \quad 1+X^{22}$ \\[0.5ex]
\hline \noalign{\vskip 0.3ex}
% N = 53
53 & $[ \![106,2,9]]$ & 3 & $1+X$ & $1+X^8, \quad 1+X^{12}, \quad 1+X^{23}$ \\[0.5ex]
\hline \noalign{\vskip 0.3ex}
% N = 54
54 & $[ \![108,2,10]]$ & 1 & $1+X$ & $1+X^{15}$ \\[0.5ex]
\hline \noalign{\vskip 0.3ex}
% N = 55
55 & $[ \![110,2,10]]$ & 3 & $1+X$ & $1+X^{10}, \quad 1+X^{12}, \quad 1+X^{21}$ \\[0.5ex]
\hline \noalign{\vskip 0.3ex}
% N = 56
56 & $[ \![112,2,10]]$ & 1 & $1+X$ & $1+X^{10}$ \\[0.5ex]
\hline \noalign{\vskip 0.3ex}
% N = 57
57 & $[ \![114,2,10]]$ & 1 & $1+X$ & $1+X^{24}$ \\[0.5ex]
\hline \noalign{\vskip 0.3ex}
% N = 58
58 & $[ \![116,2,10]]$ & 4 & $1+X$ & $1+X^9, \quad 1+X^{16}, \quad 1+X^{17}, \quad 1+X^{22}$ \\[0.5ex]
\hline \noalign{\vskip 0.3ex}
% N = 59
59 & $[ \![118,2,10]]$ & 1 & $1+X$ & $1+X^9$ \cite{PWsimu} \\[0.5ex]
\hline \noalign{\vskip 0.3ex}
% N = 60
\multirow{5}{*}{60} & \multirow{5}{*}{$[ \![120,2,10]]$} & \multirow{5}{*}{6} & $1+X$ & $1+X^9, \quad 1+X^{11}, \quad 1+X^{13}, \quad 1+X^{25}$ \\[2.5ex]
    & & & $1+X^3$ & $1+X^8$ \\[2.5ex]
    & & & $1+X^5$ & $1+X^6$ \\[0.5ex]
\hline \noalign{\vskip 0.3ex}

% N = 61
{61} & {$[ \![122,2,11]]$} & {1} & {$1+X$} & $1+X^{11}$ (optimal surface code \cite{BD07}) \\[0.5ex]
\hline \noalign{\vskip 0.3ex}
% N = 62
62 & $[ \![124,2,11]]$ & 1 & $1+X$ & $1+X^{26}$ \\[0.5ex]
\hline \noalign{\vskip 0.3ex}
% N = 63
63 & $[ \![126,2,11]]$ & 1 & $1+X$ & $1+X^{24}$ \\[0.5ex]
\hline \noalign{\vskip 0.3ex}
% N = 64
64 & $[ \![128,2,11]]$ & 1 & $1+X$ & $1+X^{14}$ \\[0.5ex]
\hline \noalign{\vskip 0.3ex}
% N = 65
65 & $[ \![130,2,11]]$ & 2 & $1+X$ & $1+X^{10}, \quad 1+X^{18}$ \\[0.5ex]
\hline \noalign{\vskip 0.3ex}
% N = 66
\multirow{3}{*}{66} & \multirow{3}{*}{$[ \![132,2,11]]$} & \multirow{3}{*}{3} & $1+X$ & $1+X^{10}, \quad 1+X^{12}$ \\[2.5ex]
    & & & $1+X^2$ & $1+X^9$ \\[0.5ex]
\hline \noalign{\vskip 0.3ex}
% N = 67
67 & $[ \![134,2,11]]$ & 1 & $1+X$ & $1+X^{12}$ \cite{PWsimu} \\[0.5ex]
\hline \noalign{\vskip 0.3ex}
% N = 68
68 & $[ \![136,2,11]]$ & 2 & $1+X$ & $1+X^{20}, \quad 1+X^{26}$ \\[0.5ex]
\hline \noalign{\vskip 0.3ex}
% N = 69
69 & $[ \![138,2,11]]$ & 2 & $1+X$ & $1+X^{15}, \quad 1+X^{19}$ \\[0.5ex]
\hline \noalign{\vskip 0.3ex}
% N = 70
70 & $[ \![140,2,11]]$ & 1 & $1+X^2$ & $1+X^{25}$ \\[0.5ex]
\hline \noalign{\vskip 0.3ex}
% N = 71
71 & $[ \![142,2,11]]$ & 3 & $1+X$ & $1+X^{11}, \quad 1+X^{16}, \quad 1+X^{21}$ \\[0.5ex]
\hline \noalign{\vskip 0.3ex}
% N = 72
72 & $[ \![144,2,12]]$ & 1 & $1+X$ & $1+X^{11}$ \\[0.5ex]
\hline \noalign{\vskip 0.3ex}
% N = 73
73 & $[ \![146,2,11]]$ & 4 & $1+X$ & $1+X^{11}, \quad 1+X^{13}, \quad 1+X^{16}, \quad 1+X^{27}$ \\[0.5ex]
\hline \noalign{\vskip 0.3ex}
% N = 74
74 & $[ \![148,2,12]]$ & 1 & $1+X$ & $1+X^{31}$ \\[0.5ex]
\hline \noalign{\vskip 0.3ex}
% N = 75
\multirow{2}{*}{75}   & \multirow{2}{*}{$[ \![150,2,11]]$}  & \multirow{2}{*}{4}  & $1+X$ & $1+X^{53}, \quad 1+X^{10}, \quad 1+X^{33}$ \\[0.5ex]
&&& \\
 &  & & $1+X^5$ & $1+X^{36}$ \\[0.5ex]
\hline \noalign{\vskip 0.3ex}
% N = 76
76 & $[ \![152,2,12]]$ & 1 & $1+X$ & $1+X^{47}$ \\[0.5ex]
\hline \noalign{\vskip 0.3ex}
% N = 77
\multirow{2}{*}{77} & \multirow{2}{*}{$[ \![154,2,11]]$} & \multirow{2}{*}{5} & \multirow{2}{*}{$1+X$} & $1+X^{12}, \quad 1+X^{14}, \quad 1+X^{23},$ \\
&&& & $1+X^{43}, \quad 1+X^{56}$ \\[0.5ex]
\hline \noalign{\vskip 0.3ex}
% N = 78
\multirow{3}{*}{78} & \multirow{3}{*}{$[ \![156,2,12]]$} & \multirow{3}{*}{4} & $1+X$ & $1+X^{12}, \quad 1+X^{14}, \quad 1+X^{55}$ \\[2.5ex]
    & & & $1+X^3$ & $1+X^{62}$ \\[0.5ex]
\hline \noalign{\vskip 0.3ex}
% N = 79
79 & $[ \![158,2,12]]$ & 1 & $1+X$ & $1+X^{12}$ \\[0.5ex]
\hline \noalign{\vskip 0.3ex}
% N = 80
\multirow{3}{*}{80} & \multirow{3}{*}{$[ \![160,2,12]]$} & \multirow{3}{*}{3} & $1+X$ & $1+X^{45}, \quad 1+X^{58}$ \\[2.5ex]
    & & & $1+X^4$ & $1+X^5$ \\[0.5ex]
\hline \noalign{\vskip 0.3ex}
% N = 81
81 & $[ \![162,2,12]]$ & 2 & $1+X$ & $1+X^{24}, \quad 1+X^{31}$ \\[0.5ex]
\hline \noalign{\vskip 0.3ex}
% N = 82
82 & $[ \![164,2,12]]$ & 2 & $1+X$ & $1+X^{36}, \quad 1+X^{67}$ \\[0.5ex]
\hline \noalign{\vskip 0.3ex}
% N = 83
83 & $[ \![166,2,12]]$ & 2 & $1+X$ & $1+X^{11} $ \cite{PWsimu}, $\quad 1+X^{23}$ \\[0.5ex]
\hline \noalign{\vskip 0.3ex}
% N = 84
\multirow{5}{*}{84} & \multirow{5}{*}{$[ \![168,2,12]]$} & \multirow{5}{*}{8} & \multirow{2}{*}{$1+X$} & $1+X^{19}, \quad 1+X^{25}, \quad 1+X^{32}, \quad 1+X^{35}$ \\[0.5ex]
&&&& $1+X^{69}, \quad 1+X^{71}, \quad 1+X^{73}$ \\[0.5ex]
&&&& \\
    & & & $1+X^7$ & $1+X^{30}$ \\[0.5ex]
\hline \noalign{\vskip 0.3ex}
% N = 85
85 & $[ \![170,2,13]]$ & 1 & $1+X$ & $1+X^{13}$ (optimal surface code \cite{BD07}) \\[0.5ex]
\hline \noalign{\vskip 0.3ex}
% N = 86
86 & $[ \![172,2,12]]$ & 1 & $1+X$ & $1+X^{73}$ \\[0.5ex]
\hline \noalign{\vskip 0.3ex}
% N = 87
87 & $[ \![174,2,13]]$ & 1 & $1+X$ & $1+X^{24}$ \\[0.5ex]
\hline \noalign{\vskip 0.3ex}
% N = 88
88 & $[ \![176,2,13]]$ & 1 & $1+X$ & $1+X^{26}$ \\[0.5ex]
\hline \noalign{\vskip 0.3ex}
% N = 89
89 & $[ \![178,2,13]]$ & 2 & $1+X$ & $1+X^{16}, \quad 1+X^{55}$ \\[0.5ex]
\hline \noalign{\vskip 0.3ex}
% N = 90
90 & $[ \![180,2,13]]$ & 1 & $1+X$ & $1+X^{78}$ \\[0.5ex]
\hline \noalign{\vskip 0.3ex}
% N = 91
91 & $[ \![182,2,13]]$ & 4 & $1+X$ & $1+X^{12}, \quad 1+X^{14}, \quad 1+X^{27}, \quad 1+X^{40}$ \\[0.5ex]
\hline \noalign{\vskip 0.3ex}
% N = 92
92 & $[ \![184,2,13]]$ & 2 & $1+X$ & $1+X^{14}, \quad 1+X^{72}$ \\[0.5ex]
\hline \noalign{\vskip 0.3ex}
% N = 93
93 & $[ \![186,2,13]]$ & 1 & $1+X$ & $1+X^{54}$ \\[0.5ex]
\hline \noalign{\vskip 0.3ex}
% N = 94
94 & $[ \![188,2,13]]$ & 2 & $1+X$ & $1+X^{26}, \quad 1+X^{36}$ \\[0.5ex]
\hline \noalign{\vskip 0.3ex}
% N = 95
95 & $[ \![190,2,13]]$ & 2 & $1+X$ & $1+X^{28}, \quad 1+X^{35}$ \\[0.5ex]
\hline \noalign{\vskip 0.3ex}
% N = 96
\multirow{3}{*}{96} & \multirow{3}{*}{$[ \![192,2,13]]$} & \multirow{3}{*}{2} & $1+X$ & $1+X^{42}$ \\[2.5ex]
    & & & $1+X^4$ & $1+X^{57}$ \\[0.5ex]
\hline \noalign{\vskip 0.3ex}
% N = 97
97 & $[ \![194,2,13]]$ & 3 & $1+X$ & $1+X^{15}, \quad 1+X^{21}, \quad 1+X^{22}$ \\[0.5ex]
\hline \noalign{\vskip 0.3ex}
% N = 98
98 & $[ \![196,2,14]]$ & 2 & $1+X$ & $1+X^{41}, \quad 1+X^{69}$ \\[0.5ex]
\hline \noalign{\vskip 0.3ex}
% N = 99
99 & $[ \![198,2,13]]$ & 2 & $1+X$ & $1+X^{15}, \quad 1+X^{61}$ \\[0.5ex]
\hline \noalign{\vskip 0.3ex}
% N = 100
\multirow{3}{*}{100} & \multirow{3}{*}{$[ \![200,2,13]]$} & \multirow{3}{*}{5} & $1+X$ & $1+X^{18}, \quad 1+X^{42}, \quad 1+X^{44}$ \\[2.5ex]
    & & & $1+X^5$ & $1+X^{68}, \quad 1+X^{72}$
\label{table: classification GB(2,2) extremaux}
\end{longtable}

\subsection{Classification of all (2,2)-GB Codes (Lengths up to 200)}

For a complete classification of all (2,2)-GB codes —specifically those of the form $GB(1 + X^a, 1 + X^b, n)$ with $\gcd(a, b, n) = 1$— along with the computational code used for this research,  we refer the reader to our GitHub repository, where the complete classification is available: \url{https://github.com/NicolasSaussay/weight-4_GB-Codes_Classification} \cite{SaussayClassification}. 

\section{Summary of Top-Performing 2D Weight-4 Surface Codes}
\label{sec: summary of TP 2D weight-4 surface codes}

We also provide a concise summary table, designed to quickly inform readers about the optimal parameters $[[2N,2,d]]$ achievable with 2D (lattice-based) weight-4 surface codes for lengths under 200.

Contrary to the previous table (Table \ref{table: classification GB(2,2) extremaux}), Table  \ref{tab: summary of optimal weight-4 surface codes} compiles data on optimal weight-4 2D surface codes from diverse sources, including non (2,2)-GB codes. Specifically, these sources encompass the standard Kitaev code, optimal 2D weight-4 surface codes  (detailed in Section \ref{sec: optimal surface codes}), the Pryadko and Wang's Github repository \cite{PWsimu}, our own two infinite families of GB codes (constructed in Section \ref{section: application of the main theorem}), and the broader set of, previously unreported, GB codes catalogued in Table \ref{table: classification GB(2,2) extremaux}.

\begin{longtable}{|c|c|c|c|}
\caption{Summary of Optimal Weight-4 2D Surface Code Parameters for Lengths N from 2 to 100}
\label{tab: summary of optimal weight-4 surface codes} \\
\toprule
\textbf{N} & \thead{\textbf{Optimal Parameters} \\ $[ \![2N, 2, d]]$} & \thead{ \textbf{Number of} \\ \textbf{Non-CGP-Equivalent} \\ \textbf{Surface Codes} } & \thead{\textbf{References to} \\ \textbf{Construction Codes}} \\
\midrule
\endfirsthead
\toprule
\textbf{N} & \thead{\textbf{Optimal Parameters} \\ $[ \![2N, 2, d]]$} & \thead{ \textbf{Number of} \\ \textbf{Non-CGP-Equivalent} \\ \textbf{Surface Codes} } & \thead{\textbf{References to} \\ \textbf{Construction Codes}} \\
\midrule
\endhead
\bottomrule
\endfoot
\bottomrule
\endlastfoot

% N = 2
{2} &  {$[[4,2,2]]$} & {1} & Optimal surface code \cite{BD07} \\
\hline \noalign{\vskip 0.2ex}
% N = 3
3 & $[[6,2,2]]$ & 1 & Table \ref{table: classification GB(2,2) extremaux}\\
\hline \noalign{\vskip 0.2ex}
% N = 4
\multirow{2}{*}{4} & \multirow{2}{*}{$[[8,2,2]]$} & \multirow{2}{*}{2} & Kitaev code\\
&&& Table \ref{table: classification GB(2,2) extremaux} \\
\hline \noalign{\vskip 0.2ex}
% N = 5
{5} & {$[[10,2,3]]$} & {1} & Optimal surface code \cite{BD07} \\
\hline \noalign{\vskip 0.2ex}
% N = 6
6 & $[[12,2,3]]$ & 1 & Table \ref{table: classification GB(2,2) extremaux}\\
\hline \noalign{\vskip 0.2ex}
% N = 7
7 & $[[14,2,3]]$ & 1 & Table \ref{table: classification GB(2,2) extremaux}\\
\hline \noalign{\vskip 0.2ex}
% N = 8
\multirow{2}{*}{8}  & \multirow{2}{*}{$[[16,2,4]]$}   &  \multirow{2}{*}{2}  & Optimal surface code \cite{BD07} \\ 
&&& Our Code (see Section \ref{sec: Study of GB(1 + X, 1 + X^{2r - 1}, 2r^2) minimum distance})\\
\hline \noalign{\vskip 0.2ex}
% N = 9
\multirow{2}{*}{9} & \multirow{2}{*}{$[[18,2,3]]$} & \multirow{2}{*}{3} & Standard Kitaev code \\
&&& Table \ref{table: classification GB(2,2) extremaux}\\
\hline \noalign{\vskip 0.2ex}
% N = 10
10 & $[[20,2,4]]$ & 2 & Table \ref{table: classification GB(2,2) extremaux}\\
\hline \noalign{\vskip 0.2ex}
% N = 11
\multirow{2}{*}{11} & \multirow{2}{*}{$[[22,2,4]]$} & \multirow{2}{*}{1} & Pryadko and Wang repository \cite{PWsimu} \\
&&& (and Table \ref{table: classification GB(2,2) extremaux})\\
\hline \noalign{\vskip 0.2ex}
% N = 12
12 & $[[24,2,4]]$ & 3 & Table \ref{table: classification GB(2,2) extremaux}\\
\hline \noalign{\vskip 0.2ex}
% N = 13
{13}   &  {$[[26,2,5]]$}  & {1}  & Optimal surface code \cite{BD07}\\
\hline \noalign{\vskip 0.2ex}
% N = 14
14 & $[[28,2,5]]$ & 1 & Table \ref{table: classification GB(2,2) extremaux}\\
\hline \noalign{\vskip 0.2ex}
% N = 15
15 & $[[30,2,5]]$ & 2 & Table \ref{table: classification GB(2,2) extremaux}\\
\hline \noalign{\vskip 0.2ex}
% N = 16
16 & $[[32,2,5]]$ & 1 & Table \ref{table: classification GB(2,2) extremaux}\\
\hline \noalign{\vskip 0.2ex}
% N = 17
17 & $[[34,2,5]]$ & 2 & Table \ref{table: classification GB(2,2) extremaux}\\
\hline \noalign{\vskip 0.2ex}
% N = 18
\multirow{2}{*}{18}   & \multirow{2}{*}{$[[36,2,6]]$}    & \multirow{3}{*}{2}  & Optimal surface code \cite{BD07} \\
&&& Our Code (see Section \ref{sec: Study of GB(1 + X, 1 + X^{2r - 1}, 2r^2) minimum distance})\\
\hline \noalign{\vskip 0.2ex}
% N = 19
\multirow{2}{*}{19} & \multirow{2}{*}{$[[38,2,5]]$} & \multirow{2}{*}{2} & Pryadko and Wang repository \cite{PWsimu} \\
&&& Table \ref{table: classification GB(2,2) extremaux}\\
\hline \noalign{\vskip 0.2ex}
% N = 20
20 & $[[40,2,5]]$ & 3 & Table \ref{table: classification GB(2,2) extremaux}\\
\hline \noalign{\vskip 0.2ex}
% N = 21
21 & $[[42,2,6]]$ & 2 & Table \ref{table: classification GB(2,2) extremaux}\\
\hline \noalign{\vskip 0.2ex}
% N = 22
22 & $[[44,2,6]]$ & 2 & Table \ref{table: classification GB(2,2) extremaux}\\
\hline \noalign{\vskip 0.2ex}
% N = 23
23 & $[[46,2,6]]$ & 1 & Table \ref{table: classification GB(2,2) extremaux}\\
\hline \noalign{\vskip 0.2ex}
% N = 24
24 & $[[48,2,6]]$ & 5 & Table \ref{table: classification GB(2,2) extremaux}\\
\hline \noalign{\vskip 0.2ex}
% N = 25
{25} &  {$[[50,2,7]]$} & {1} & Optimal surface code \cite{BD07}\\
\hline \noalign{\vskip 0.2ex}
% N = 26
26 & $[[52,2,7]]$ & 1 & Table \ref{table: classification GB(2,2) extremaux}\\
\hline \noalign{\vskip 0.2ex}
% N = 27
27 & $[[54,2,7]]$ & 1 & Table \ref{table: classification GB(2,2) extremaux}\\
\hline \noalign{\vskip 0.2ex}
% N = 28
28 & $[[56,2,7]]$ & 2 & Table \ref{table: classification GB(2,2) extremaux}\\
\hline \noalign{\vskip 0.2ex}
% N = 29
29 & $[[58,2,7]]$ & 2 & \cite{PWsimu}, Table \ref{table: classification GB(2,2) extremaux}\\
\hline \noalign{\vskip 0.2ex}
% N = 30
30 & $[[60,2,7]]$ & 1 & Table \ref{table: classification GB(2,2) extremaux}\\
\hline \noalign{\vskip 0.2ex}
% N = 31
31 & $[[62,2,7]]$ & 2 & Table \ref{table: classification GB(2,2) extremaux}\\
\hline \noalign{\vskip 0.2ex}
% N = 32
\multirow{2}{*}{32} & \multirow{2}{*}{$[[64,2,8]]$} & \multirow{3}{*}{2} & Optimal surface code \cite{BD07} \\
&&& Our Code (see Section \ref{sec: Study of GB(1 + X, 1 + X^{2r - 1}, 2r^2) minimum distance})\\
\hline \noalign{\vskip 0.2ex}
% N = 33
33 & $[[66,2,7]]$ & 3 & Table \ref{table: classification GB(2,2) extremaux}\\
\hline \noalign{\vskip 0.2ex}
% N = 34
34 & $[[68,2,8]]$ & 1 & Table \ref{table: classification GB(2,2) extremaux}\\
\hline \noalign{\vskip 0.2ex}
% N = 35
35 & $[[70,2,7]]$ & 4 & Table \ref{table: classification GB(2,2) extremaux}\\
\hline \noalign{\vskip 0.2ex}
% N = 36
36 & $[[72,2,8]]$ & 3 & Table \ref{table: classification GB(2,2) extremaux}\\
\hline \noalign{\vskip 0.2ex}
% N = 37
37 & $[[74,2,8]]$ & 1 & Pryadko and Wang repository \cite{PWsimu} \\
&&& (and Table \ref{table: classification GB(2,2) extremaux})\\
\hline \noalign{\vskip 0.2ex}
% N = 38
38 & $[[76,2,8]]$ & 2 & Table \ref{table: classification GB(2,2) extremaux}\\
\hline \noalign{\vskip 0.2ex}
% N = 39
39 & $[[78,2,8]]$ & 2 & Table \ref{table: classification GB(2,2) extremaux}\\
\hline \noalign{\vskip 0.2ex}
% N = 40
40 & $[[80,2,8]]$ & 5 & Table \ref{table: classification GB(2,2) extremaux}\\
\hline \noalign{\vskip 0.2ex}
% N = 41
{41} & {$[[82,2,9]]$} & {1} & Optimal surface code \cite{BD07}\\
\hline \noalign{\vskip 0.2ex}
% N = 42
42 & $[[84,2,9]]$ & 1 & Table \ref{table: classification GB(2,2) extremaux}\\
\hline \noalign{\vskip 0.2ex}
% N = 43
43 & $[[86,2,9]]$ & 1 & Table \ref{table: classification GB(2,2) extremaux}\\
\hline \noalign{\vskip 0.2ex}
% N = 44
44 & $[[88,2,9]]$ & 1 & Table \ref{table: classification GB(2,2) extremaux}\\
\hline \noalign{\vskip 0.2ex}
% N = 45
45 & $[[90,2,9]]$ & 3 & Table \ref{table: classification GB(2,2) extremaux}\\
\hline \noalign{\vskip 0.2ex}
% N = 46
46 & $[[92,2,9]]$ & 1 & Table \ref{table: classification GB(2,2) extremaux}\\
\hline \noalign{\vskip 0.2ex}
% N = 47
47 & $[[94,2,9]]$ & 1 & Table \ref{table: classification GB(2,2) extremaux}\\
\hline \noalign{\vskip 0.2ex}
% N = 48
48 & $[[96,2,9]]$ & 3 & Table \ref{table: classification GB(2,2) extremaux}\\
\hline \noalign{\vskip 0.2ex}
% N = 49
49 & $[[98,2,9]]$ & 1 & Table \ref{table: classification GB(2,2) extremaux}\\
\hline \noalign{\vskip 0.2ex}
% N = 50
\multirow{3}{*}{50} & \multirow{3}{*}{$[[100,2,10]]$} & \multirow{3}{*}{3} & Table \ref{table: classification GB(2,2) extremaux} \\
&&&  Optimal surface code \cite{BD07}\\
&&&  Our Code (see Section \ref{sec: Study of GB(1 + X, 1 + X^{2r - 1}, 2r^2) minimum distance}) \\
\hline \noalign{\vskip 0.2ex}
% N = 51
51 & $[[102,2,9]]$ & 4 & Table \ref{table: classification GB(2,2) extremaux}\\
\hline \noalign{\vskip 0.2ex}
% N = 52
52 & $[[104,2,9]]$ & 3 & Table \ref{table: classification GB(2,2) extremaux}\\
\hline \noalign{\vskip 0.2ex}
% N = 53
53 & $[[106,2,9]]$ & 3 & Pryadko and Wang repository \cite{PWsimu} \\
&&& Table \ref{table: classification GB(2,2) extremaux}\\
\hline \noalign{\vskip 0.2ex}
% N = 54
54 & $[[108,2,10]]$ & 1 & Table \ref{table: classification GB(2,2) extremaux}\\
\hline \noalign{\vskip 0.2ex}
% N = 55
55 & $[[110,2,10]]$ & 3 & Table \ref{table: classification GB(2,2) extremaux}\\
\hline \noalign{\vskip 0.2ex}
% N = 56
56 & $[[112,2,10]]$ & 1 & Table \ref{table: classification GB(2,2) extremaux}\\
\hline \noalign{\vskip 0.2ex}
% N = 57
57 & $[[114,2,10]]$ & 1 & Table \ref{table: classification GB(2,2) extremaux}\\
\hline \noalign{\vskip 0.2ex}
% N = 58
58 & $[[116,2,10]]$ & 4 & Table \ref{table: classification GB(2,2) extremaux}\\
\hline \noalign{\vskip 0.2ex}
% N = 59
59 & $[[118,2,10]]$ & 1 & Pryadko and Wang repository \cite{PWsimu} \\ 
\hline \noalign{\vskip 0.2ex}
% N = 60
60 & $[[120,2,10]]$ & 6 & Table \ref{table: classification GB(2,2) extremaux}\\
\hline \noalign{\vskip 0.2ex}
% N = 61
{61} & {$[[122,2,11]]$} & {1} & Optimal surface code \cite{BD07}\\
\hline \noalign{\vskip 0.2ex}
% N = 62
62 & $[[124,2,11]]$ & 1 & Table \ref{table: classification GB(2,2) extremaux}\\
\hline \noalign{\vskip 0.2ex}
% N = 63
63 & $[[126,2,11]]$ & 1 & Table \ref{table: classification GB(2,2) extremaux}\\
\hline \noalign{\vskip 0.2ex}

% N = 64
64 & $[[128,2,11]]$ & 1 & Table \ref{table: classification GB(2,2) extremaux}\\
\hline \noalign{\vskip 0.2ex}
% N = 65
65 & $[[130,2,11]]$ & 2 & Table \ref{table: classification GB(2,2) extremaux}\\
\hline \noalign{\vskip 0.2ex}
% N = 66
66 & $[[132,2,11]]$ & 3 & Table \ref{table: classification GB(2,2) extremaux}\\
\hline \noalign{\vskip 0.2ex}
% N = 67
\multirow{2}{*}{67} & \multirow{2}{*}{$[[134,2,11]]$} & \multirow{2}{*}{1} & Pryadko and Wang repository \cite{PWsimu} \\
&&& (and Table \ref{table: classification GB(2,2) extremaux})\\
\hline \noalign{\vskip 0.2ex}
% N = 68
68 & $[[136,2,11]]$ & 2 & Table \ref{table: classification GB(2,2) extremaux}\\
\hline \noalign{\vskip 0.2ex}
% N = 69
69 & $[[138,2,11]]$ & 2 & Table \ref{table: classification GB(2,2) extremaux}\\
\hline \noalign{\vskip 0.2ex}
% N = 70
70 & $[[140,2,11]]$ & 1 & Table \ref{table: classification GB(2,2) extremaux}\\
\hline \noalign{\vskip 0.2ex}
% N = 71
71 & $[[142,2,11]]$ & 3 & Table \ref{table: classification GB(2,2) extremaux}\\
\hline \noalign{\vskip 0.2ex}
% N = 72
\multirow{2}{*}{72} & \multirow{2}{*}{$[[144,2,12]]$} & \multirow{3}{*}{2} & Optimal surface code \cite{BD07} \\
&&& Our Code (see Section \ref{sec: Study of GB(1 + X, 1 + X^{2r - 1}, 2r^2) minimum distance})\\
\hline \noalign{\vskip 0.2ex}
% N = 73
73 & $[[146,2,11]]$ & 4 & Table \ref{table: classification GB(2,2) extremaux}\\
\hline \noalign{\vskip 0.2ex}
% N = 74
74 & $[[148,2,12]]$ & 1 & Table \ref{table: classification GB(2,2) extremaux}\\
\hline \noalign{\vskip 0.2ex}
% N = 75
75 & $[[150,2,11]]$ & 4 & Table \ref{table: classification GB(2,2) extremaux}\\
\hline \noalign{\vskip 0.2ex}
% N = 76
76 & $[[152,2,12]]$ & 1 & Table \ref{table: classification GB(2,2) extremaux}\\
\hline \noalign{\vskip 0.2ex}
% N = 77
77 & $[[154,2,11]]$ & 5 & Table \ref{table: classification GB(2,2) extremaux}\\
\hline \noalign{\vskip 0.2ex}
% N = 78
78 & $[[156,2,12]]$ & 4 & Table \ref{table: classification GB(2,2) extremaux}\\
\hline \noalign{\vskip 0.2ex}
% N = 79
79 & $[[158,2,12]]$ & 1 & Table \ref{table: classification GB(2,2) extremaux}\\
\hline \noalign{\vskip 0.2ex}
% N = 80
80 & $[[160,2,12]]$ & 2 & Table \ref{table: classification GB(2,2) extremaux}\\
\hline \noalign{\vskip 0.2ex}
% N = 81
81 & $[[162,2,12]]$ & 2 & Table \ref{table: classification GB(2,2) extremaux}\\
\hline \noalign{\vskip 0.2ex}
% N = 82
82 & $[[164,2,12]]$ & 2 & Table \ref{table: classification GB(2,2) extremaux}\\
\hline \noalign{\vskip 0.2ex}
% N = 83
\multirow{2}{*}{83} & \multirow{2}{*}{$[[166,2,12]]$} & \multirow{2}{*}{2} & Pryadko and Wang repository \cite{PWsimu} \\
&&& Table \ref{table: classification GB(2,2) extremaux}\\
\hline \noalign{\vskip 0.2ex}
% N = 84
84 & $[[168,2,12]]$ & 8 & Table \ref{table: classification GB(2,2) extremaux}\\
\hline \noalign{\vskip 0.2ex}
% N = 85
{85} & {$[[170,2,13]]$} & {1} & Optimal surface code \cite{BD07} \\
\hline \noalign{\vskip 0.2ex}
% N = 86
86 & $[[172,2,12]]$ & 1 & Table \ref{table: classification GB(2,2) extremaux}\\
\hline \noalign{\vskip 0.2ex}
% N = 87
87 & $[[174,2,13]]$ & 1 & Table \ref{table: classification GB(2,2) extremaux}\\
\hline \noalign{\vskip 0.2ex}
% N = 88
88 & $[[176,2,13]]$ & 1 & Table \ref{table: classification GB(2,2) extremaux}\\
\hline \noalign{\vskip 0.2ex}
% N = 89
89 & $[[178,2,13]]$ & 2 & Table \ref{table: classification GB(2,2) extremaux}\\
\hline \noalign{\vskip 0.2ex}
% N = 90
90 & $[[180,2,13]]$ & 1 & Table \ref{table: classification GB(2,2) extremaux}\\
\hline \noalign{\vskip 0.2ex}
% N = 91
91 & $[[182,2,13]]$ & 4 & Table \ref{table: classification GB(2,2) extremaux}\\
\hline \noalign{\vskip 0.2ex}
% N = 92
92 & $[[184,2,13]]$ & 2 & Table \ref{table: classification GB(2,2) extremaux}\\
\hline \noalign{\vskip 0.2ex}
% N = 93
93 & $[[186,2,13]]$ & 1 & Table \ref{table: classification GB(2,2) extremaux}\\
\hline \noalign{\vskip 0.2ex}
% N = 94
94 & $[[188,2,13]]$ & 2 & Table \ref{table: classification GB(2,2) extremaux}\\
\hline \noalign{\vskip 0.2ex}
% N = 95
95 & $[[190,2,13]]$ & 2 & Table \ref{table: classification GB(2,2) extremaux}\\
\hline \noalign{\vskip 0.2ex}
% N = 96
96 & $[[192,2,13]]$ & 2 & Table \ref{table: classification GB(2,2) extremaux}\\
\hline \noalign{\vskip 0.2ex}
% N = 97
97 & $[[194,2,13]]$ & 3 & Table \ref{table: classification GB(2,2) extremaux}\\
\hline \noalign{\vskip 0.2ex}
% N = 98
\multirow{2}{*}{$98$} & \multirow{2}{*}{$[[196,2,14]]$} & \multirow{2}{*}{4} & Optimal surface code \cite{BD07} \\
&&& Table \ref{table: classification GB(2,2) extremaux}\\
\hline \noalign{\vskip 0.2ex}
% N = 99
99 & $[[198,2,13]]$ & 2 & Table \ref{table: classification GB(2,2) extremaux}\\
\hline \noalign{\vskip 0.2ex}
% N = 100
100 & $[[200,2,13]]$ & 4 &  Table \ref{table: classification GB(2,2) extremaux} \\
\end{longtable}

\section*{Acknowledgment}
All authors acknowledge support from the Plan France 2030 through the NISQ2LSQ project (ANR-22-PETQ-0006). We also thank Gilles Zémor for his insightful discussions.

\section{Conclusion}
\label{sec:conclusion}

In this work, we significantly advance the understanding and practical construction of (2,2)-GB codes for quantum error correction. Our primary contribution is the establishment of a lower bound on the minimum distance, directly tied to the shortest Manhattan norm of non-zero vectors in associated $\mathbb{Z}^2$-sublattices. This bound proved to be an indispensable theoretical tool, systematically guiding the explicit construction of high-performing codes, a feat that would be challenging, if not impossible, without it.  

Leveraging this bound, we successfully constructed three novel infinite families of optimal (2,2)-GB codes: $ [[ 2n^2, 2, n ]] $, $ [[ 4r^2, 2, 2r ]] $, and $ [[ 2(t^2 + (t + 1)^2), 2, 2t + 1 ]] $. Notably, the family with parameters $ [[ 4r^2, 2, 2r ]] $ fills a previously perceived gap, as these optimal even-distance codes were widely thought to have no GB code equivalent. Through a specialized CSS code equivalence relation (CGP-equivalence) tailored to the inherent structure of Cayley-graph based CSS codes, we rigorously demonstrated that two of our families ($ [[ 2n^2, 2, n ]] $ and $ [[ 4r^2, 2, 2r ]] $) are indeed non-CGP-equivalent to established Kitaev and optimal even-distance 2D surface codes. Our third family, $ [[ 2(t^2 + (t + 1)^2), 2, 2t + 1 ]] $, was shown to be equivalent (in the standard quantum code sense) to the optimal odd-distance 2D surface code, offering an alternative construction.

Finally, this work delivers a comprehensive classification of all top-performing non-equivalent (2,2)-GB codes with lengths under 200 and a detailed comparison table with existing notable 2D weight-4 surface codes. For a deeper dive into the full classification of (2,2)-GB, our GitHub repository \cite{SaussayClassification} offers full access to the comprehensive classification of (2,2)-GB codes with lengths under 200.

This culminating effort provides a rich dataset for future practical quantum code implementations, and its structured nature makes it a crucial resource for researchers seeking codes with excellent decoding characteristics. \\

This research opens several promising avenues for future investigation. A direct extension involves classifying GB and broader CSS codes under larger equivalence relations to further validate distinctness and explore links to decoding algorithms. We also envision adapting our graph-theoretic methodology to construct superior GB codes with higher-weight generators, potentially by extending our framework to hypergraphs or more complex unoriented Cayley graph.

\end{document}